\documentclass[sigconf]{acmart}

\AtBeginDocument{%
  \providecommand\BibTeX{{
    \normalfont B\kern-0.5em{\scshape i\kern-0.25em b}\kern-0.8em\TeX}}}

\renewcommand\footnotetextcopyrightpermission[1]{}

\usepackage{balance}
\usepackage{xspace}
\usepackage[T1]{fontenc}
\usepackage{xspace}
\usepackage{listings}
\usepackage{subfigure}
\usepackage{mdwlist}
\usepackage{graphicx}
\usepackage{gensymb}
\usepackage{enumitem}[inline]

\usepackage{mathtools}
\DeclarePairedDelimiter\floor{\lfloor}{\rfloor}

\usepackage[ruled,vlined,linesnumbered,commentsnumbered,noend]{algorithm2e}

\lstdefinelanguage{persistent}
{
 morekeywords={PERSISTENT, unsigned, int},
 morecomment=[s]{/*}{*/},
 sensitive=false,
 escapeinside={*}{*}
}

\newlist{mylist}{enumerate*}{1}
\setlist[mylist]{label=(\roman*)}

\usepackage{makecell}

\newcommand{\revision}[1]{\textcolor{black}{#1}}

\newcommand\name{\textsc{ALFRED}\xspace}
\newcommand\sceptic{\textsc{ScEpTIC}\xspace}

\newcommand{\baselineV}{\textsc{Volatile}\xspace}
\newcommand{\baselineNV}{\textsc{NonVolatile}\xspace}
\newcommand{\trigger}{\textsc{Probe}\xspace}
\newcommand{\execute}{\textsc{Execute}\xspace}
\newcommand{\latch}{\textsc{Loop-Latch}\xspace}
\newcommand{\freturn}{\textsc{Function-Return}\xspace}
\newcommand{\idempotent}{\textsc{IdempotentBoundaries}\xspace}

\newcommand{\fakepar}[1]{\vspace{0.5mm}\noindent\textbf{#1.}}
\newcommand\figref[1]{Fig.\,\ref{#1}}
\newcommand\secref[1]{Sec.\,\ref{#1}} \renewcommand\eqref[1]{Eq.\,\ref{#1}}
\newcommand\step[1]{$\langle$#1$\rangle$}

\newcommand{\capt}[1]{\mdseries{\emph{#1}}}
\newcommand{\code}[1]{\texttt{\textbf{#1}}}

\usepackage{titlesec}
\titleformat{\subsubsection}{\normalfont\large\bfseries}{\thesubsubsection}{1em}{}

\author{Andrea Maioli}
\affiliation{%
    \institution{Politecnico di Milano, Italy}
}
\email{andrea1.maioli@polimi.it}

\author{Luca Mottola}
\affiliation{%
    \institution{Politecnico di Milano, Italy and RI.SE Computer Science and Uppsala University, Sweden}
}
\email{luca.mottola@polimi.it}
\begin{document}

\title{\name: Virtual Memory for Intermittent Computing}

\begin{abstract}
  We present \name: a virtual memory abstraction that resolves the dichotomy between volatile and non-volatile memory in intermittent computing.
  Mixed-volatile microcontrollers allow programmers to allocate part of the application state onto non-volatile main memory.
  Programmers are therefore to explore manually the trade-off between simpler management of persistent state against the energy overhead for non-volatile memory operations and intermittence anomalies due to re-execution of non-idempotent code.
  This approach is laborious and yields sub-optimal performance.
  We take a different stand with \name: we provide programmers with a virtual memory abstraction detached from the specific volatile nature of memory and automatically determine an efficient mapping from virtual to volatile or non-volatile memory.
  Unlike existing works, \name does not require programmers to learn a new programming model or language syntax, while the mapping is entirely resolved at compile-time, reducing the run-time energy overhead.
  We implement \name through a series of program machine-level code transformations.
  Compared to existing systems, we demonstrate that \name reduces energy consumption by up to \emph{two orders of magnitude} given a fixed workload.  
  This enables the workloads to finish sooner, as the use of available energy shifts from ensuring forward progress to useful application processing.
\end{abstract}

\maketitle
\pagestyle{plain}

\section{Introduction}
\label{sec:introduction}

Ambient energy harvesting~\cite{harvesting-survey} enables deployments of battery-less sensing devices~\cite{batteryless-future, pible, soil-termoelectric, sensys20deployment, bridge-sensor, traffic-flow-sensor}, reducing environment impact and maintenance costs.
However, harvested energy is erratic and generally not sufficient to power devices continuously. 
Devices thus experience frequent power failures that cause executions to become \textit{intermittent}, with periods of active operation interleaved by periods where the device is off recharging its energy buffers. 
Power failures cause devices to shut down and lose the program state, restarting all over again when energy is newly available.
Forward progress of programs is therefore compromised.

\fakepar{Problem} Several systems exist to ensure forward progress, as we discuss in \secref{sec:background}.
Common to these solutions is the insertion of state-saving operations within the execution flow.
A state-saving operation offers the opportunity to create a replica of the program state, including main memory, register files, and program counter, onto non-volatile memory.
The program state is eventually restored from non-volatile memory when power returns, ensuring forward progress across power failures.
The placement of state-saving operations in the program may be either decided in a (semi-)automatic fashion~\cite{Mementos, Hibernus, Hibernus++, HarvOS, QuickRecall, chinchilla, ratchet} or implicitly determined by programmers with custom programming abstractions~\cite{alpaca, chain, DINO, Coati, ink, coala}. 

Mixed-volatile microcontroller platforms also exist, which offer the ability to store slices of the program state directly to non-volatile memory.
This is achieved using specific pragma statements when declaring a variable, as in~\cite{msp430fr5969}:
\begin{lstlisting}[language=persistent, aboveskip=8pt, belowskip=5pt]
#pragma PERSISTENT(x) 
unsigned int x = 5; 
\end{lstlisting}
Program state allocated on non-volatile memory is automatically retained across power failures and may be excluded from state-saving operations, simplifying the management of persistent state.

Using mixed-volatile microcontroller platforms comes at the cost of increased energy consumption: non-volatile memory operations may require up to $247$\% the energy of their volatile counterpart~\cite{msp430fr5969, maioli20enssys}.
Storing only parts of the program state on non-volatile memory may also yield intermittence anomalies~\cite{brokenTM,maioli21ewsn}, which require further energy to be corrected.
These occur when non-idempotent code is re-executed that manipulates variables on non-volatile memory, producing results that are different than a continuous execution. 
Using mixed-volatile platforms, quantifying the advantages in simpler management of persistent state  against the corresponding energy overhead is complex, as these depend on factors including energy patterns and execution flow.

\fakepar{\name} We take a different stand.
Rather than requiring programmers to manually determine when to use non-volatile memory for what slice of the program state, we promote a higher-level of abstraction through a concept of \emph{virtual memory}.
Programmers write intermittent code without explicitly mapping variables to volatile or non-volatile memory.
Given a placement of state-saving operations in the code, we \emph{automatically} decide \emph{what} slice of the program state must be allocated onto non-volatile memory, and \emph{at what point} in the execution.
Programmers are therefore relieved from deciding the mapping between program state and memory.
Moreover, the mapping is not fixed at variable level, but is \emph{automatically adjusted} at different places in the code for the same data item, based on read/write patterns and program structure.

\name\footnote{\textbf{A}utomatic a\textbf{L}location o\textbf{F} non-volatile memo\textbf{R}y for transi\textbf{E}ntly-powered \textbf{D}evices.} is our implementation of virtual memory for intermittent computing, based on two key features:
\begin{enumerate*}
\item it is \emph{transparent to programmers}: no dedicated syntax is to be learned, and programmers write code in the familiar sequential manner without the need to tag variables.
\item the mapping from virtual to volatile or non-volatile memory is entirely \emph{resolved at compile-time}, reducing the energy overhead that represents the cost of using the abstraction.
\end{enumerate*}

\revision{The virtual memory abstraction we conceive  does not provide virtualization in the same sense as mainstream OSes.}
\revision{Instead, it offers an abstraction that shields programmers from the need to statically determine a specific memory allocation schema.}
\name is therefore sharply different compared to mainstream virtual memory systems~\cite{vmem1,vmem2}.
These usually provide an idealized abstraction of storage resources, so that software processes operate as if they had access to a contiguous memory area, possibly even larger than the one physically available.
Address translation hardware maps virtual addresses to physical addresses at run-time.
In \name, we target resource-constrained energy-harvesting devices that compute intermittently~\cite{batteryless-future}.
The abstraction we offer provides programmers with a higher-level view on the persistency properties of different memory areas, and automatically determines the mapping from the virtual memory to the volatile or non-volatile one.
Because of resource constraints, we determine this mapping at compile-time.

\name determines this mapping using three key program transformation techniques, illustrated in \secref{sec:memory-principles}.
Their ultimate goal is simple, yet challenging to achieve, especially at compile-time:
\begin{center}
  \emph{Use the energy-efficient volatile memory as much as possible, \\while enabling forward progress using non-volatile memory\\
    with reduced energy consumption compared to existing solutions.} 
\end{center}
This entails that we need to promote the use of volatile memory whenever convenient, for example, to compute intermediate results or to store temporary data that need not survive a power failure, while reallocating the data that does require to be persistent onto non-volatile memory in anticipation of a possible power failure.
By doing so, we decrease energy consumption by taking the best of both worlds: we benefit from the lower energy consumption of volatile memory whenever possible, and rely on the persistency features of non-volatile memory whenever required.

Applying program transformations at compile-time is, however, challenging because of the lack of run-time information.
\secref{sec:uncertainty} illustrates how we address the uncertainty that may correspondingly arises, using a set of dedicated program normalization passes.
The result of the transformations require a specific memory layout to operate correctly and a solution to the intermittence anomalies that possibly arise.
We describe in \secref{sec:impl} how we deal with these issues, using an approach that is co-designed with our program transformation techniques.

We build a prototype implementation of \name based on \sceptic~\cite{maioli21ewsn, sceptic-repo}, an extensible open-source emulation environment for intermittent programs.
Given fixed workloads generated from staple benchmarks in the field~\cite{ratchet, clank, Hibernus, Hibernus++, Mementos, QuickRecall, maioli21ewsn}, we measure \name performance in energy consumption, number of clock cycles, memory accesses, and restore operations.
We compare this performance against several baselines obtained by abstracting the key design dimensions of existing systems in a framework that allows us to instantiate baselines that correspond to existing works, while retaining the freedom to explore alternative configurations.
We find that, depending on the benchmark, \name can provide \emph{several-fold improvements} in energy consumption, which allow the system to shift the energy budget to useful application computations.
This ultimately allows the system to achieve similar improvements in the time to complete the fixed workload.

\section{Related Work}
\label{sec:background}

Ensuring forward progress is arguably the focus of most existing works in intermittent computing~\cite{batteryless-future}.
Common to these is the use of some form of persistent state stored on non-volatile memory.

A significant fraction of existing solutions employ a form checkpointing to cross
power failures~\cite{Mementos,Hibernus++,chinchilla,HarvOS, DICE}.
This consists in replicating the content of main memory, special
registers, and program counter onto non-volatile memory at specific points in the code.
Whenever the device resumes with new energy, these are retrieved back from non-volatile memory and computations restart from a point close to the power failure.
Systems such as Hibernus~\cite{Hibernus, Hibernus++} operate in a reactive manner: an interrupt is fired from a hardware device that prompts the application to take a checkpoint, for example, whenever the energy buffer falls below a threshold.
Differently, systems exist that place explicit function calls to proactively perform  checkpoints~\cite{Mementos,HarvOS,ratchet,chinchilla}.
The specific placement is a function of program structure and energy patterns.

Other approaches offer abstractions that programmers use to define and manage persistent state~\cite{DINO,chain,alpaca,ink} and time profiles~\cite{mayfly}.
For example, DINO~\cite{DINO} allows programmers to split the sequential execution of a program by inserting specific task boundaries and ensuring transactional semantics between consecutive boundaries.
Alpaca~\cite{alpaca} goes a step further and provides dedicated abstractions to defines tasks as individual execution units that run with transactional semantics against power failures and subsequent reboots, and channels to exchange data across tasks.

\begin{figure*}[t]
    \resizebox{.9\textwidth}{!}{\includegraphics{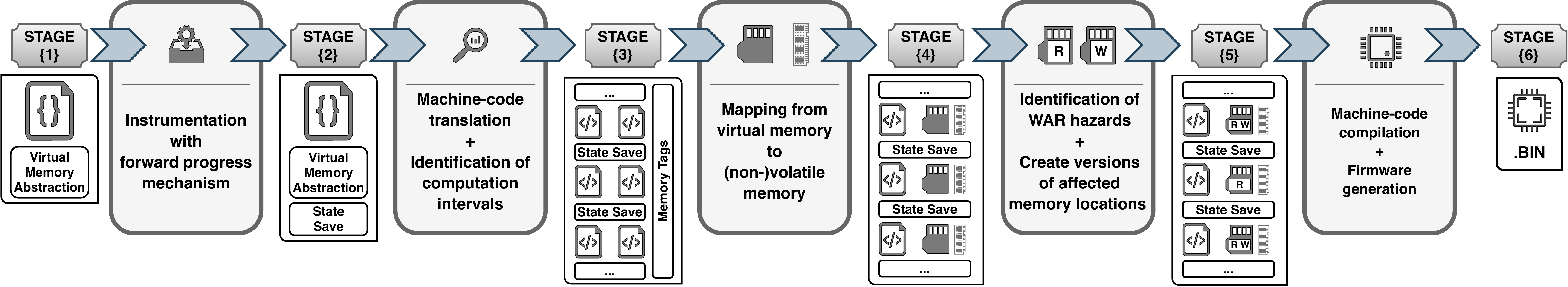}}
    \vspace{-2mm}
    \caption{\name compile-time pipeline.}
    \vspace{-3mm}
    \label{fig:pipeline}
\end{figure*}

Using mixed-volatile platforms, intermittence anomalies potentially occur due to repeated executions of non-idempotent code~\cite{brokenTM,maioli21ewsn}.
These are unexpected program behaviors that make executions differ from their continuous counterparts.
Systems are available that address intermittence anomalies with dedicated checkpoint placement strategies~\cite{ratchet} or custom programming abstractions~\cite{DINO,chain,alpaca,ink}.
In the latter, the abstraction semantics is designed to cater for the possible occurrence of intermittence anomalies. 
Tools also exist for testing their occurrence~\cite{maioli21ewsn, sceptic-repo}, and approaches are available that conversely take advantage of them to realize intermittence-aware control flows, effectively promoting the occurrence of power failures to an additional program input~\cite{maioli20enssys}.
Additional issues in intermittent computing include performing general testing of intermittent programs~\cite{EDB,EKHO,CleanCut,SIREN}, profiling their energy consumption~\cite{epic,EDB,SIREN}, and handling peripheral states across power failures~\cite{sytare,restop,samoyed,karma}.

Our work offers a different standpoint.
Unlike the works above, we take the decision about what part of the application state to allocate on non-volatile memory away from programmers, and offer a uniform abstraction that does not entail any specific allocation of data to memory facilities.
A set of program transformation techniques automatically determines an energy-efficient allocation at compile time, as a function of program structure and read/write patterns.
Most importantly, such an allocation is not fixed once and for all at variable-level as in existing solutions, but is possibly adjusted at different places in the code for the same data item. 

Closest to our work are \revision{ TICS~\cite{TICS} and the system of Jayakumar et al.~\cite{jayakumar-hybrid-nvm-mapping}}.
\revision{TICS~\cite{TICS} limits the size of persistent state by solely saving the active stack frame and modified memory locations outside of it, which is conceptually similar to our approach.
  However, TICS primarily helps programmers deal with time across power failures, whereas we specifically target energy efficiency independent of data age.
  TICS also exclusively uses non-volatile memory for global data and undo-logging~\cite{chinchilla} to avoid intermittence anomalies~\cite{brokenTM, maioli21ewsn}.
  In contrast, we opportunistically promote slices of main memory onto the faster and more efficient volatile memory to reduce energy consumption, and employ program transformation techniques that ensure memory idempotency~\cite{ratchet}.}

Jayakumar et al.~\cite{jayakumar-hybrid-nvm-mapping} technique operates at run-time by adjusting the mapping of global variables, program code, and stack frames between volatile and non-volatile memory, doing so at the granularity of individual functions.
They \revision{rely on hardware interrupts generated by an external component to trigger state-saving operations at runtime and} tentatively allocate everything to non-volatile memory first, then incrementally move data or code to volatile memory until forward progress is compromised.
At that point, they backtrack to the latest functioning configuration.
Besides working at the granularity of single data items at compile-time, rather than individual functions at run-time, our design is fundamentally different, as \revision{the allocation of data to volatile or non-volatile memory we determine is thought to \emph{systematically} improve energy consumption.}
\revision{Therefore, if forward progress is possible before applying \name, it remains so afterwards.
\name is thus never detrimental to the application's ability to do useful work.}

\section{Virtual Memory Mapping}
\label{sec:memory-principles}

The program transformation techniques of \name determine the mapping from virtual to volatile or non-volatile memory.
\revision{They are independent of the target architecture, as they are applied on an architecture-independent intermediate representation of the input program commonly used in compilers~\cite{llvm}.}
We illustrate the compile-time pipeline in \secref{sec:overview}, followed by an explanation of the individual techniques in \secref{sec:implicit-memory-save} to \secref{sec:reads}.

\subsection{Overview}
\label{sec:overview}

\figref{fig:pipeline} shows the compile-time pipeline of \name.
The input at stage~\step{1} is a program written using the virtual memory abstraction; therefore, variables in the program are not explicitly mapped to either volatile or non-volatile memory.

The program is first processed through the compile-time of an existing checkpoint system~\cite{Mementos, Hibernus, Hibernus++, HarvOS, QuickRecall, chinchilla, ratchet} or task-based programming abstraction~\cite{alpaca, chain, DINO, Coati, ink, coala}.
Either way, at stage~\step{2} the program includes state-save operations inlined in the execution flow as calls to a checkpointing subsystem or placed at task boundaries.
These operations are meant to dump program state onto non-volatile memory prior to a power failure and to restore the program state from non-volatile memory when energy is newly available after a power failure.
The techniques we explain next are orthogonal to how state-save operations are placed in the code.

Unlike existing programming systems for intermittent computing, our techniques work at the level of machine-code.
At this level, memory operations are visible as they are actually executed on the target platform.
At stage~\step{3} in \figref{fig:pipeline} we translate the program into \revision{an intermediate representation of the source code} and initially map every memory operation \emph{to volatile memory}.
If we were to execute the code this way, state-save operations would need to dump the entire main memory to the non-volatile one when executing.

Further, at the same stage we partition the code into logical units we call \textit{computation intervals}.
A computation interval consists in a sequence of machine-code instructions executed between two state-save operations.
For programs using checkpoint mechanisms~\cite{Mementos, Hibernus, Hibernus++, HarvOS, QuickRecall, chinchilla, ratchet}, computation intervals correspond to sequences of instructions placed between two checkpoints.
Instead, for programs using task-based programming abstractions~\cite{alpaca, chain, DINO, Coati, ink, coala}, computation intervals essentially correspond to tasks.

From now on, the three program transformations we illustrate next are applied in the order we present them.
We focus on the intuition and general application of each transformation, whereas we postpone the discussion about dealing with compile-time uncertainty to \secref{sec:uncertainty}.
Our techniques operate on every memory target that appears in the program; these may include not just memory targets that the compiler uses to map variables in source code, but also the memory used by operations that are normally transparent to programmers, such as \code{PUSH} or \code{POP} instructions.
We detail how we identify the memory addresses of data items possibly involved in a transformation in \secref{sec:uncertainty} and how to compute their addresses after the transformations are applied in \secref{sec:impl}.

As we hinted before, the mapping we want to achieve is one where volatile memory is used as much as possible to store data that does not require to be persistent, for example, intermediate results or temporary data, as it is more energy efficient than its non-volatile counterpart.
By the same token, we want to make sure to use the latter, and to pay the corresponding energy overhead, whenever persisting data to survive power failures is necessary.
Intuitively, the transformations generate a mapping from virtual to volatile or non-volatile memory where the former acts as a volatile cache of sorts where intermediate results are computed.

The snippets we show next include both source and machine code for clarity.
Line numbers refer to source code.
As mentioned already, however, \name operates entirely on machine code.

\begin{figure}[t]
    \subfigure[Before the transformation.]{
      \resizebox{0.48\columnwidth}{!}{\includegraphics{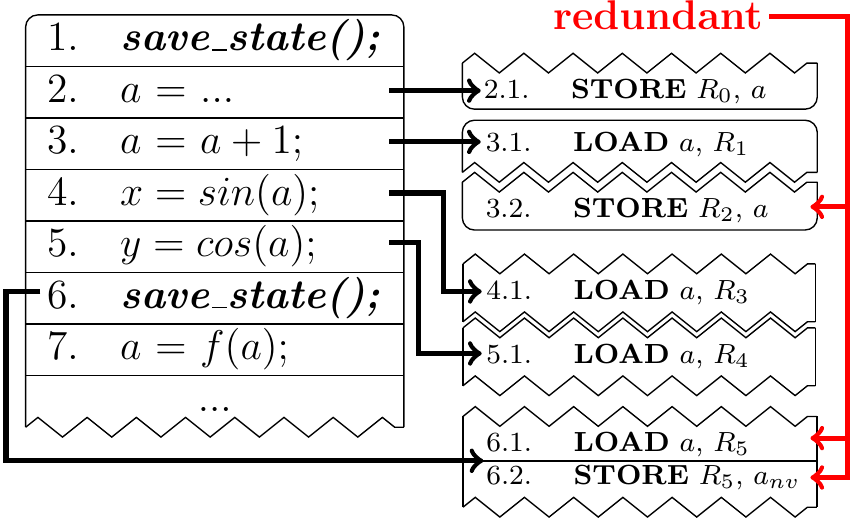}}
      \label{fig:example-checkpoint}
    }\hspace{-1mm}
    \subfigure[After the transformation.]{ \resizebox{0.48\columnwidth}{!}{\includegraphics{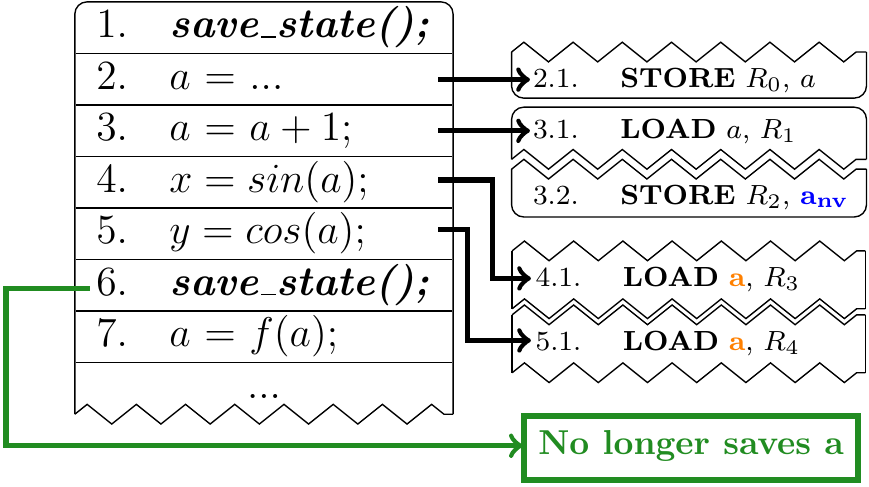}}
      \label{fig:example-implicit-checkpoint}
    }
    \vspace{-3mm}
    \caption{Example of mapping write operations.}
    \vspace{-3mm}
\end{figure}

\subsection{Mapping Write Operations}
\label{sec:implicit-memory-save}

The first transformation we apply is based on a key intuition: \emph{a memory write operation should target non-volatile memory as soon as the written data is final compared to the next state-save operation, so it relieves the latter from the corresponding processing.}

The notion of \emph{final} describes situations where the program does not alter the data anymore before the next state-save operation. 
Our intuition essentially corresponds to \emph{anticipating} the actions that the state-save operation would perform anyways.
This allows these operations to spare the overhead for saving data that can be considered final earlier: after the transformation the data is already on non-volatile memory when the state-save operation executes.

\fakepar{Example} Consider the program of \figref{fig:example-checkpoint} and let us focus on the instructions before the state-save operation of line $6$.
These instructions correspond to a computation interval.
We find two \code{STORE} instructions that target the volatile memory location that variable $a$ is initially mapped to.
Note that the second \code{STORE} instruction writes the same value that the state-save operation of line $6$ stores for variable $a$, because the latter is initially allocated onto volatile memory and must be preserved across power failures.
This is the case because the data for variable $a$ is \emph{final} at line $3$. 

To save the overhead of redundant memory operations, we make the \code{STORE} instruction of line $3$ immediately target non-volatile memory.
This transformation allows us to remove the instructions that are necessary to save variable $a$ at the state-save operation of line $6$, along with the corresponding energy overhead, as line $3$ already saves the content variable $a$ onto non-volatile memory.

\figref{fig:example-implicit-checkpoint} shows the resulting program, which has reduced energy overhead because the state-save operation is no longer concerned with variable $a$ that is made persistent already at line $3$.
Conceptually, this corresponds to move the \code{STORE} instruction that would normally be part of the state-save operation to the last point in the program where variable $a$ is actually written.

This transformation does not alter the target of the \code{STORE} instruction of line $2$, where the data is not final yet.
Doing so would incur an unnecessary energy overhead due to a write operation on non-volatile memory for non-final data, which is going to be over-written soon after.
In fact, the \code{STORE} instruction of line $2$ produces an intermediate result for variable $a$, which we need not persist.

\fakepar{Generalization} We apply this technique to an arbitrary computation interval as follows.
For each memory location $x$, we consider the possibly empty set of memory write instructions $I_w = (I_{w1}, ..., I_{wn})$ that manipulate $x$ and are included in the computation interval; $I_{wn}$ is thus the last such instruction and there is no other memory write instruction before the next state-save operation.

We relocate the target of $I_{wn}$ to non-volatile memory, as whatever data $I_{wn}$ stores is final.
The targets of all other write instructions $I_{w1}, ..., I_{w(n-1)}$ stay on volatile memory, as they produce intermediate result that $I_{wn} $ eventually overwrites.
Note that this transformation is sufficient to preserve the value of the memory location $x$ across power failures, while reducing the number of instructions targeting non-volatile memory.

By applying this transformation to all computation intervals and all memory locations, state-saving operations at stage \step{4} in \figref{fig:pipeline} are left with \emph{only} register file and special registers to dump on non-volatile memory, and accordingly modified.
If a memory location is altered in a computation interval, our technique identifies when such a change is final and persists the data there.
Otherwise, if $I = \emptyset$ there is no need to persist the data, as some previous state-save operation already did that the last time the data changed.

This processing not only reduces the operations on non-volatile memory, but also reduces the overhead of state-saving operations. 
A regular checkpoint mechanism would save the entire content of volatile memory onto the non-volatile one~\cite{Mementos, Hibernus, Hibernus++, HarvOS}, including unmodified memory locations.
In our case, memory locations not modified in a computation interval are excluded from processing.
We thus achieve differential checkpointing~\cite{DICE} with \emph{zero} run-time overhead in both energy and memory consumption.

Next, consider the read instructions possibly included in the computation interval between $I_{wn}$ and the state-save operation.
As the data is now on non-volatile memory, in principle, they should also be relocated to non-volatile memory.
Whether this is the most efficient choice, however, is not as simple.
The third transformation, described in \secref{sec:reads}, addresses the related trade-offs.

\subsection{Mapping Read Operations}
\label{sec:implicit-memory-restore}

The second transformation is dual to the first one and based on the corresponding intuition: \emph{when resuming after a power failure, restore routines may be limited to register file and special registers, while memory read operations from non-volatile memory should be postponed to whenever the data is needed, if at all.}

\begin{figure}[t]
    \subfigure[Before the transformation.]{
      \resizebox{0.48\columnwidth}{!}{\includegraphics{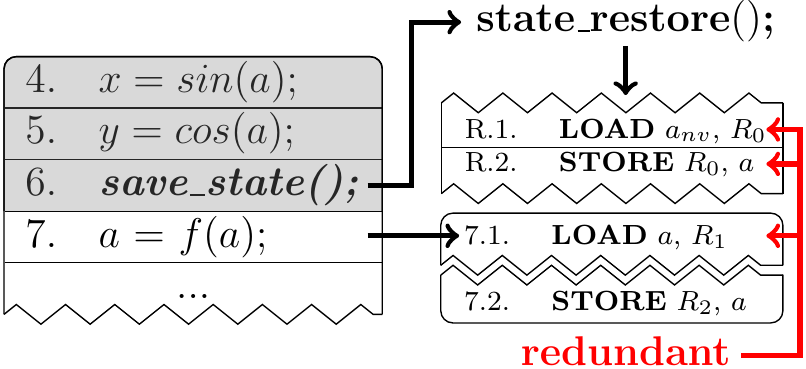}}
      \label{fig:example-restore}
    } \hspace{-1.5mm}
    \subfigure[After the transformation.]{
      \resizebox{0.48\columnwidth}{!}{\includegraphics{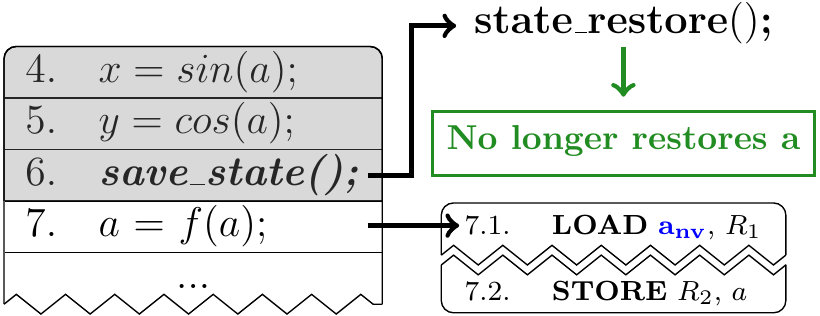}}
      \label{fig:example-implicit-restore}
    }    
    \vspace{-2mm}
    \caption{Example of mapping read operations.}
    \vspace{-3mm}
\end{figure}

This transformation effectively corresponds to \emph{postponing} the restore operation to when the data is actually used and a read operation would execute anyways.
By doing so, we spare the instructions in the restore routines that would load the data back to volatile memory from the non-volatile one.
This is the case after applying the first transformation, which makes state-save operations be limited to restoring the register file and special registers.
The content of main memory is persisted earlier, when it becomes final.

\fakepar{Example} Consider the program of \figref{fig:example-restore}.
Following a power failure, the execution resumes from line $6$ as the restore routines loads the value of the program counter from non-volatile memory, along with register file, other special registers, and the slice of main memory that was persisted prior to the power failure.
However, note that the \code{LOAD} instruction of line $7$ reads the same value for variable $a$ that is loaded earlier as part of the restore routine.

Dually to the first transformation, a more efficient strategy is to limit the restore routine to register file and special registers, and make the \code{LOAD} instruction of line $7$ target the non-volatile memory where the data resides.
Compared to the plain application of a checkpoint mechanism, for example, this transformation allows us to remove the instructions that restore variable $a$ from checkpoint data, as the first read instruction that is actually part of the program is relocated to the right address on non-volatile memory.

\figref{fig:example-implicit-restore} shows the program after this transformation, which bears reduced energy overhead because the restore routine is no longer concerned with variable $a$, as it is loaded straight from non-volatile memory if and when necessary.
Conceptually, this corresponds to move the \code{LOAD} instruction that would normally be part of the restore to routine for variable $a$ to the first point in the program where variable $a$ is actually read.

\fakepar{Generalization} Similar to the previous transformation, we apply this technique to an arbitrary computation interval as follows.
First, we limit restore routines to load back register file and special registers from non-volatile memory.
Next, for each concerned memory location $x$, we consider the possibly empty set of memory read instructions  $I_r = (I_{r1}, ..., I_{rn})$ that manipulate $x$ and are included in the computation interval.
Dually to the first transformation, $I_{r1}$ is the first such instruction and there is no other memory read instruction after the state-save operation that marks the start of the computation interval.
We relocate the target of $I_{r1}$ to non-volatile memory, as that is where the data is to be loaded from.

Whether the remaining $n-1$ read operations $I_{r2}, ..., I_{rn}$ in a computation interval are to target volatile or non-volatile memory is determined by applying the program transformation that follows.

\subsection{Consolidating Read Operations}
\label{sec:reads}

\begin{figure}[t]
    \resizebox{0.49\columnwidth}{!}{\includegraphics{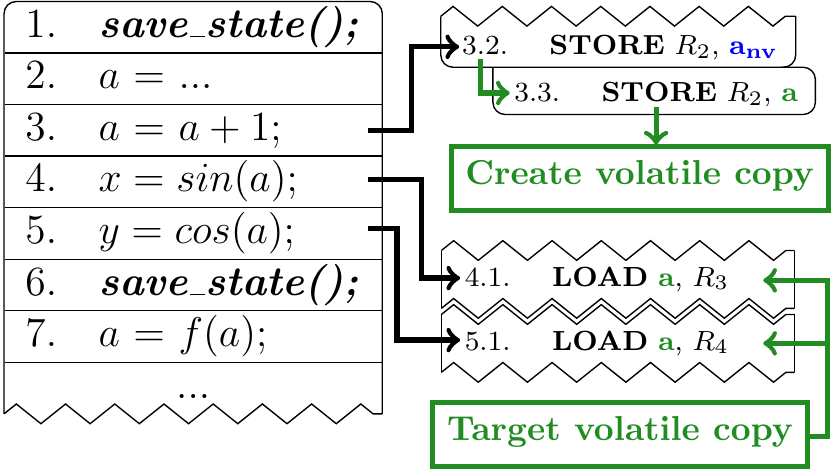}}
    \vspace{-2mm}
    \caption{Consolidating read operations.}
    \label{fig:additional-store}
    \vspace{-3mm}
\end{figure}

Starting with a program that exclusively uses volatile memory at stage \step{3} in \figref{fig:pipeline}, the first two transformation techniques relocate the target of selected read or write operations to non-volatile memory.
As data now resides on non-volatile memory in the vicinity of state-save operations, further relocations to non-volatile memory may be required for other read operations in a computation interval.
This is the case, for example, for read operations following the last write operation that makes data final on non-volatile memory, as mentioned in \secref{sec:implicit-memory-save}.
Whether this is the most efficient choice, however, is not straightforward to determine.

The third transformation we apply is based on the intuition that \emph{whenever memory operations are relocated to non-volatile memory, it may be convenient to create a volatile copy of data to benefit from lower energy consumption for subsequent read operations.}  

\fakepar{Example} The program in \figref{fig:example-implicit-checkpoint} includes further read operations after line $3$ and memory location $a$ is on non-volatile memory as a result of the first transformation.
In principle, we should relocate the read instructions on line $4$ and $5$ to non-volatile memory, as that is where the relevant data resides.
Because of the higher energy consumption of non-volatile memory, doing so may possibly backfire, outweighing the gains of the first transformation.

We must thus determine whether it is worth paying the penalty for creating a volatile copy of variable $a$ to benefit from the more energy efficient operations there.
Such a penalty is essentially represented by an additional \code{STORE} instruction required, right after the \code{STORE} of line $3$, to create a copy of the data on volatile memory, as shown in \figref{fig:additional-store}.
The new \code{STORE} uses the same source register, hence it represents the only added overhead.
The benefit is the improved energy consumption obtained by making the instructions of line $4$ and $5$ target volatile memory, instead of the non-volatile one.
Note that the exact same situation occurs for read instructions following the first \code{LOAD} instruction in \figref{fig:example-implicit-restore}. 

Consider the frequently used MSP430-FR5969~\cite{msp430fr5969, DINO, alpaca, chain, maioli20enssys}, which features an internal FRAM as non-volatile memory, and say the microcontroller runs at $16 MHz$, where FRAM accesses require an extra clock cycle.
Based on datasheet information~\cite{msp430fr5969}, we calculate that if read operations in line $4$ and $5$ target non-volatile memory, the program consumes $1.522 nJ$ for these operations.
In contrast, if we pay the penalty of the additional \code{STORE} instruction, but use volatile memory for all other read operations, the program consumes $1.376 nJ$ to achieve the same processing.
This is a $10.6\%$ improvement.
We accordingly insert an additional \code{STORE} instruction after line $3$ to copy $a$ to volatile memory and we keep the read operations of line $4$ and $5$ target volatile memory.

\fakepar{Generalization}
For each memory location $x$, we consider the $n$ read instructions $I_{r1}, ..., I_{rn}$ in a computation interval that we need to consolidate, thus excluding those altered by the second transformation.
We compute the energy consumption of a single non-volatile memory read instruction as
\begin{equation}
    E_{read\_nv} = E_{nv\_read\_cc} * (1 + CC_{read}),
    \label{eq:nv-energy}
\end{equation}
where $E_{nv\_read\_cc}$ is the energy consumption per clock cycle of the non-volatile memory read instruction and $CC_{read}$ are the extra clock cycles that the instruction execution possibly requires.
Based on operating frequency, mixed-volatile microcontrollers may incur in extra clock cycles when operating on the slower non-volatile memory.
These clock cycles consume the same energy as a regular non-volatile read operation.

The break-even point between paying the penalty of an additional \code{STORE} instruction to benefit from more energy-efficient volatile read operations, versus the cost of allocating all read operations to non-volatile memory is determined according to inequality
\begin{equation}
  E_{read\_nv} * n < E_{write} + E_{read} * n,
    \label{eq:transf}
\end{equation}
where $E_{read\_nv}$ is the one of \eqref{eq:nv-energy}, $n$ is the number of the considered memory read instructions in the computation interval,
and $E_{read}$ and $E_{write}$ represent the energy consumption of a volatile memory read and write instruction, respectively.
This can be rewritten as
\begin{equation}
    0 < E_{write} - n * (E_{nv\_read\_cc} * (1 + CC_{read}) - E_{read}).
    \label{eq:energy-diff}
\end{equation}
As the energy figures are fixed for a given microcontroller, \eqref{eq:energy-diff} is exclusively a function of $n$, that is, the number of memory read instructions to consolidate in the computation interval.
We can accordingly state that creating a volatile copy of the considered memory location is beneficial as long as 
\begin{equation}
    n > n_{min} \text{,  } \text{ with }\text{  } n_{min} = \floor*{\frac{E_{write}}{E_{nv\_read\_cc} * (1 + CC_{read}) - E_{read}}},
    \label{eq:energy-diff-n}
\end{equation}
where $n_{min}$ represents the minimum number of memory read instructions necessary to ensure that creating a volatile copy of the considered memory location incurs in lower overall energy consumption.
If the condition of \eqref{eq:energy-diff-n} is not met, we make the $n$ read operations target non-volatile memory.

Interestingly, $n_{min}$ is independent of the specific read/write memory patterns and of program structure.
It only depends on hardware characteristics.
As an example, $n_{min}$ is $0$ (2) for the MSP430-FR5969 at a clock frequency of $16 MHz$ ($8 Mhz$).
This means, for example, that if the MCU is to run at $16 MHz$, it is \emph{always} beneficial to create a volatile copy of the relevant memory locations.

\section{Compile-time Uncertainty}
\label{sec:uncertainty}

The transformation techniques of \secref{sec:memory-principles} rely on program information, such as the order of instruction execution and accessed memory addresses, that may not not be completely available at compile-time.
Constructs altering the control flow, such as conditional statements or loops, and memory accesses through pointers make these information dependent on the program state.
We describe next how we resolve this uncertainty, making it possible to apply the techniques of \secref{sec:memory-principles} to arbitrary programs.

We distinguish between two types of compile-time uncertainty.
\textit{Memory uncertainty} occurs when the exact memory address that a read/write operation targets cannot be determined.
We resolve this uncertainty using virtual memory tags, as described in \secref{sec:uncertainty-memory}.
\textit{Instruction uncertainty} occurs when the order of instruction execution is not certain.
Addressing this issue requires different techniques depending on program structure.
Starting from \secref{sec:uncertainty-instruction-intro}, we show how to resolve instruction uncertainty in case of loops, conditional statements, and function calls.
Finally, in \secref{sec:uncertainty-checkpoints}, we address a particular case of instruction uncertainty that affects computation intervals boundaries when the execution of state-save operations is uncertain at compile-time.

Here again, the code snippets include both source and machine code for easier illustration, with line numbers pointing to the former, yet \name operates entirely on machine code.

\subsection{Memory Uncertainty}
\label{sec:uncertainty-memory}

\begin{figure}[t]
    \resizebox{\columnwidth}{!}{\includegraphics{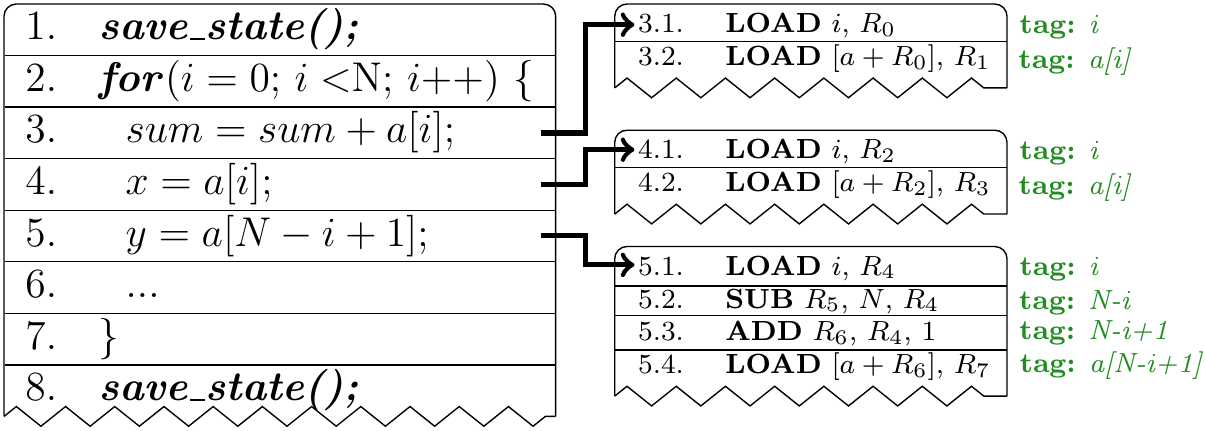}}
    \caption{Example of the same group of instructions accessing multiple memory locations.}
\vspace{-3mm}
    \label{fig:instructions-tag}
\end{figure}

Our key observation here is that the techniques of \secref{sec:memory-principles} do not necessarily require exact memory addresses to operate; rather, they need to identify the groups of instructions accessing the same memory location, whatever that may be.

\fakepar{Example}
\figref{fig:instructions-tag} shows an example.
Lines $3$, $4$, and $5$ target multiple memory locations across different iterations of the loop.
The corresponding physical addresses in memory change at every iteration.

To apply the techniques of \secref{sec:memory-principles}, however, exact knowledge of the physical addresses in memory is not required.
We rather need to determine that, at any given iteration of the loop, lines $3$ and $4$ target the same memory location, whereas line $5$ targets a different one.
Note that the information originally available in machine code is insufficient to this end: from that, we can only conclude that lines $3$, $4$, and $5$ access all the addresses in the range $(a[0], a[N-1])$.

We automatically associate a \emph{virtual memory tag} to every memory locations an instruction targets, as shown in \figref{fig:instructions-tag}.
A virtual memory tag is an abstraction of physical memory that aids the application of the techniques of \secref{sec:memory-principles} by succinctly capturing what memory locations are \emph{the same} in a computation interval.

In the program of \figref{fig:instructions-tag}, we attach the tag $a[i]$ to the memory locations read in lines $3$ and $4$.
Instead, we attach the tag \mbox{$a[N-i+1]$} to the memory location read in line $5$.
This information is sufficient for the technique mapping read operations, described in \secref{sec:implicit-memory-restore}, to understand that line $3$ and $4$ are to be considered as one sequence $I_r'$, whereas line $5$ is to be considered as a different sequence $I_r''$.  

Virtual memory tags are, in a way, similar to debug symbols attached to machine code.
They are obtained by inspecting the source code ahead of the corresponding translation, through multiple passes of a dedicated pre-processor.
The transformations of \secref{sec:memory-principles} look at these information, instead of the memory locations as represented in machine code.
Even in the case of pointers, we can combine virtual memory tags with memory alias analysis~\cite{alias-analysis1, alias-analysis2} to identify cases of indirect access to the same memory location.
Unlike debug symbols, however, these information is removed from the program at stage \step{5} before generating the final binary. 

\subsection{Instruction Uncertainty}
\label{sec:uncertainty-instruction-intro}
Key to the application of the program transformations in \secref{sec:implicit-memory-save} and \secref{sec:implicit-memory-restore} is the identification of the last (first) memory write (read) instruction in a computation interval.
This may be affected by loops, conditional statements, and function calls that alter the order of instruction execution.
Further, whenever the execution of state-save operations depends on run-time information, for example, when a checkpoint call is inserted in a loop body, the span of computation intervals is also undefined at compile time. 
We describe in the next sections how we address these issues.

\subsection{Instruction Uncertainty  $\rightarrow$ Loops}
\label{sec:uncertainty-loops}
Loops control the execution of a subset of instructions.
In this section we describe how we address the instruction uncertainty that loops may introduce.

\begin{figure}[t]
    \subfigure[Example of a compile-time uncertainty in a loop.]{
        \resizebox{\columnwidth}{!}{\includegraphics{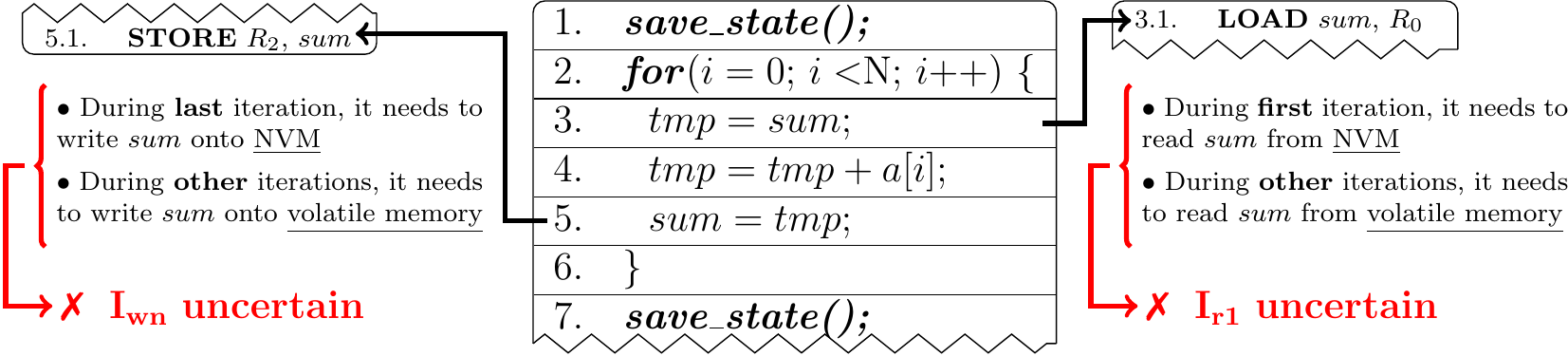}}
        \label{fig:uncertain-last-write}
    }
    \subfigure[Normalized form of the loop that removes the compile-time uncertainty.]{
        \resizebox{\columnwidth}{!}{\includegraphics{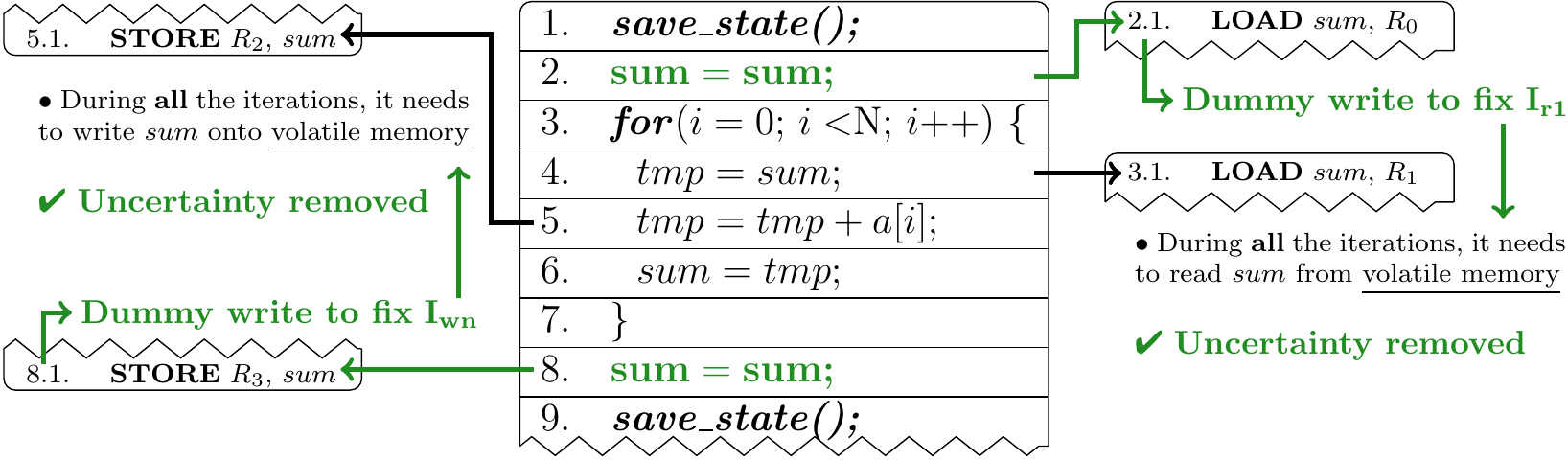}}
        \label{fig:certain-last-write}
    }
    \vspace{-2mm}
    \caption{Example of compile-time uncertainty with loops.}
    \vspace{-3mm}
\end{figure}

\fakepar{Example} \figref{fig:uncertain-last-write} exemplifies the situation.
Say we are to apply the mapping of write operations, described in \secref{sec:implicit-memory-save}.
Doing so requires to identify the last memory write instruction $I_{wn}$ before the state-save operation.
Depending on the value of $i$ compared to $N$, the  write operation in line $5$ may or may not be the one that makes the data final for variable $\mathit{sum}$.
The symmetric reasoning is valid when we are to apply the mapping of read operations, described in \secref{sec:implicit-memory-restore}.
Depending on the value of $i$ compared to $N$, the  read operation in line $3$ may or may not be the first for variable $\mathit{sum}$ after the state restore.
As a matter of fact, $i$ and $N$ are in control of what write (read) instruction is the $I_{wn}$ ($I_{r1}$). 

One may act pessimistically and make both the \code{LOAD} on line $3$ and the \code{STORE} on line $5$ target non-volatile memory.
This choice may be inefficient, because for all values of $i$ that are neither $0$ nor $N-1$, the loop computes intermediate results that are going to be overwritten anyways, so the cost of non-volatile memory operations is unnecessary.
To complicate matters, the value of $N$ itself may vary across different executions of the same fragment of code, as it may depend on program inputs and runtime state.

\fakepar{Generalization}
In general, such uncertainty arises whenever one of the following conditions are satisfied:
\begin{mylist}
    \item a loop controls the execution of memory write instruction $I_{wn}$ that may execute as last write in the computation interval, or
    \item a loop controls the execution of a memory read instruction $I_{r}$ that may execute before any memory write in the computation interval
\end{mylist}.

\fakepar{Normalization}
We apply techniques of \emph{program normalization}~\cite{normalizationControl,normalizationVariants} to resolve this uncertainty, as well as all others that possibly arise when the order of instruction execution depends on run-time information.
We normalize all the instructions $I_{wn}$ ($I_{r}$) that meet the above conditions, so that they no longer can execute as last memory write (read).
Program normalization refers to a set of established program transformations techniques designed to facilitate program analysis and automatic parallelization.
Many compiler techniques~\cite{cetus} for multi-core processors, for example, include multiple program normalization passes.

To resolve the uncertainty in \figref{fig:uncertain-last-write}, we need to be in the position to persist the value of $\mathit{sum}$ once we are sure the loop is over and \emph{before} the state-save operation. 
\figref{fig:certain-last-write} shows one way to achieve this.
We add a \textit{dummy write} consisting in a pair of \code{LOAD} and \code{STORE} instructions for variable $\mathit{sum}$ \emph{after} the loop\footnote{Note that these modifications to machine code occur after the compiler already applied code elimination steps.}.
These instructions are inserted after code elimination steps  and bear no impact on program semantics and, but fix where in the code the data for $\mathit{sum}$ is final, regardless of the value of $i$ and $N$.
We add a similar instruction \emph{prior} to the loop to fix where the first read operation for $\mathit{sum}$ occurs.
Different than before, we can make both \code{STORE} on line $8$ and the \code{LOAD} on line $2$ target non-volatile memory without unnecessary overhead.
All other operations now concern intermediate results that may be stored on volatile memory.
As a result, $i$ and $N$ are effectively no longer in control of what is the $I_{wn}$ ($I_{r1}$) write (read) instruction that the transformation in \secref{sec:implicit-memory-save} and \secref{sec:implicit-memory-restore} would allocate to non-volatile memory.

The normalization step does introduce an overhead.
To keep that at minimum, whenever possible we leverage information cached in registers.
For example, in \figref{fig:certain-last-write}, when line~$6$ executes it updates variable $\mathit{sum}$ with the value of variable $\mathit{tmp}$, which is stored in a register.
If the value is still available in a register, the operation in line $8$ may simply access that instead of re-loading the value from main memory.
Applying this kind of optimization is, however, not always possible, as the content of registers may be overwritten by other instructions that execute in between.
In \secref{sec:eval} we provide evidence that, despite the overhead of the normalization passes, \name programs are more energy-efficient than their regular counterparts. 

\subsection{Instruction Uncertainty  $\rightarrow$ Conditionals and Memory Mapping}
\label{sec:uncertainty-conditionals-map}
Conditional statements, such as the $if$ statement, controls the execution of a subset of instructions.
In this section we describe how we address the instruction uncertainty that conditional statements may introduce.

\begin{figure}[t]
    \subfigure[Program]{
        \resizebox{0.3\columnwidth}{!}{\includegraphics{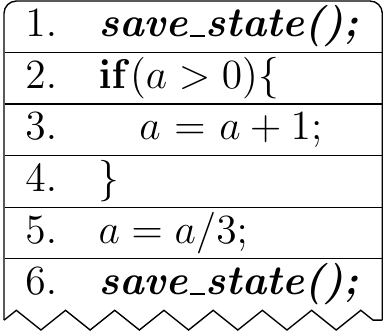}}
        \label{fig:uncertain-conditional-program}
    }
    \subfigure[Control Flow Graph]{
        \resizebox{0.65\columnwidth}{!}{\includegraphics{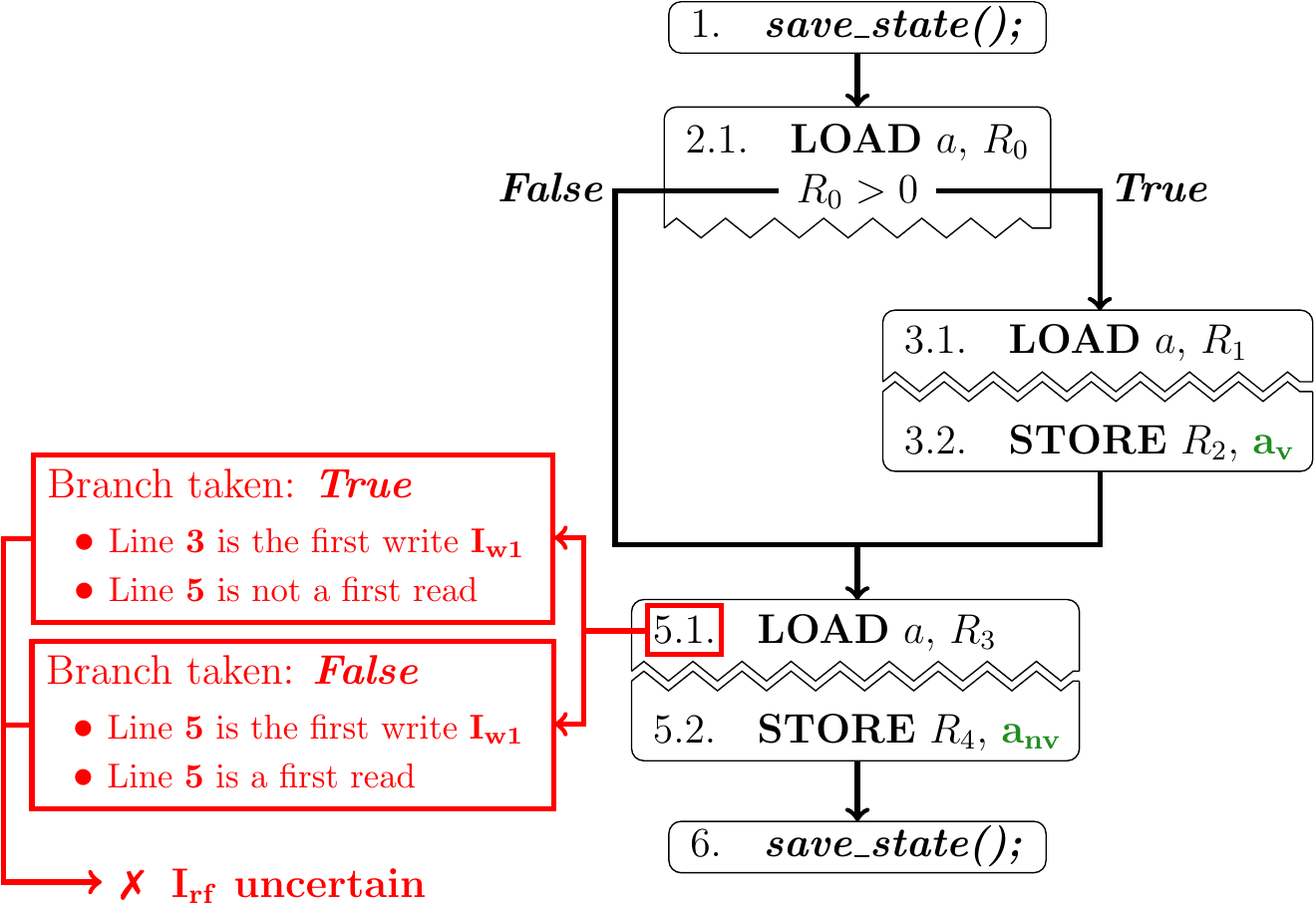}}
        \label{fig:uncertain-conditional-cfg}
    }
    \caption{Example of compile-time uncertainty with conditional statements when mapping read operations.}
    \label{fig:uncertain-conditional}
\end{figure}

\fakepar{Example}
\figref{fig:uncertain-conditional-program} exemplifies the situation, for which \figref{fig:uncertain-conditional-cfg} shows the program Control Flow Graph (CFG).
Note that we already applied the mapping of memory write operations, described in \secref{sec:implicit-memory-save}, which makes the \code{STORE} of line $3$ target volatile memory, as it does not write final data for $a$.
Say that we are to apply the mapping of read operations, described in \secref{sec:implicit-memory-restore}.
Doing so requires to identify all the first memory read instructions $I_{rf} = I_{r1}, ..., I_{rn}$ that execute before any memory write instruction $I_{w1}$.
As \figref{fig:uncertain-conditional-program} shows, when the $if$ statement of line $2$ evaluates to \textit{true}, the \code{STORE} of line $3$ executes as $I_{w1}$ and the \code{LOAD} of lines $2$ and $3$ are part of $I_{rf}$, but not the \code{LOAD} of line $5$.
Instead, when it evaluates to \textit{false}, the \code{STORE} of line $5$ executes as $I_{w1}$ and the \code{LOAD} of lines $2$ and $5$ are part of $I_{rf}$.
As a matter of fact, the $if$ statement of line $2$ controls the instruction that executes as $I_{w1}$ and consequently controls whether the \code{LOAD} of line $5$ is part of $I_{rf}$.
Therefore, it controls its mapping onto volatile/non-volatile memory.
In fact, when the $if$ statement of line $2$ makes the \code{STORE} of line $3$ execute, the \code{LOAD} of line $5$ must target volatile memory, as this is where the current value of $a$ resides.
Otherwise, it must target non-volatile memory.
As a consequence, we are unable to decide the mapping of the \code{LOAD} of line $5$, as an instruction can target one memory location at a time and mapping such \code{LOAD} to volatile (non-volatile) memory would cause incorrect results in the case it executes before (after) $I_{w1}$.

\begin{figure}[t]
    \subfigure[Program]{
        \resizebox{0.3\columnwidth}{!}{\includegraphics{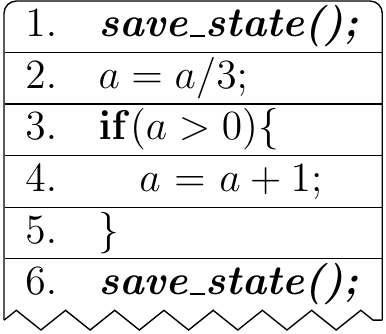}}
        \label{fig:uncertain-conditional-write-program}
    }
    \subfigure[Control Flow Graph]{
        \resizebox{0.65\columnwidth}{!}{\includegraphics{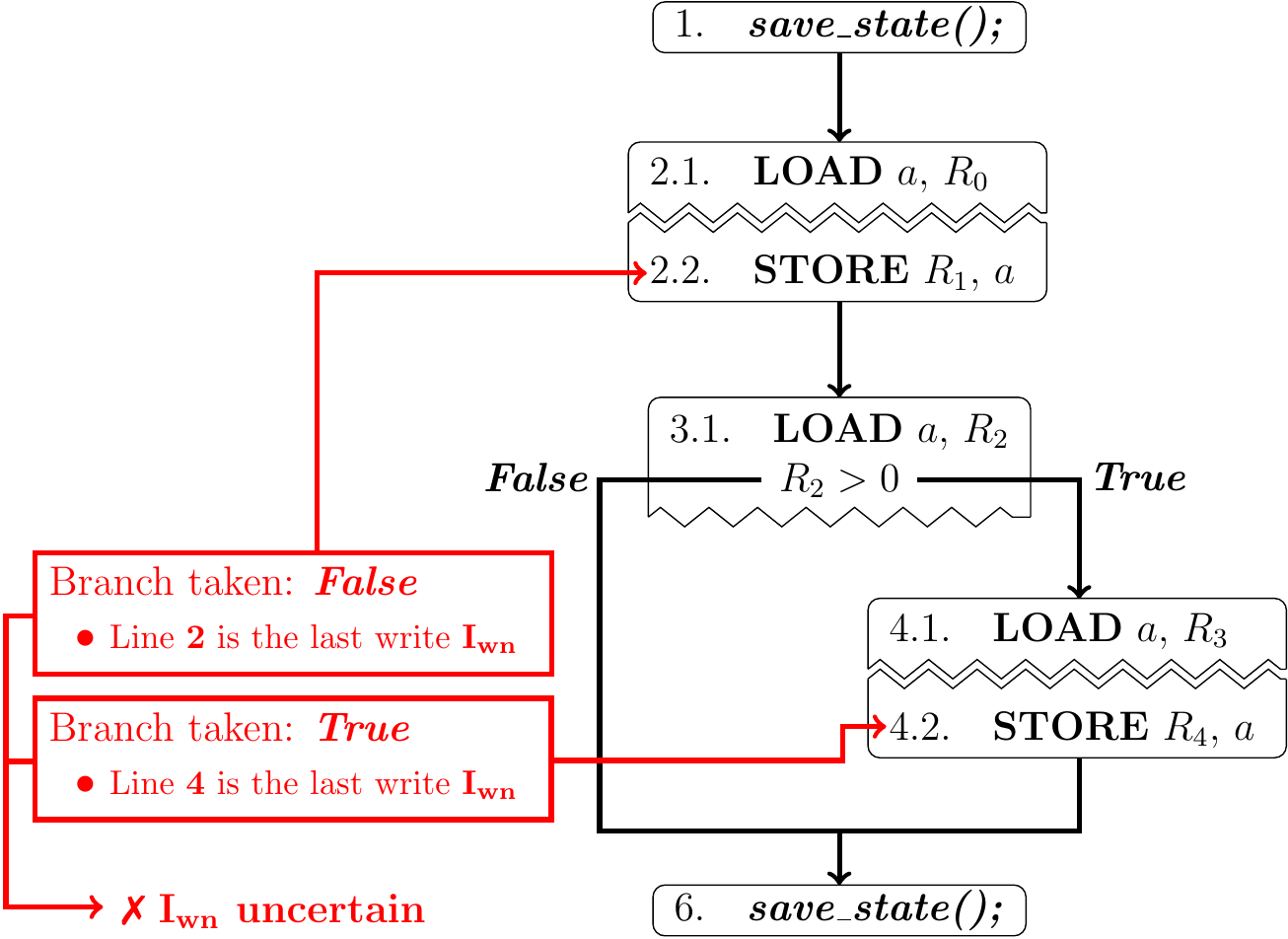}}
        \label{fig:uncertain-conditional-write-cfg}
    }
    \caption{Example of compile-time uncertainty with conditional statements when mapping write operations.}
    \label{fig:uncertain-conditional-write}
\end{figure}

Note that, as \figref{fig:uncertain-conditional-write} shows, a similar case can happen when we apply the mapping of write operations, described in \secref{sec:implicit-memory-save}.
In such a case, the conditional statement of line $2$ controls the instruction $I_{wn}$ that may execute as last write.

\fakepar{Generalization}
Depending on the type of operation that we are mapping, the described instruction uncertainty happens under different conditions.
When we map memory read operations, the uncertainty arises whenever
\begin{mylist}
    \item a conditional statement $C$ controls the execution of a memory write instruction $I_{w1}$ targeting a memory location with tag $x$ that may execute as the first write of the computation interval and
    \item there exists a memory read instruction $I_{rx}$ targeting a memory location with tag $x$ that, depending on the outcome of $C$, may or may not execute after $I_{w1}$
\end{mylist}.
In the example of \figref{fig:uncertain-conditional}, the \code{STORE} of line $3$ is $I_{w1}$ and the \code{LOAD} of line $5$ is $I_{rx}$.
As we previously describe, such uncertainty makes us unable to identify a mapping for $I_{rw}$ that ensures results consistency.

Instead, when we map memory write operations, the uncertainty arises whenever a conditional statement controls the execution of a memory write instruction $I_{wn}$ that may execute as last write of the computation interval.
In the example of \figref{fig:uncertain-conditional-write}, the \code{STORE} of line $4$ is such $I_{wn}$.

\fakepar{Addressing uncertainty}
Two different strategies allows us to address the compile-time uncertainty that conditional statements cause when mapping memory read and write operations: \textit{conservative} and \textit{non-conservative}.

\begin{figure}[t]
    \subfigure[Conservative strategy]{
        \resizebox{0.41\columnwidth}{!}{\includegraphics{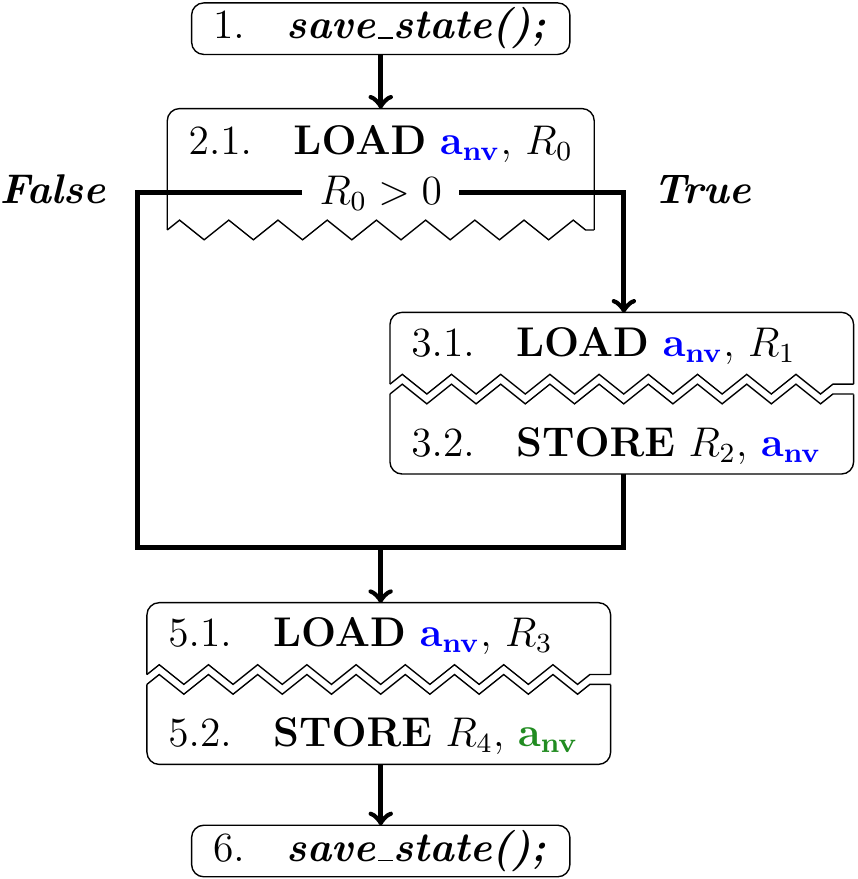}}
        \label{fig:uncertain-conditional-fix-conservative}
    }
    \subfigure[Non-conservative strategy]{
        \resizebox{0.54\columnwidth}{!}{\includegraphics{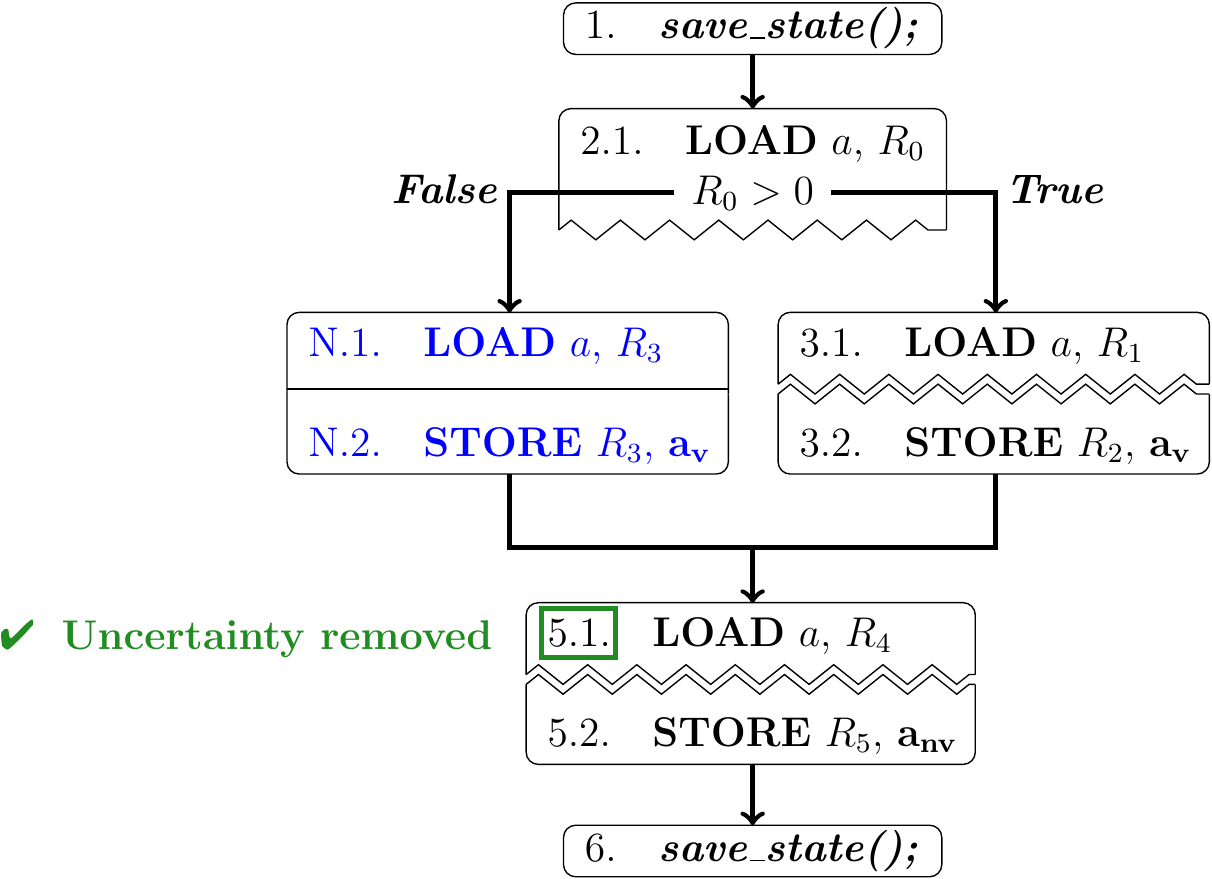}}
        \label{fig:uncertain-conditional-fix-non-conservative}
    }
    \caption{Example of the two strategies to address the uncertainty of \figref{fig:uncertain-conditional}.}
    \label{fig:uncertain-conditional-fix}
\end{figure}

The \textbf{conservative} strategy has the effect of mapping to non-volatile memory all the instructions involved in the compile-time uncertainty.
It considers as first (last) read (write) $I_{fr}$ ($I_{wn}$) all the memory read (write) instructions $I_{r}$ ($I_{w}$) for which there exists a path in the program CFG where $I_{r}$ ($I_{w}$) executes before (after) any memory write instruction in the computation interval.
Moreover, when mapping write operations, it ignores all the memory write instructions $I_{w1}$ controlled by a conditional statement that may execute as first write.
For example, in \figref{fig:uncertain-conditional}, we consider as part of $I_{rf}$ the \code{LOAD} instructions of lines $2$, $3$, and $5$.
Moreover, when mapping write operations, the \code{STORE} of line $3$ is ignored and therefore targets non-volatile memory.
\figref{fig:uncertain-conditional-fix-conservative} shows the result.
Note that with such strategy, the $if$ statement of line $2$ still controls the instruction that executes as $I_{w1}$.
However, as we make all the uncertain instructions target non-volatile memory, this is no longer a source of uncertainty.

The \textbf{non-conservative} strategy instead normalizes the instructions that can execute as first (last) read (write).
For doing so, such strategy relies on the same dummy-write operations that we describe in \secref{sec:uncertainty-loops} for addressing the uncertainty in loops.
Note that to normalize the first memory reads, such strategy uses a dummy-write operation to fix where the first memory write instruction execute.
For example, in \figref{fig:uncertain-conditional}, we place a dummy-write targeting $a$ in the $false$ branch of the $if$ statement of line $2$.
This makes the \code{LOAD} of line $5$ execute after a memory write instruction, ensuring that it no longer can execute both as first and non-first read.
\figref{fig:uncertain-conditional-fix-non-conservative} shows the result.
Note that with such strategy, the $if$ statement of line $2$ still controls which instruction among the two branches executes as $I_{w1}$.
However, as only one of them can act as $I_{w1}$, this is no longer a source of uncertainty and we can consider both as $I_{w1}$.

\fakepar{Strategy selection}
The non-conservative strategy normalizes also the energy consumption of the branches, as it starts from the configuration of the conservative strategy and reduces the energy consumption of one branch at the expenses of the other.
Say that our target platform is the MSP430-FR5969~\cite{msp430fr5969} at $16Mhz$.
In the example of \figref{fig:uncertain-conditional-fix}, the conservative strategy makes the program to consume $1.52nJ$ when the $false$ branch executes and $3.04nJ$ when the $true$ executes.
Assuming that no additional jump instruction need to be placed in the program, the non-conservative strategy increases the energy consumption of the $false$ branch to $2.14nJ$ and decreases the energy consumption of the $true$ branch to $2.14nJ$.

Conversely from the loop case, we are unable to establish if the normalization of the non-conservative stategy provides a lower energy consumption with respect to the conservative stategy, as this requires us to know the frequency of execution of each branch.
Not only such information is not available at compile-time, but it may also not always be possible to predict or identify, as branch execution depends on various factors, such as program inputs and computational state.
For this reason, we discard the non-conservative stategy, as it may have a negative effect on the program's energy consumption.
However, to avoid intermittence anomalies, we may need to apply the non-conservative strategy to normalize the instructions that may execute as first memory write targeting a non-volatile memory location.
We address such case in \secref{sec:impl}.

Hence, to address the uncertainty introduced by conditional statements, we apply the conservative strategy to identify a memory mapping that grants no compile-time uncertainty.
Then, we rely on our technique to consolidate memory read operations, as described in \secref{sec:reads}, to identify the most efficient memory mapping that minimizes the energy consumption of all the branches.
The next section describes how we account for uncertainty during the application of such technique.

\subsection{Instruction Uncertainty  $\rightarrow$ Conditionals and reads consolidation}
\label{sec:uncertainty-conditionals-reads}
As we describe in \secref{sec:reads}, consolidating memory read operations requires to identify the memory read instructions $I_r$ that happen after a memory read or write instruction $I_{nv}$ targeting non-volatile memory.
The presence of conditional statements that controls the execution of any such $I_r$ or $I_{nv}$ is a source of multiple compile-time uncertainties: the number $n$ of memory reads $I_r$ to consider and the memory read or write instruction $I_{nv}$ to consider.
To complicate matters, the conservative policy of \secref{sec:uncertainty-conditionals-map} that we apply to address uncertainty when mapping memory write instructions may introduce multiple conditionally-executed memory write instructions $I_{w}$ that target non-volatile memory.
Such conditionally-executed instructions are source of uncertainty when identifying the memory reads instructions $I_r$, preventing us to consolidate them.
We describe next each one of the three cases.

\begin{figure}[t]
    \subfigure[Program] {
        \resizebox{0.3\columnwidth}{!}{\includegraphics{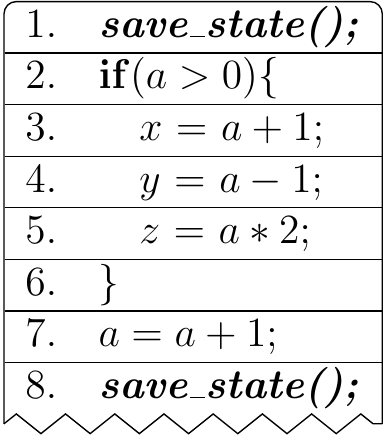}}
        \label{fig:uncertain-conditional-n-program}
    }
    \subfigure[Control Flow Graph] {
        \resizebox{0.6\columnwidth}{!}{\includegraphics{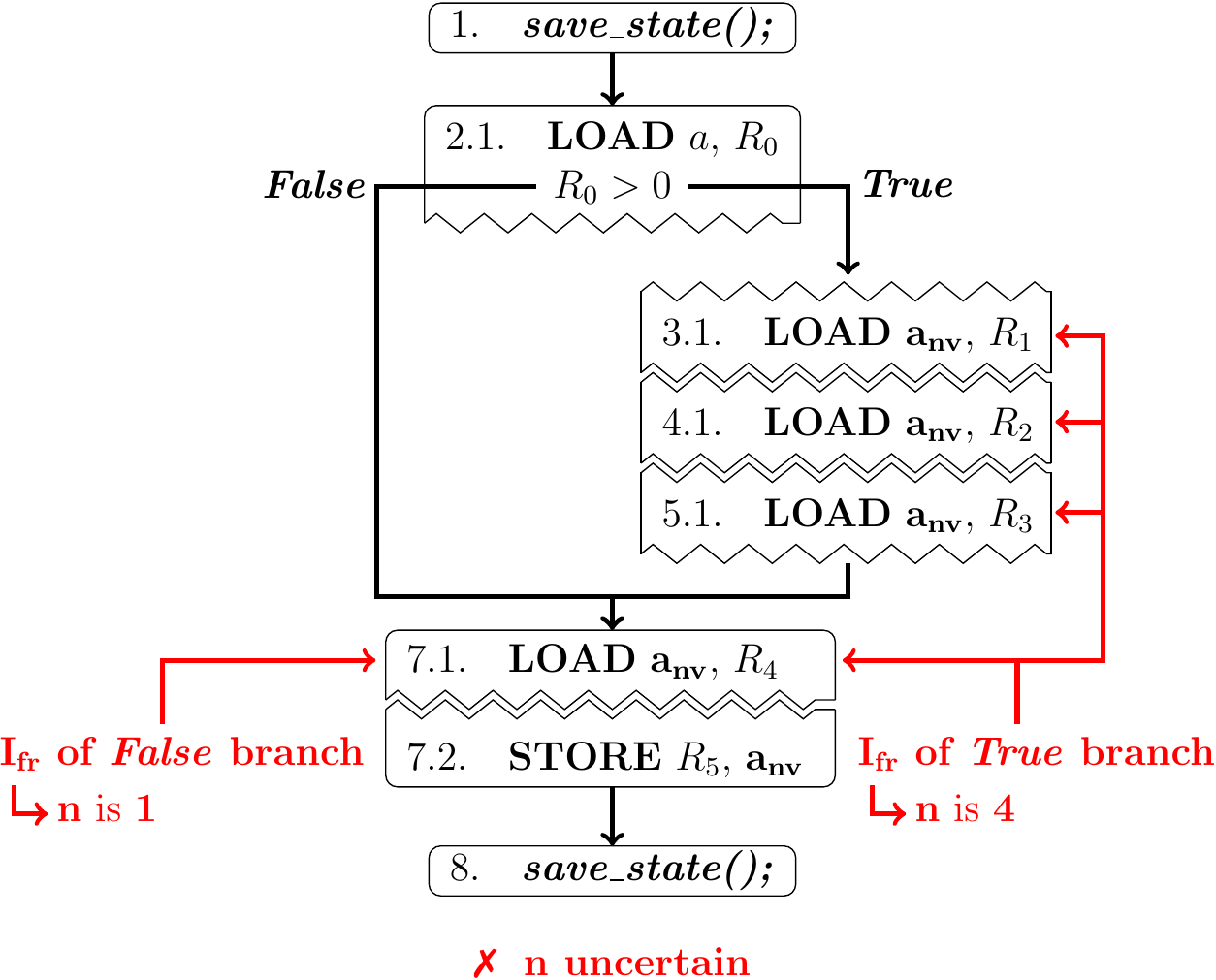}}
        \label{fig:uncertain-conditional-n-cfg}
    }
    \caption{Example of compile-time uncertainty with conditional statements when consolidating read operations. \capt{The uncertainty affects the number $n$ of memory reads to consider.}}
    \label{fig:uncertain-conditional-n}
\end{figure}

\subsubsection{Number of reads}
The presence of conditional statements that control the execution of memory read instructions $I_r$ that execute after the first (last) memory read (write) instruction $I_{nv}$ targeting non-volatile memory is a source of uncertainty on the number $n$ of $I_r$ operations, which we use to evaluate if such $I_r$ need to target volatile memory.

\fakepar{Example}
\figref{fig:uncertain-conditional-n} exemplifies the situation, where we already mapped read and write instructions to non-volatile memory.
Note that the \code{LOAD} of lines $2$-$7$ and the \code{STORE} of line $7$ target non-volatile memory as a result of the initial mapping or read and write operations.
Say that we are to apply the consolidation of the read operations.
We identify the \code{LOAD} of line $2$ as the first memory read instruction $I_{nv}$ that targets non-volatile memory.
Hence, we now need to identify the number $n$ of memory reads that execute after such $I_{nv}$.
Let us suppose that $n_{min}$ is $1$, that is, the lower bound of the number of memory reads required for applying the consolidation.
As \figref{fig:uncertain-conditional-n-cfg} shows, when the $if$ statement of line $2$ evaluates to $true$, $n$ is $4$ and we need to consolidate such reads, as $n > n_{min}$.
Otherwise, when it evaluates to $false$, $n$ is $1$ and we need not to consolidate such reads, as $n <= n_{min}$.
As a matter of fact, the $if$ statement of line $2$ controls the number $n$ of memory reads that executes after $I_{nv}$, which are the targets of the consolidation step.

In general, this problem happens whenever a conditional statement controls the execution of any memory read instruction $I_r$ that we make target non-volatile memory when mapping read and write instructions.

\fakepar{Addressing uncertainty}
To establish the number $n$ that we need to consider, there are two possible strategies, \textit{optimistic} and \textit{pessimistic}.
Note that such strategies are symmetric to the \textit{conservative} and \textit{non-conservative} strategies that we consider in \secref{sec:uncertainty-conditionals-map}.

The \textbf{optimistic} strategy considers the maximum possible number of memory read instructions, that is, the number of memory reads in the execution path with the highest possible energy consumption.
In \figref{fig:uncertain-conditional-n}, such strategy considers $n$ equal to $4$.

Instead, the \textbf{pessimistic} strategy considers the minimum possible number of memory read instructions, that is, the number of memory reads in the execution path with the lowest possible energy consumption.
In \figref{fig:uncertain-conditional-n}, such strategy considers $n$ equal to $1$.

Note that with both strategies, the $if$ statement of line $2$ still controls the number $n$ of memory read instructions that execute after $I_{nv}$.
However, such strategies address the uncertainty by fixing the number of memory reads $n$.

Choosing the optimistic stategy over the pessimistic one has the same effect of choosing the non-conservative strategy that we consider for removing the uncertainty from conditional operations when we map memory read and write instructions.
For this reason, we select the more conservative \textit{pessimistic} strategy.

\begin{figure}[t]
    \subfigure[Program]{
        \resizebox{0.3\columnwidth}{!}{\includegraphics{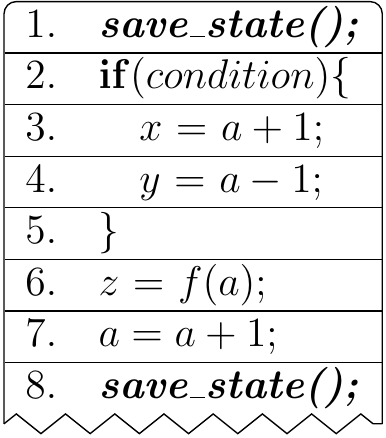}}
        \label{fig:uncertain-conditional-first-read-program}
    }
    \subfigure[Control Flow Graph]{
        \resizebox{0.6\columnwidth}{!}{\includegraphics{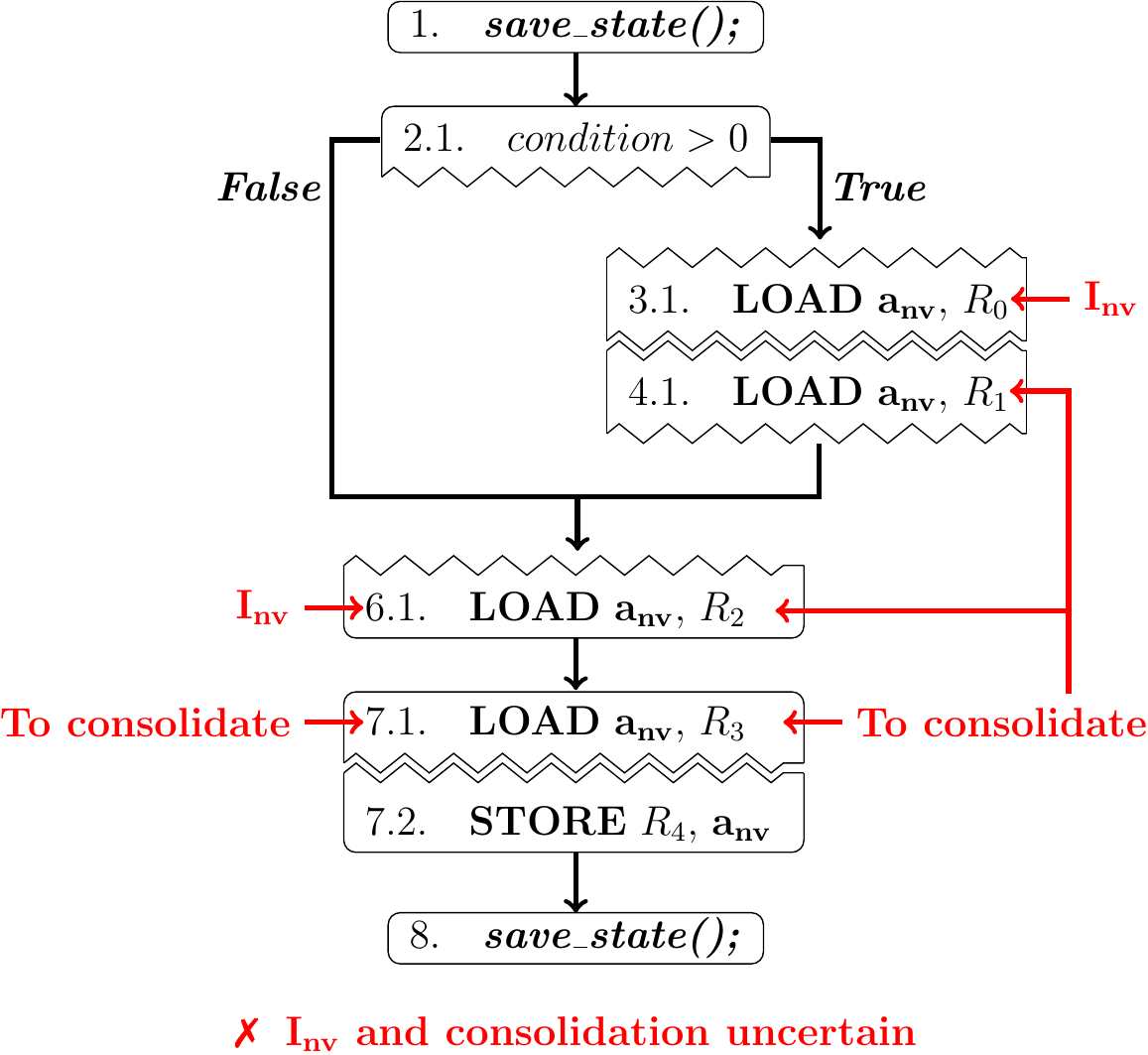}}
        \label{fig:uncertain-conditional-first-read-cfg}
    }
    \caption{Example of compile-time uncertainty with conditional statements when consolidating read operations. \capt{The uncertainty affects the instruction $I_{nv}$ from which we start consolidating read operations.}}
    \label{fig:uncertain-conditional-first-read}
\end{figure}

\subsubsection{Consolidating initial instruction}
The presence of conditional statements that control the execution the first (last) memory read (write) instruction $I_{nv}$ targeting non-volatile memory is a source of uncertainty, as $I_{nv}$ is the instruction from which we create a volatile copy of the targeted memory location.

\fakepar{Example}
\figref{fig:uncertain-conditional-first-read} exemplifies the situation, where we already mapped read and write instructions to non-volatile memory.
Note that the \code{LOAD} of lines $3$-$7$ and the \code{STORE} of line $7$ target non-volatile memory as a result of the initial mapping.
Say that we are to apply the consolidation of read operations.
We need to identify the first instruction $I_{nv}$ to execute that targets non-volatile memory, as well as the number $n$ of memory read instructions that follows $I_{nv}$.
As \figref{fig:uncertain-conditional-first-read-cfg} shows, when the $if$ statement of line $2$ evaluates to $true$, $I_{nv}$ is the \code{LOAD} of line $3$.
Otherwise, $I_{nv}$ is the \code{LOAD} of line $7$.
As a matter of fact, the $if$ statement of line $2$ controls the instruction $I_{nv}$ from which we start the consolidation of read operations.

In general, such uncertainty may happen whenever a conditional statement controls the execution of an instruction $I_{x}$ that we make target non-volatile memory as a consequence of the mapping step.
The uncertainty always arises when $I_{x}$ is a memory write instruction.
Instead, when $I_{x}$ is a read instruction, an uncertainty arises only when there exists a path in the program CFG such that $I_{x}$ executes as first memory read among all the $I_{x}$.

\fakepar{Addressing uncertainty}
We address such uncertainty by following an iterative approach while consolidating read operations.
First, we identify all the memory read and write instructions $I_{x}$ that we mapped to non-volatile memory as a consequence of the read and write mapping step.
Then, we consider each one of such $I_{x}$ as $I_{nv}$, that is, the instruction from which we consolidate the memory read operations that follows.
Note that we process each $I_{x}$ in order of execution in the program' CFG.
Moreover, we limit the memory read instructions that we consider after any $I_{nv}$ to the ones that belongs to the same branch of $I_{nv}$, if any.
This allows us to identify an optimal solution that covers only the instructions inside such branch, without introducing any new uncertainty.

For example, in \figref{fig:uncertain-conditional-first-read}, we identify as $I_{x}$ the \code{LOAD} instructions of lines $2$-$7$ and the \code{STORE} of line $7$.
Let us suppose that $n_{min}$ is $0$, that is, the case for the MSP430-FR5969 when the clock frequency is $16MHz$.
We start considering the \code{LOAD} of line $3$ as $I_{nv}$.
We identify $n$ equal to $1$, as we limit our search of memory read instructions inside the branch of lines $3$-$4$.
Being $n > n_{min}$, we consolidate the \code{LOAD} in the branch, creating a volatile copy of $a$ after the \code{LOAD} of line $3$ and making the \code{LOAD} of line $4$  target such copy.

We now proceed with the next $I_{nv}$ instruction, which is the \code{LOAD} of line $6$, as all the previous \code{LOAD} now target volatile memory.
Here we identify $n$ equal to $1$ and, being $n > n_{min}$, we proceed with the consolidation.
Hence, we crate a volatile copy of $a$ after the \code{LOAD} of line $6$ and we make the \code{LOAD} of line $7$ target such copy.

Next, we proceed with the \code{STORE} of line $7$, for which we need not any consolidation, as $n$ is $0$.
Our analysis now ends, as we now processed all the memory read and write instructions that target non-volatile memory.

Such iterative approach allows us to evaluate the consolidation of every possible subset of memory read instructions.
This allows us to identify the most energy efficient memory mapping that is specific only to single branches and that would otherwise not be efficient or applicable to an antire computation interval due to the presence of uncertainties.

\subsubsection{Accounting for conditional writes}
The presence of conditional statements that control the execution of any memory write instruction $I_{nv}$ targeting non-volatile memory is a source of uncertainty that prevents us to establish if the subsequent memory read needs to target volatile or non-volatile memory.

\begin{figure}[t]
    \subfigure[Program]{
        \resizebox{0.3\columnwidth}{!}{\includegraphics{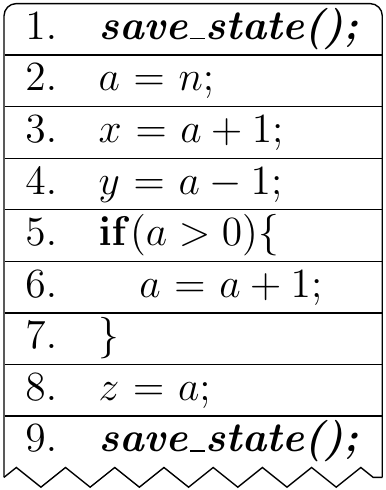}}
        \label{fig:uncertain-conditional-last-write-program}
    }
    \subfigure[Control Flow Graph]{
        \resizebox{0.6\columnwidth}{!}{\includegraphics{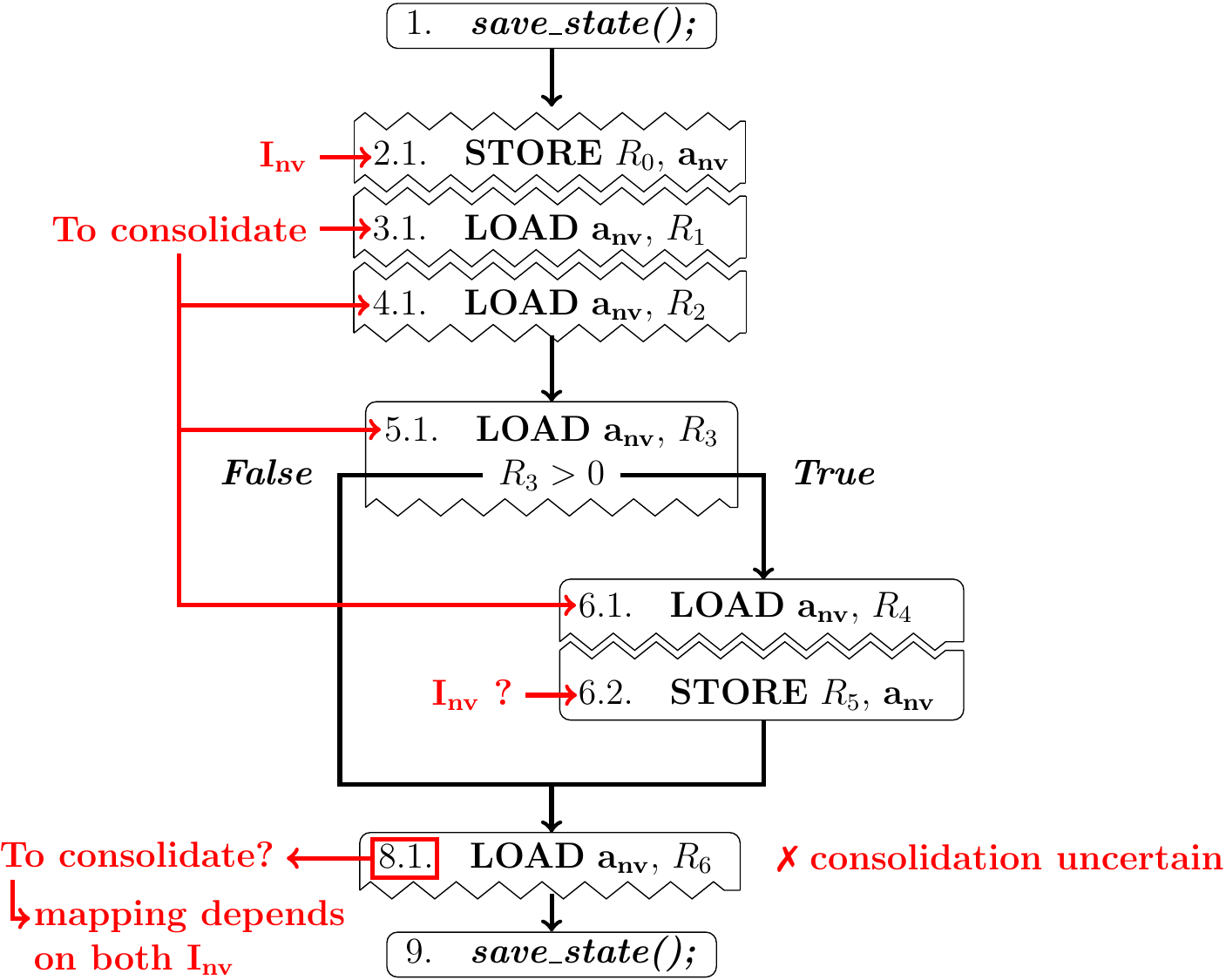}}
        \label{fig:uncertain-conditional-last-write-cfg}
    }
    \caption{Example of compile-time uncertainty with conditional statements when consolidating read operations. \capt{The uncertainty affects a memory write instruction $I_{nv}$ from which we create a volatile copy of a memory location.}}
    \label{fig:uncertain-conditional-last-write}
\end{figure}

\fakepar{Example}
\figref{fig:uncertain-conditional-last-write} exemplifies the situation, where we already mapped read and write instructions to non-volatile memory.
Say that we are to apply the consolidation of read operations.
We start applying the iterative algorithm that we previously describe and we consider as the \code{STORE} of line $2$ as $I_{nv}$.
To proceed with the consolidation of read operations, we need to identify all the memory read instructions $I_r$ that execute after such \code{STORE}, and before any other \code{STORE} targeting non-volatile memory.
As \figref{fig:uncertain-conditional-last-write-cfg} shows, when the $if$ of line $5$ evaluates to $true$, the \code{STORE} of line $6$ executes and we do not consider the \code{LOAD} of line $8$ as $I_r$, as a prior write executes.
Instead, when the $if$ of line $5$ evaluates to $false$, we consider the \code{LOAD} of line $8$ as $I_r$.
As a matter of fact, the $if$ statement of line $5$ controls the \code{STORE} of line $6$ and controls whether we need to consider as $I_r$ all the instructions that may execute after it.

One could proceed and apply the conservative policy of making the \code{LOAD} of line $8$ target non-volatile memory.
However, such option may not be the most energy efficient and may fail to identify the most efficient memory mapping.
Say that $n_{min}$ is $0$, that is, the case for the MSP430-FR5969 when the clock frequency is $16MHz$.
The number of $I_r$ instructions that execute after the \code{STORE} of line $6$ is $1$, which is higher than $n_{min}$.
As a result, creating a volatile copy of $a$ after the \code{STORE} of line $6$ provides a more efficient solution that making the \code{LOAD} of line $8$ target non-volatile memory.

\begin{figure}[t]
    \subfigure[Before consolidation]{
        \resizebox{0.45\columnwidth}{!}{\includegraphics{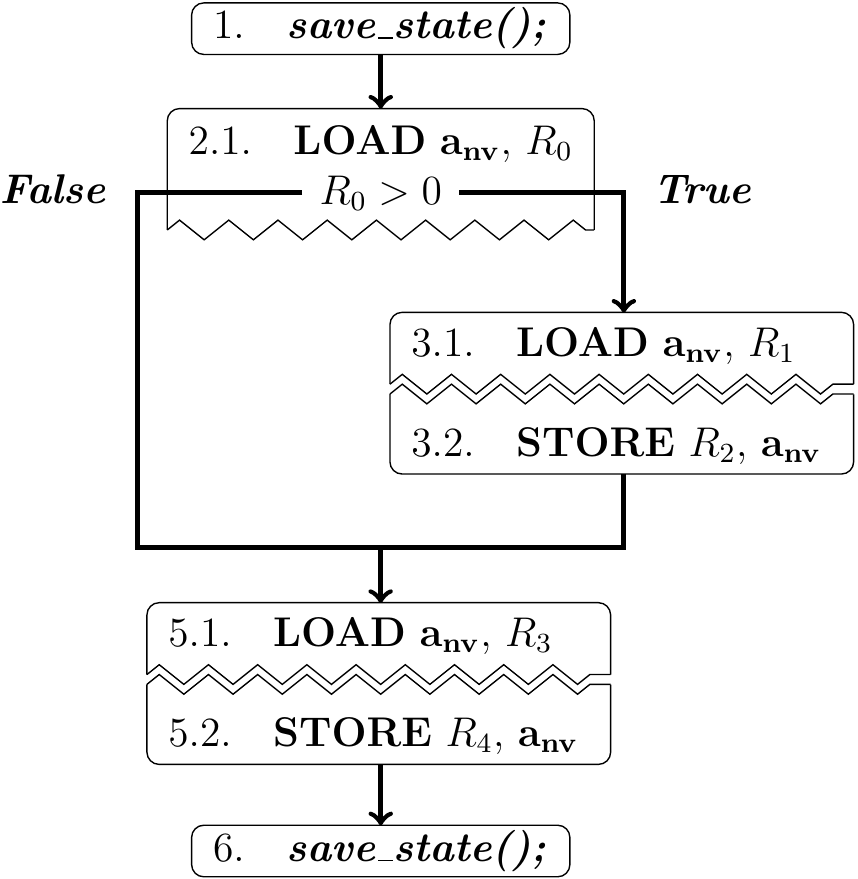}}
        \label{fig:uncertain-conditional-before}
    }
    \subfigure[After consolidation]{
        \resizebox{0.45\columnwidth}{!}{\includegraphics{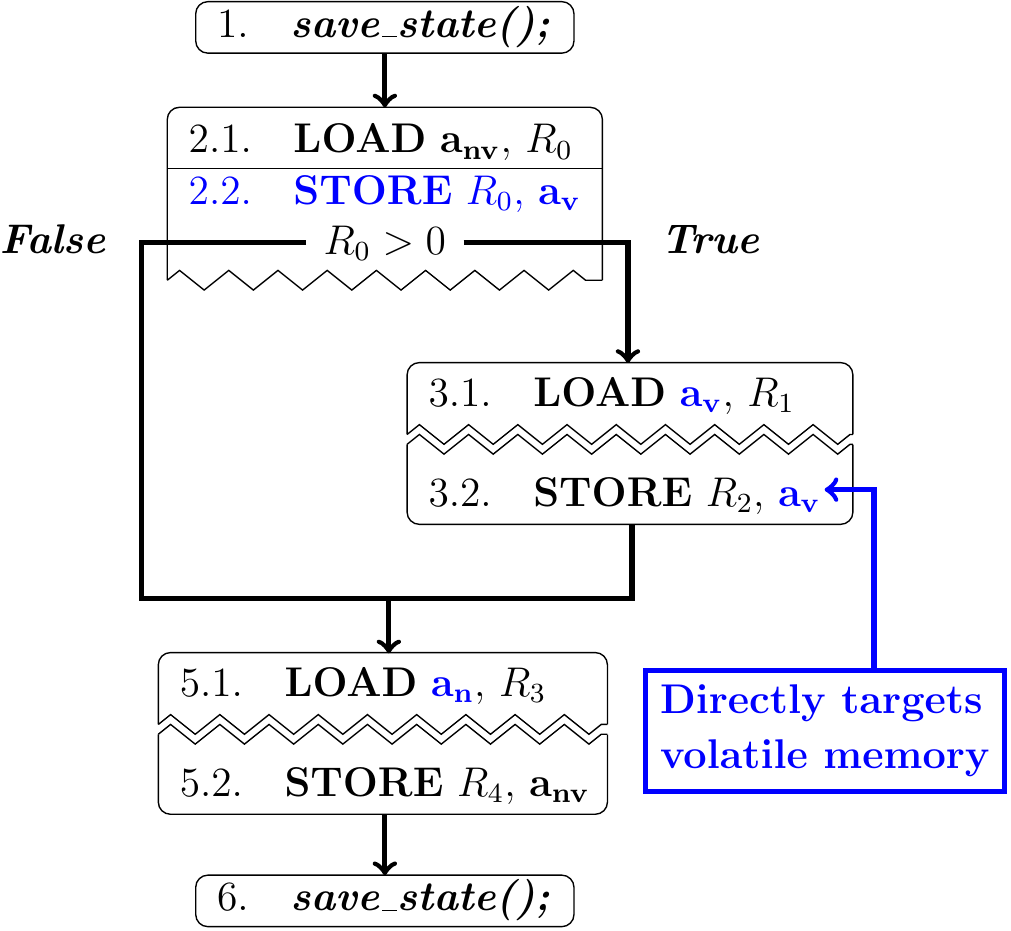}}
        \label{fig:uncertain-conditional-after}
    }
    \caption{Example of a consolidation of memory reads that removes the overhead introduced by the conditional strategy when addressing the compile-time uncertainty of conditional statements.}
    \label{fig:uncertain-conditional-consolidated}
\end{figure}

\fakepar{Addressing uncertainty}
To address such kind of uncertainty we proceed iteratively, using a similar approach to the one we previously describe.
Note that we cannot extend the instructions that we consider outside the branch, as we otherwise may introduce new compile-time uncertainties or wrongly make instructions target volatile/non-volatile memory.
First, we start by evaluating the number of memory read instructions $n$ by ignoring all the memory read $I_r$ that may e
Next, we gradually include the memory reads instructions happening after the next conditionally-executed memory write instruction, until we reach a non-conditionally executed memory write.
Whenever $n > n_{min}$, we consolidate the identified $I_r$.

In the example of \figref{fig:uncertain-conditional-last-write-cfg}, we consider as $I_{nv}$ the \code{STORE} of line $2$.
First, we identify as $I_r$ the \code{LOAD} of lines $3$-$6$.
Note that $n$ is $3$, as our pessimistic strategy forces us to not account for the \code{LOAD} of line $6$ when counting $n$, as it is conditionally-executed.
Say that $n_{min}$ is $0$, that is, the case for the MSP430-FR5969 when the clock frequency is $16MHz$.
Being $n > n_{min}$, we consolidate such $I_r$ instructions by creating a volatile copy of $a$ after the \code{STORE} of line $2$ and making them target volatile memory.
Next, we now consider the read that executes after the \code{STORE} of line $6$, with $n$ equal to $1$.
We consider the \code{LOAD} of line $8$ as the only $I_r$, as the \code{LOAD} of lines $3$-$6$ were already consolidated during the previous step.
Being $n$ equal to $1$, that is, $n > n_{min}$, we consolidate such $I_r$ instruction.

These iterative steps ensure that we identify the most efficient mapping for all the memory read instructions in the computation interval, as we evaluate the consolidation of all the possible subsets of instructions.
Note that at each step, we identify as $I_r$ only the memory read instructions that target non-volatile memory, consisting in the memory read instructions that the previous iteration did not consolidate.
This is necessary as we would otherwise account for the energy consumption of the memory read operations that we already map to volatile memory, potentially causing the identification of an inefficient mapping for the memory read instructions that still target non-volatile memory.

As we iteratively account for conditionally-executed memory write instructions, an iteration may consider more then one memory write instruction for which a volatile copy is not created, that is, a memory write instruction for which we did not consolidate the following reads during the previous iteration.
Such case affects the application criteria for the read consolidation technique, as we may need to create more than one volatile copy.
Hence, to account for the extra energy consumption required for creating such copies, we need to introduce a new parameter, $n_{w}$, in the $n_{min}$ formula that we introduce in \secref{sec:reads}:
\begin{equation}
    n_{min} = \floor*{\frac{E_{write}*(n_{w})}{E_{nv\_read\_cc} * (1 + CC_{read}) - E_{read}}},
    \label{eq:new-n}
\end{equation}
The $n_{w}$ parameter corresponds to the number of memory write instructions that we need to insert in the program to create a volatile copy of the considered memory location.

\fakepar{Accounting for conservative strategy}
At each iteration, we may consider as additional write a memory write instruction $I_{wi}$ that does not produce final data, yet targets non-volatile memory due to the presence of an uncertainty that we avoid by applying the conservative strategy described in \secref{sec:uncertainty-conditionals-map}.
When we consider such memory write $I_{wi}$ and we end up in consolidating the memory reads that follows, we need not to create a volatile copy after $I_{wi}$.
Instead, we directly make $I_{wi}$ target volatile memory, as the consolidation of the following reads removes the uncertainty that forced us to make $I_{wi}$ target non-volatile memory in the first place.

That is the case of the aforementioned \figref{fig:uncertain-conditional}, for which \figref{fig:uncertain-conditional-before} shows the resulting control flow graph after the mapping of read and write operations.
As we previously describe, we make the \code{STORE} of line $3$ target non-volatile memory to avoid compile-time uncertainty on the mapping of the \code{LOAD} of line $5$.
Say that we are to consolidate memory reads and that $n_{min}$ is $0$, that is, the case for the MSP430-FR5969 when the clock frequency is $16MHz$.
We first consider the \code{LOAD} of line $2$ as $I_{nv}$ and we identify the \code{LOAD} of line $3$ as the only $I_r$, as the $if$ of line $2$ contains a \code{STORE} instruction.
Being such $I_r$ conditionally-executed, we apply the pessimistic strategy and we set $n$ equal to $0$.
Thus, we do not consolidate any read, as $n <= n_{min}$.

Next, we consider also the \code{STORE} of line $3$ and the memory reads that execute after.
$I_r$ now consists in the \code{LOAD} of lines $3$ and $5$.
This leads to $n$ equal to $1$, as the \code{LOAD} of line $3$ is conditionally-executed.
We now need to compute $n_{min}$ using \eqref{eq:new-n}, as we are considering an additional conditionally-executed write operation.
The \code{STORE} of line $3$ does not produce final data and targets non-volatile memory to avoid a compile-time uncertainty, which we remove upon consolidation of memory reads.
In fact, after its consolidation, the \code{LOAD} of line $5$ targets volatile memory and the compile-time uncertainty is no longer present.
As a consequence, the \code{STORE} of line $3$ need not to target non-volatile memory and we can make it target volatile memory.
For this reason, the number $n_{w}$ of memory write instructions required to create a volatile copy of $a$ is $1$, that is, the additional \code{STORE} that we need to place after the \code{LOAD} of line $2$.
As such, $n_{min}$ is still $0$ and being $n > n_{min}$ we proceed with the consolidation of read operations.
\figref{fig:uncertain-conditional-after} shows the result.

\subsection{Instruction Uncertainty  $\rightarrow$ Function Calls}
\label{sec:uncertainty-function-calls}

\begin{figure}[t]
    \subfigure[Example of a compile-time uncertainty in function calls]{
        \resizebox{\columnwidth}{!}{\includegraphics{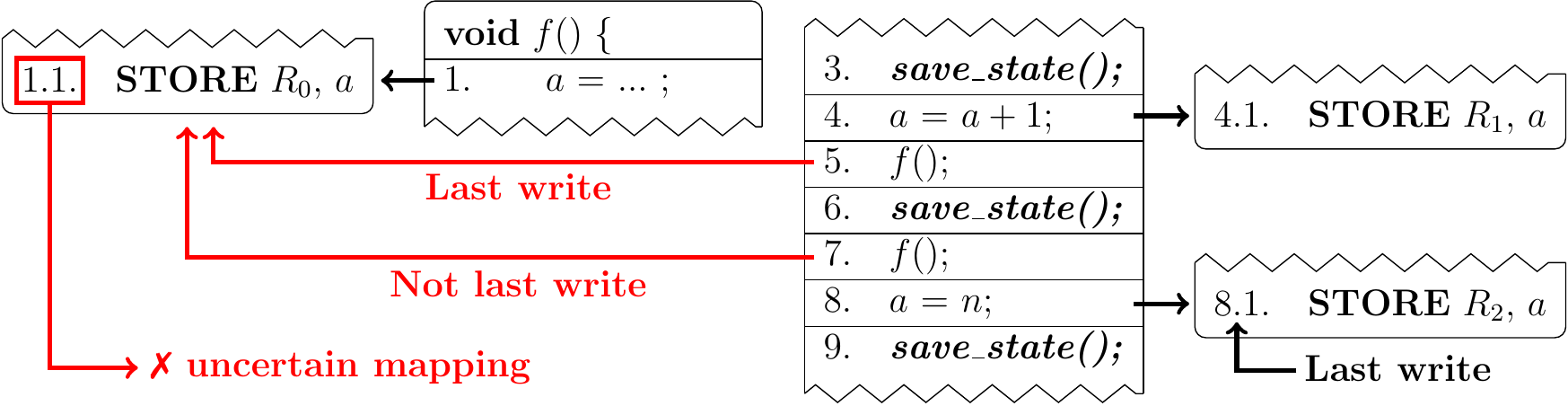}}
        \label{fig:uncertain-f-call}
    }
    \subfigure[Normalized form of the program that removes the compile-time uncertainty]{
        \resizebox{\columnwidth}{!}{\includegraphics{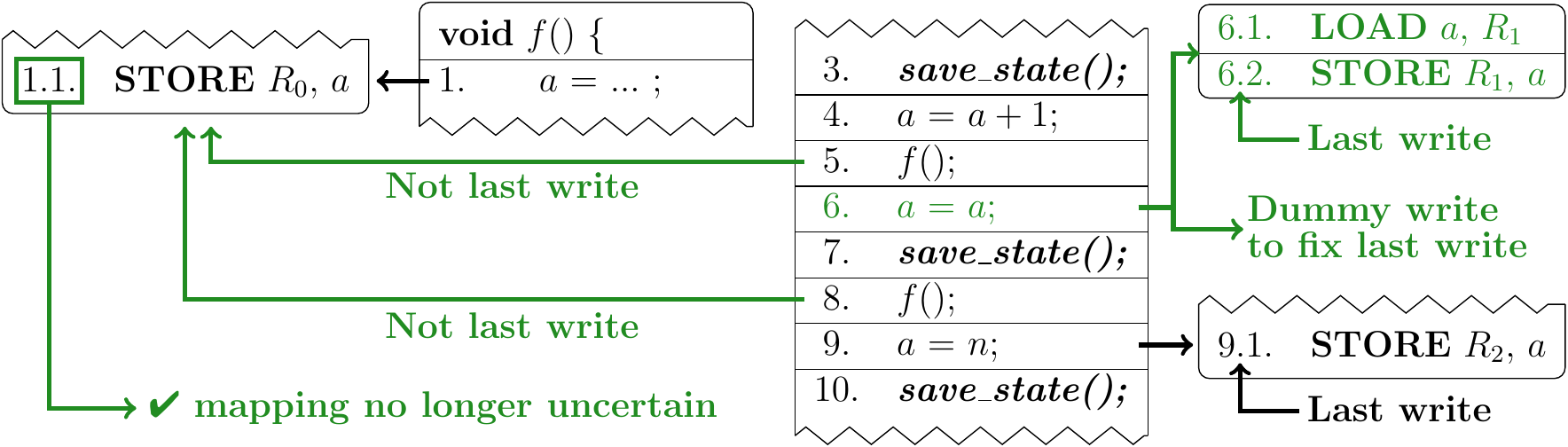}}
        \label{fig:uncertain-f-call-fix}
    }
    \caption{Example of a compile-time uncertainty with function calls.}
\end{figure}

Developers usually split programs into several functions, or subroutines, to logically divide the different program funcionalities.
Being each function a separate computation unit that reads and writes a private slice of main memory, that is, the stack frame of the function, we consider each function as a separate computation interval when applying the techniques of \secref{sec:memory-principles}.

\fakepar{Mapping memory reads and writes}
From a memory standpoint, the results of functions execution always reside outside the function' stack frame, as a function can produce results in two different ways: it writes a memory location outside its stack frame, or, it returns the computed result to its caller.
We map the memory read and write operations targeting outside the function' frame following the same principles that we describe in \secref{sec:implicit-memory-save}, \secref{sec:implicit-memory-restore}, and \secref{sec:reads}, accounting for function calls and their outside-frame accesses when identifying the first (last) memory read (write) instructions.

Differently, we map the stack frame of functions to volatile or non-volatile memory depending on whether a function includes a state-saving operation.
Functions not containing any state-saving operation execute within a single computation interval.
As such, their stack frame contains only intermediate results, which we need not to preserve across power failures.
Hence, we map all the memory read and write instructions that target the stack frame of such functions to volatile memory.
Instead, we normally apply our technique for mapping the read and write operations to functions containing one or more state-saving operations, as such functions execute across multiple computation intervals and their stack frame contains data that need to be preserve across power failures to correctly complete the function execution.
Note that this is usually the case of the program entry point, that is, the main function.

\fakepar{Compile-time uncertainty}
The presence of multiple calls to a function that executes outside frame accesses is a source of compile-time uncertainty on the mapping of such accesses.

\figref{fig:uncertain-f-call} exemplifies the situation, where the function $f$ executes a memory write operation that writes the global variable $a$, consisting in an outside frame access.
Say that we are to apply the mapping of write operations, described in \secref{sec:implicit-memory-save}.
Doing so requires to identify the last memory write instruction $I_{wn}$ of each computation interval.
Depending on where a call to $f$ executes, the \code{STORE} of line $1$ may or may not be the one that makes the data final for the global variable $a$.
In the computation interval of lines $4$-$5$ the call to $f$ of line $3$ executes and the \code{STORE} of line $1$ makes the data final for the global variable $a$.
Hence, such \code{STORE} is the $I_{wn}$ we need to make target non-volatile memory.
Instead, when the call to $f$ of line $7$ executes in the computation interval of lines $7$-$8$, the \code{STORE} of line $1$ produces intermediate data for the global variable $a$, as the \code{STORE} of line $8$ is the one making the data final for $a$.
Here the \code{STORE} of line $8$ is the $I_{wn}$ we need to make target non-volatile memory, whereas the \code{STORE} of line $1$ need to target volatile memory.
As a matter of fact, the calls to $f$ control whether the \code{STORE} of line $1$ is the $I_{wn}$.
Note that the symmetric reasoning is valid when we apply the mapping of read operations, described in \secref{sec:implicit-memory-restore}.

In general, the described instruction uncertainty arises whenever a function $f$ contains a memory write (read) instruction $I_x$ that targets outside $f$ frame and there exists at least two function calls $c_1$ and $c_2$ to $f$ such that:
\begin{mylist}
    \item when $c_1$ executes $f$, $I_x$ is the last (first) memory write (read) and
    \item when $c_2$ executes $f$, $I_x$ is not the last (first) memory write (read)
\end{mylist}.

\fakepar{Function calls normalization}
Such compile-time uncertainty shares a similar pattern of the uncertainty of loops, described in \secref{sec:uncertainty-loops}.
Hence, we rely on the same normalization technique to address such compile-time uncertainty.
We insert a dummy write operation after (before) the function call where $f$ executes a last (first) memory write (read).
This ensures that function calls no longer control whether $f$ executes a last (first) memory write (read).

In the example of \figref{fig:uncertain-f-call}, we place a dummy write targeting $a$ after line $5$
This ensures that the \code{STORE} of line $1$ produces only intermediate data for $a$, as when the call to $f$ of line $5$ executes, the \code{STORE} of line $1$ no longer produces final data for $a$.
\figref{fig:uncertain-f-call-fix} shows the normalized program.

\subsection{Instruction Uncertainty $\rightarrow$ Computation Intervals}
\label{sec:uncertainty-checkpoints}
To apply the techniques described in \secref{sec:memory-principles} to an arbitrary program, we need to partition its instructions into computation intervals, consisting in sequences of instructions executed between two state-saving operations.
Differently from previous cases, calls to a function containing a state-saving operation do not introduce any uncertainty, as they ends a computation interval, acting as a computation interval boundary.
Instead, the presence of loops and conditional statements that control the execution of state-saving operations may introduce compile-time uncertainty on the span of computation intervals.
We describe next how we address these issues.
Note that programs instrumented with task-based mechanisms~\cite{DINO, alpaca, chain, ink, Coati, coala} are usually not affected by this problem, as state-saving operations only execute upon tasks compleition.

\begin{figure}[t]
    \subfigure[Program] {
        \resizebox{0.35\columnwidth}{!}{\includegraphics{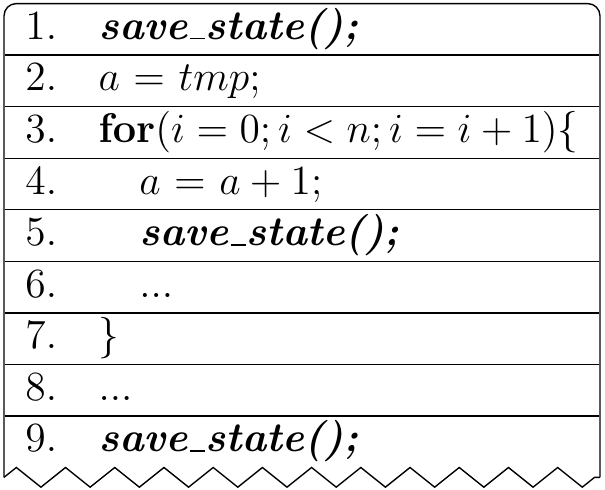}}
        \label{fig:uncertain-ci-loop-program}
    }
    \subfigure[Control flow representation] {
        \resizebox{\columnwidth}{!}{\includegraphics{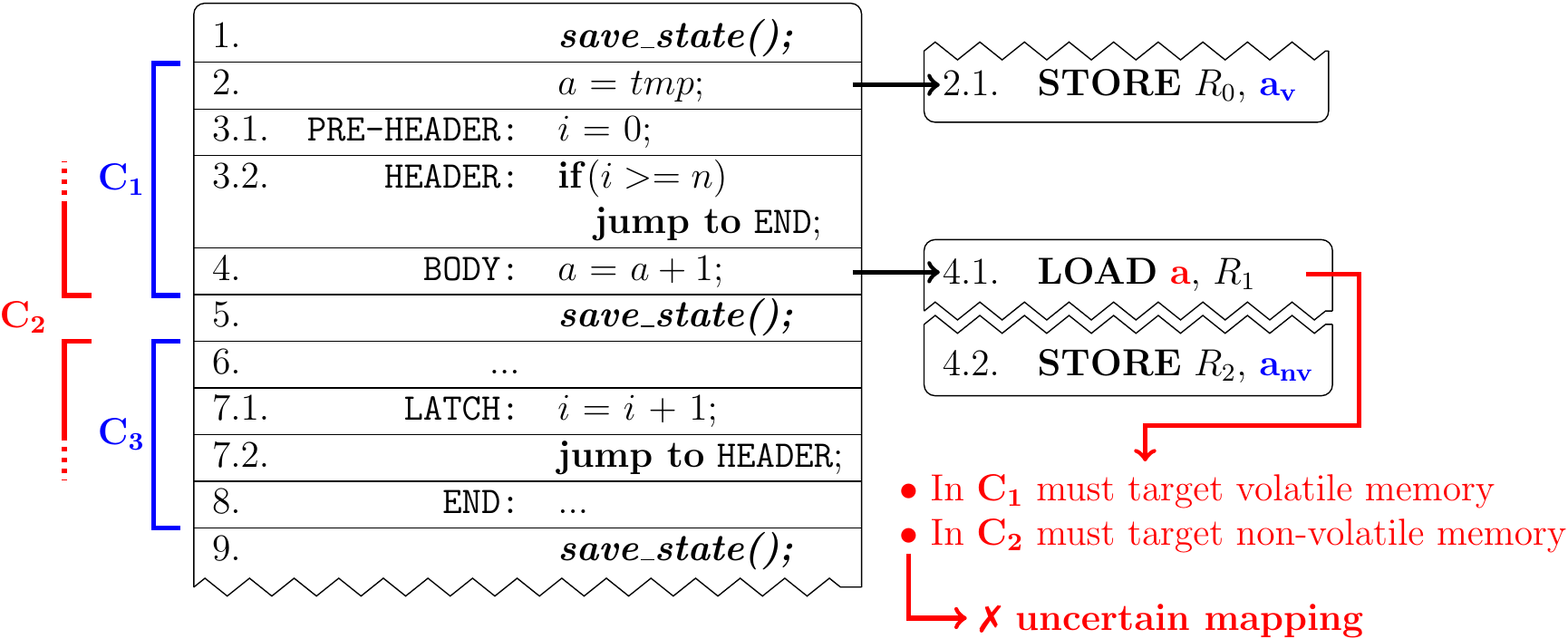}}
        \label{fig:uncertain-ci-loop-cfg}
    }
    \caption{Example of a compile-time uncertainty in computation intervals of loops.}
    \label{fig:uncertain-ci-loop}
\end{figure}

\subsubsection{Loops}
The presence of state-saving operations inside loops may introduce compile-time uncertainty on the instructions that we need to include in a computation interval.

\fakepar{Example}
\figref{fig:uncertain-ci-loop} exemplifies the situation, for which \figref{fig:uncertain-ci-loop-cfg} shows the resulting control flow when the program of \figref{fig:uncertain-ci-loop-program} is translated to machine-code.
Say that we are partitioning the program into computation intervals, required to apply the mapping of read and write operations, described in \secref{sec:memory-principles}.
Following how the loop executes, we identify three computation intervals.
The computation interval $CI_1$ of lines $2$-$4$ includes the instructions executed before the loop, that is, lines $2$-$3$, and part of the instructions executed during the first loop iteration, that is, lines $3$-$4$.
The computation interval $CI_2$ of lines $6$-$7$ and $3$-$4$ includes the remaining instructions executed during the first loop iteration, that is, lines $3$-$4$, and part of the instructions executed during the second loop iteration, that is, lines $6$-$7$.
Note that $CI_2$ covers all the iterations of the loop, except the first and last iterations.
Finally, the computation interval $CI_3$ of lines $6$-$8$ includes the instructions executed during the last loop iteration, that is, lines $6$-$7$, and the instructions executed after the loop, that is, line $8$.

The state-saving operation of line $5$ marks the end of a computation interval in the middle of a loop iteration, making the memory operations of lines $3$-$4$ part of both $CI_1$ and $CI_2$.
Say that we map the memory write operations and that we are to map the memory read operations.
When mapping $CI_1$, we make the \code{LOAD} of line $4$ target volatile memory, as it need to access the intermediate result produced by the \code{STORE} of line $2$.
Instead, when mapping $CI_2$, we make the same \code{LOAD} of line $4$ target non-volatile memory, as it represents the first instruction of $CI_2$ that reads $a$.
This represents a compile-time uncertainty, as the \code{LOAD} of line $4$ cannot target both volatile and non-volatile memory at the same time.

\begin{figure}[t]
    \resizebox{\columnwidth}{!}{\includegraphics{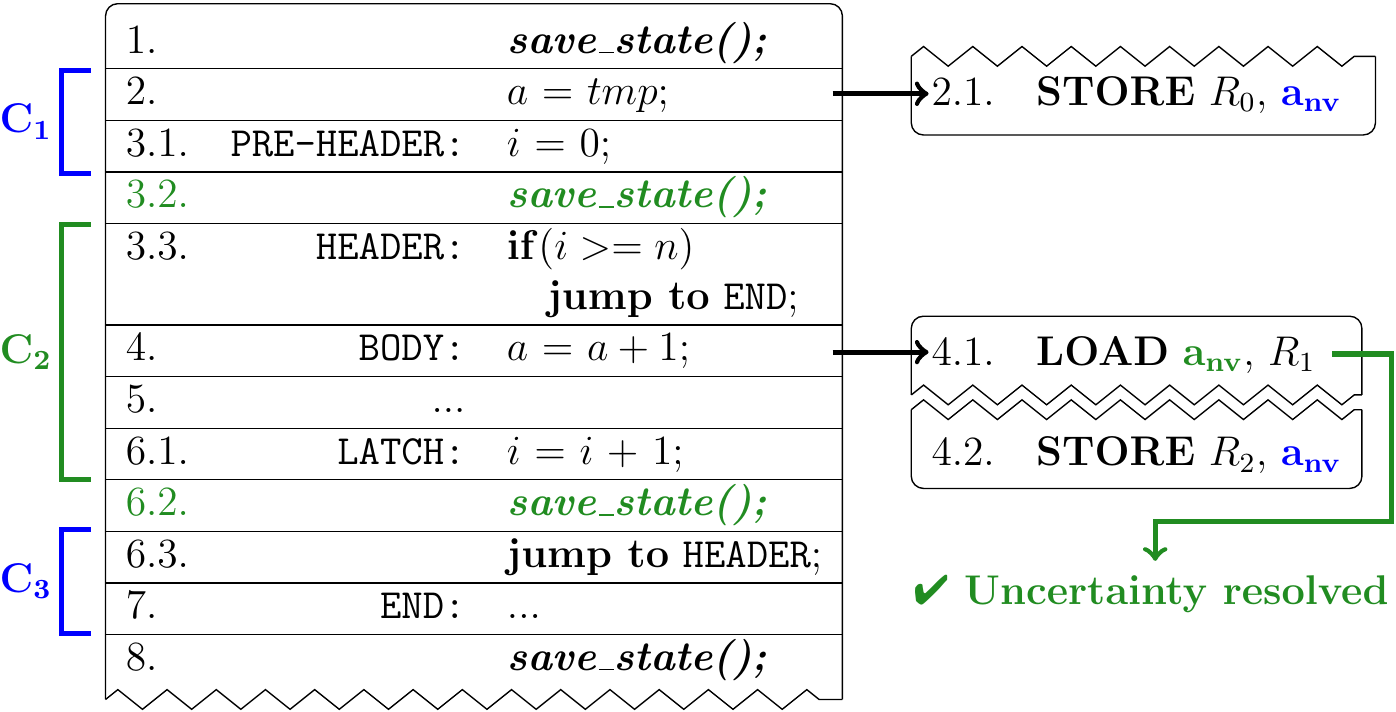}}
    \caption{Normalization of the computation interval boundaries of the example of \figref{fig:uncertain-ci-loop}.}
    \label{fig:uncertain-ci-loop-fix}
\end{figure}

In general, such type of compile-time uncertainty arises when a loop controls the execution of a state-saving operation and one of the resulting computation intervals contains both
\begin{mylist}
    \item a memory read or write operation $I_1$ on a memory location $x$ that executes during a loop iteration and
    \item a memory read or write operation $I_2$ on a memory location $x$ that executes outside a loop
\end{mylist}.
$I_1$ is the instruction that is part of multiple computation intervals and, as a consequence, has an uncertain mapping.
In the example of \figref{fig:uncertain-ci-loop}, $I_1$ is the \code{LOAD} of line $4$ and $I_2$ is the \code{STORE} of line $2$.

\fakepar{Normalization of computation intervals}
To avoid such kind of compile-time uncertainty, we normalize the computation interval boundaries, ensuring that memory read and write instructions executed during the iterations of a loop do not belong to multiple computation intervals.

First, we place a state-saving operation before the loop header.
This ensures that the memory read and write instructions of the header and body cannot belong to a computation interval that starts prior to the loop.
In the example of \figref{fig:uncertain-ci-loop}, we place a state-saving operation between the \texttt{PRE-HEADER} and \texttt{HEADER} of line $3$.

Similarly, we move the last state-saving operation of the loop at the end of the latch, before the jump instruction.
This ensures that the memory read and write instructions of the body and latch cannot belong to a computation interval that starts after to the loop.
In the example of \figref{fig:uncertain-ci-loop}, we move the state-saving operation of line $5$ at line $7$.

\figref{fig:uncertain-ci-loop-fix} shows the result.
The \code{LOAD} of line $4$ of \figref{fig:uncertain-ci-loop} is now part of only one computation interval, that is, the one of lines $3$-$7$ of \figref{fig:uncertain-ci-loop-fix}.

Note that by moving the checkpoint at the end of the latch we do not alter the number of instructions executed between two checkpoints.
For this reason, our transformation does not cause non-terminating path bugs~\cite{CleanCut}, where the program is not able to progress across power failures due to the instructions between two checkpoints consuming more energy than the device can buffer.

\begin{figure}[t]
    \subfigure[Example of a compile-time uncertainty in computation intervals of conditional statements.]{
        \resizebox{0.28\columnwidth}{!}{\includegraphics{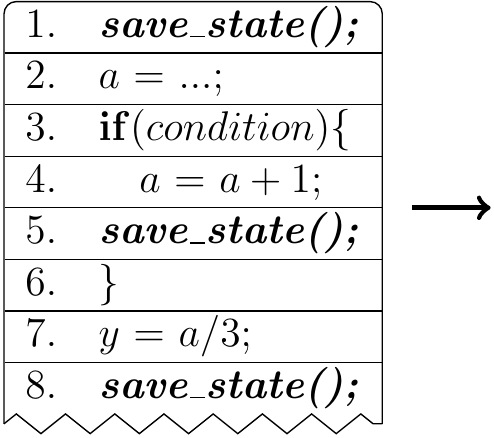}}
        \resizebox{0.67\columnwidth}{!}{\includegraphics{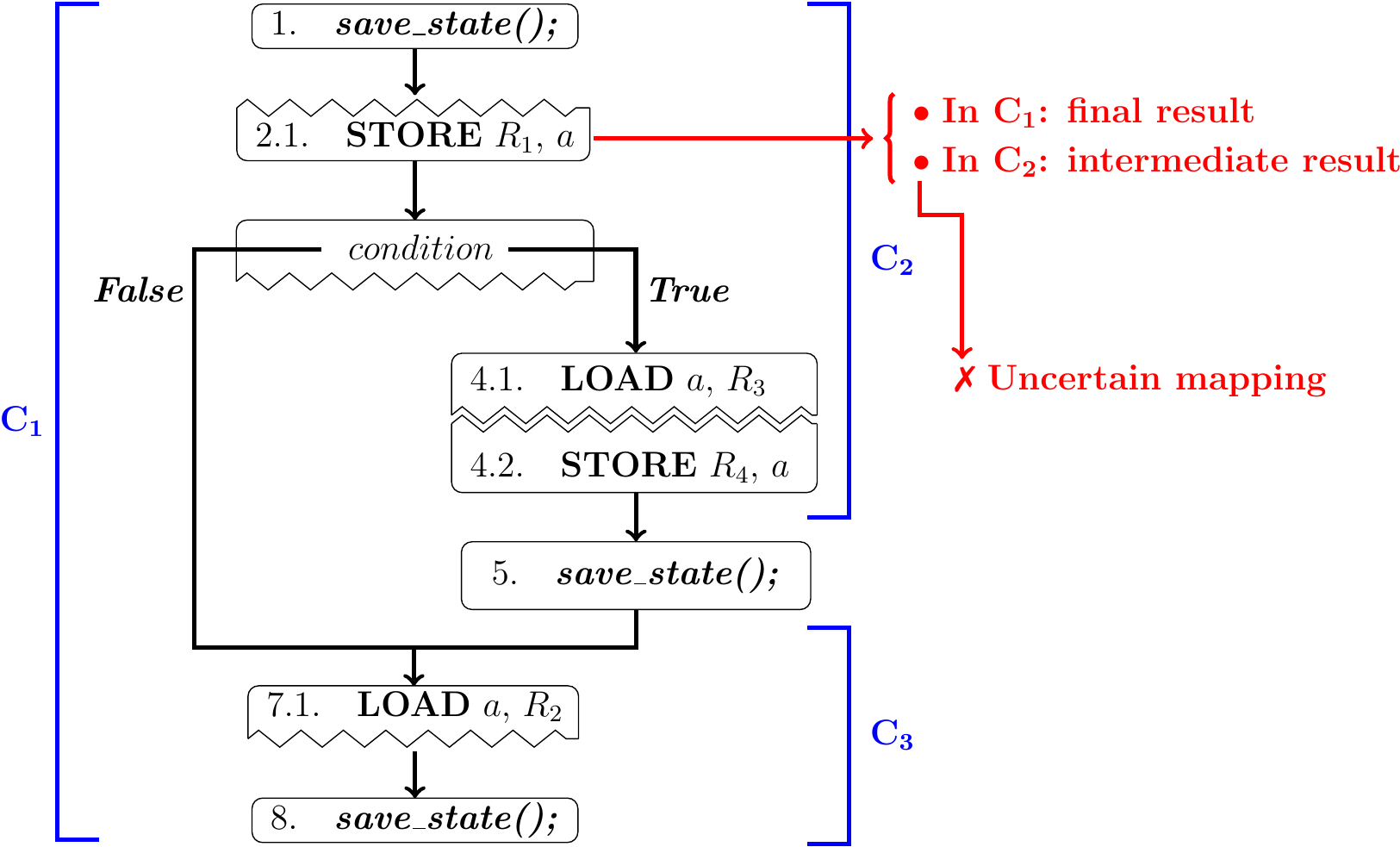}}
        \label{fig:uncertain-ci-conditional}
    }
    \subfigure[Normalization of computation intervals of conditional statements.]{
        \resizebox{0.28\columnwidth}{!}{\includegraphics{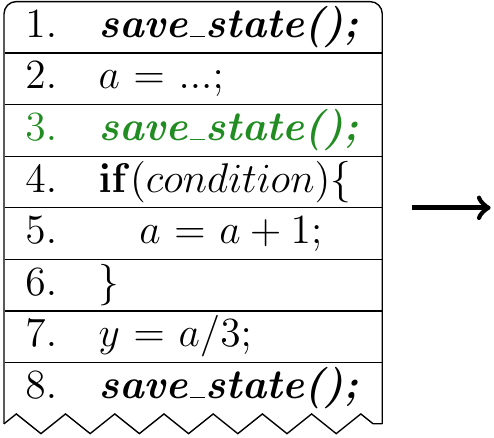}}
        \resizebox{0.67\columnwidth}{!}{\includegraphics{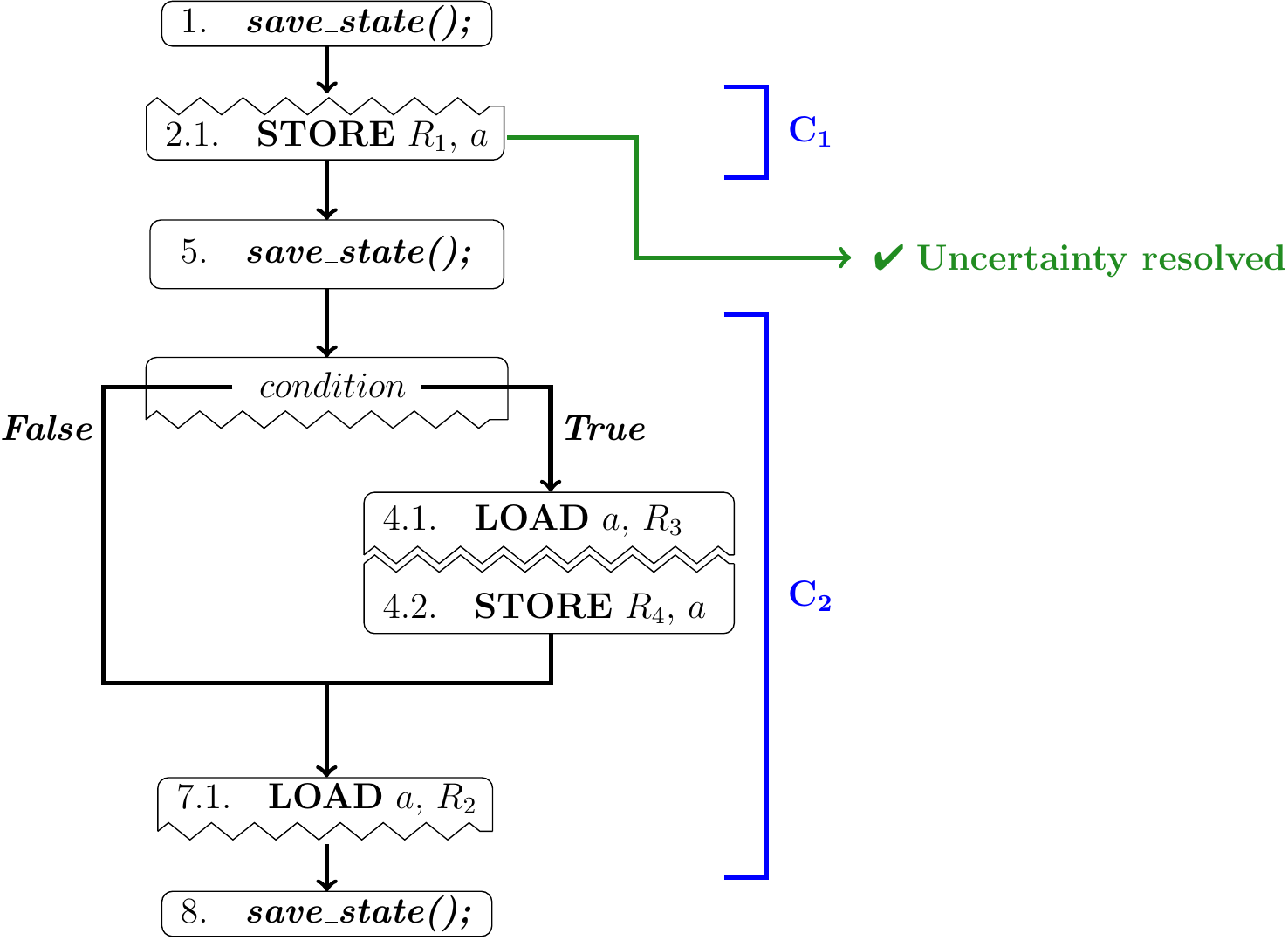}}
        \label{fig:uncertain-ci-conditional-fix}
    }
    \caption{Example of a compile-time uncertainty in computation intervals of conditional statements.}
\end{figure}

\subsubsection{Conditional Statements}
Similarly to the loop case, conditional statements that controls the execution of state-saving operations may introduce a compile-time uncertain.

\figref{fig:uncertain-ci-conditional} exemplifies the situation.
The presence of the $if$ statement of line $2$ causes the \code{STORE} of line $2$ to be part of multiple computation intervals.
This represents the same uncertainty that happens in loops, leading us to an uncertain mapping for the \code{STORE} of line $2$ when we are to map write operations.
The symmetric reasoning is also valid for the \code{LOAD} of line $7$.

\fakepar{Normalization of computation intervals}
Similarly to loops, such type of compile-time uncertainty arises when a memory read or write operation is part of multiple computation intervals.
As such, to avoid such kind of compile-time uncertainty, we rely on the same normalziation technique of loops, which normalizes the computation interval boundaries.

First, we place a state-saving operation before the conditional statement.
In the example of \figref{fig:uncertain-ci-conditional}, we place a state-saving operation before the $if$ statement of line $3$.
This fixes the end of the computation interval prior to the conditional statement.

Then, we remove from each branch the last state-saving operation.
In the example of \figref{fig:uncertain-ci-conditional}, we remove the state-saving operation of line $5$.
\figref{fig:uncertain-ci-conditional-fix} shows the result.

After such transformation, if a branch still contains a state-saving operation, we need to place a state-saving operation after the conditional statement, so to fix the boundary of the computation intervals that the conditional statement controls.
Note that this is required only when a branch contains multiple state-saving operations in the first place, as otherwise would introduce an unnecessary overhead that serves no purpose in avoiding the compile-time uncertainty.
In the example of \figref{fig:uncertain-ci-conditional}, such additional state-saving operation is not required.

\section{Memory Handling}
\label{sec:impl}

To make the techniques of \secref{sec:memory-principles} and \secref{sec:uncertainty} work correctly, we devise a custom memory layout that can be determined at compile-time and a schema to address the possible intermittence anomalies.

\subsection{Memory Layout}
\label{sec:memory-layout}

Despite virtual memory tags ensure we can group instructions that operate on the same memory location,  we still need to identify the addresses of the volatile or non-volatile versions of a memory location to correctly direct read/write operations.

We address this problem by placing the volatile and non-volatile versions of a memory location at the same offset with respect to the corresponding base address.
Note that the compiler treats the two segments as separate memory sections and makes them start at a fixed offset.
This ensures that the volatile and non-volatile versions of the same memory location are at a fixed offset, too.

We can then express the address of the non-volatile version of a memory location as a function of the address of its volatile version, and vice versa.
This allows us to allocate memory operations to either memory segment with ease, even in the presence of indirect accesses through pointers.
For instance, to make an instruction that originally operates on volatile memory now target the non-volatile one, we add the offset between volatile and non-volatile segments to its target address.
We operate the other way around when we make an instruction target volatile memory from the non-volatile one.
When the instruction executes, it retrieves the address information that are unknown at compile-time and calculates the actual target.

\subsection{Dealing with Intermittence Anomalies}
\label{sec:dealing-anomalies}

Using mixed-volatile platforms, the re-executions of non-idempotent portions of code may cause intermittence anomalies~\cite{maioli19lctes, maioli21ewsn, DINO, ratchet,brokenTM}, consisting in behaviors unattainable in a continuous execution.
The problem possibly arises regardless of whether the code is written directly by programmers~\cite{maioli19lctes, maioli21ewsn, DINO, ratchet} or is the result of the program transformations of \secref{sec:memory-principles}.

\begin{figure}[t]
    \subfigure[Example of an intermittence anomaly.]{
        \resizebox{\columnwidth}{!}{\includegraphics{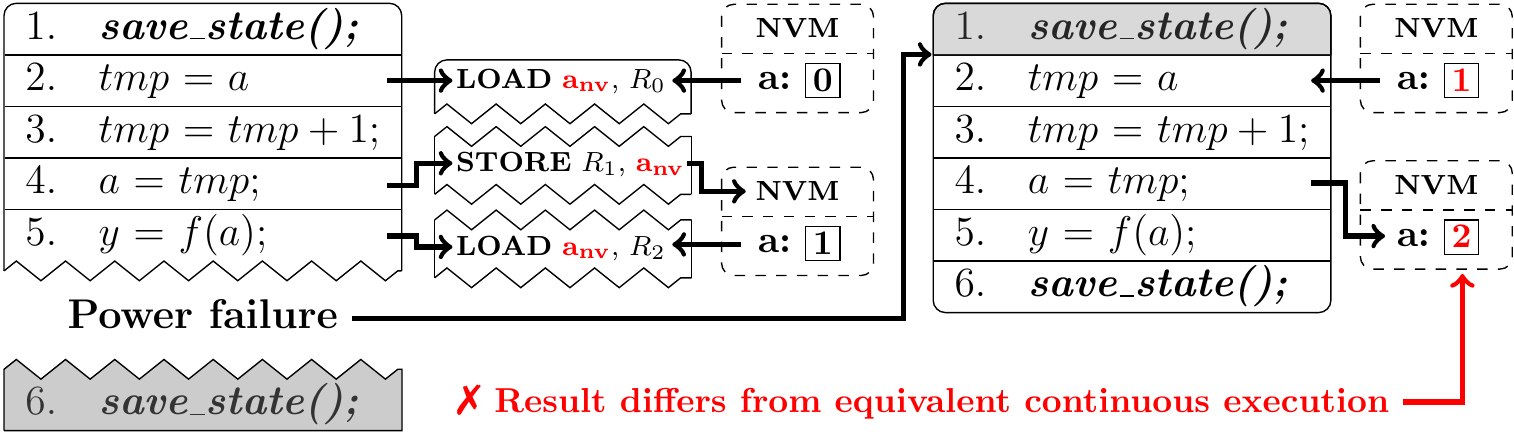}}
        \label{fig:intermittence-anomaly}
    }
    \subfigure[Example of how to avoid the intermittence anomaly with memory versioning.]{
        \resizebox{\columnwidth}{!}{\includegraphics{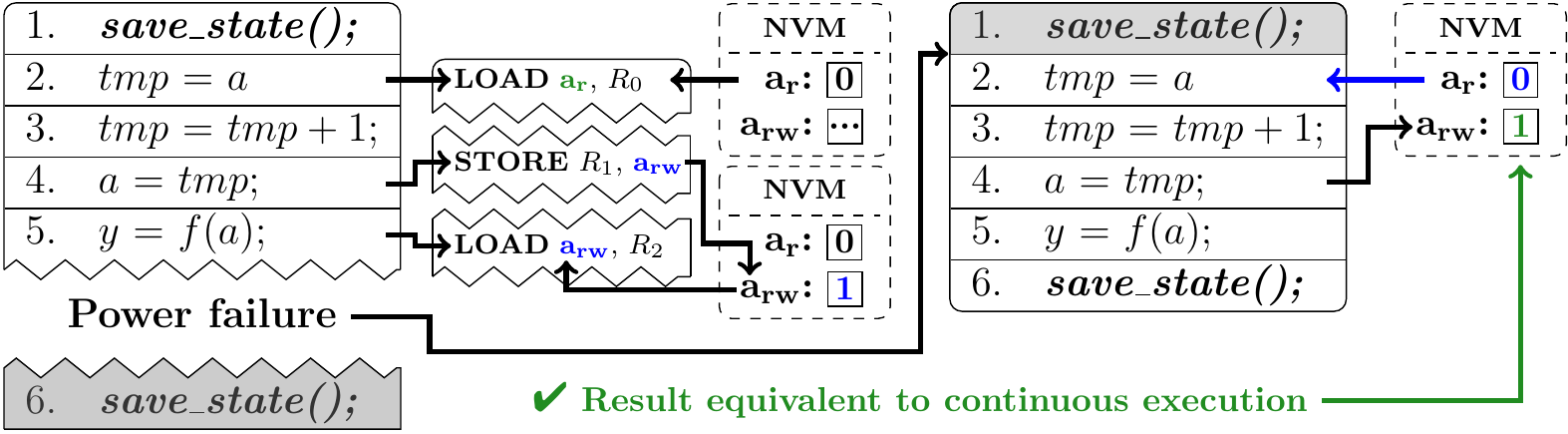}}
        \label{fig:avoiding-intermittence-anomaly}
    }
    \caption{Example of an intermittence anomaly. }
\end{figure}

\fakepar{Example} Consider the program of \figref{fig:intermittence-anomaly}.
Variable $a$ is non-volatile.
Following the state-save operation on line $1$, the current value of variable $a$, that is, $0$, is initially retrieved from non-volatile memory.
The execution continues and line $4$ updates the value of variable $a$ on non-volatile memory to $1$.
This is how a continuous execution would normally unfold.

Imagine a power failure happens right after the execution of line~$4$.
 When the device resumes as energy is back, the program restores the program state from non-volatile memory, which includes the program counter.
The program then resumes from line $2$, which is re-executed.
As variable $a$ on non-volatile memory retains the effects of the operations executed before the power failure, the value read by line $2$ is now $1$, that is, the value written in line $4$ before the power failure in the previous power cycle.
This causes line $4$ to produce a result that is unattainable in any continuous execution, as it updates the value of variable $a$ to $2$, instead of $1$.

Many such situations exist that possibly cause erratic behaviors, including memory operations on the stack and heap~\cite{ratchet,maioli19lctes, maioli21ewsn}.

\fakepar{Memory versioning} Intermittence anomalies happen whenever a power failure introduces a Write-After-Read (WAR) hazard~\cite{maioli19lctes, DINO, ratchet} on a non-volatile memory location.
In \figref{fig:intermittence-anomaly}, the memory read of line $2$ and the memory write of line $4$ represent a WAR hazard for variable $a$.
Several techniques exist to avoid the occurrence of intermittence anomalies~\cite{maioli19lctes, maioli21ewsn, DINO, ratchet, chinchilla, alpaca, clank}.
In general, it is sufficient to break the sequences of instructions involved in WAR hazards~\cite{maioli19lctes, maioli21ewsn, DINO, ratchet} so the involved instructions necessarily execute in different power cycles.
Existing solutions place additional checkpoints~\cite{ratchet} or enforce transactional semantics to specific portions of code ~\cite{alpaca, clank, DINO, chinchilla}.

We use a different approach that tightly integrates with the compile-time operation of \name.
First, to reduce the number of instructions possibly re-executed, every call to a state-save operation in \name systematically dumps the state on non-volatile memory, regardless of the current energy level.
This is different than in many checkpoint systems, where the decision to take a checkpoint is subject to current energy levels~\cite{Mementos,HarvOS,Hibernus,Hibernus++}.
The overhead we impose by doing this is very limited, as state-save operations are limited to register file and program counter after applying the transformations of \secref{sec:memory-principles}.

For each computation interval, we then create two versions of each non-volatile memory location possibly involved in a WAR hazard.
One version is a \textit{read-only copy} and contains the result produced by previous computation intervals; the other version is a \textit{read-and-write copy} and contains the result of the considered computation interval.
We direct the memory read (write) instructions to the read-only (read-and-write) copy.
This ensures that in case of a re-execution, the read operations access the values produced by the previous computation interval, as the (partial) results of the current computation interval remain invisible in the read-and-write copies.
When transitioning to the next computation interval, the read-only and read-and-write copies are swapped to allow the next computation interval to access the (now, read-only) data of the computation interval just concluded.

\figref{fig:avoiding-intermittence-anomaly} shows how this solves the intermittence anomaly of \figref{fig:intermittence-anomaly}.
Line $2$ reads variable $a$'s read-only copy, whereas line $4$ writes variable $a$'s read-and-write copy.
Line $4$ accordingly reads variable $a$'s read-and-write copy, as it needs the data that line $4$ produces.
If a power failure happens after line $4$ and line $2$ is eventually re-executed, that read operation still targets $a$ read-only copy, which correctly reports $0$.
Instead, after swapping the two copies, the next computation interval correctly accesses the copy of variable $a$ that reports value $1$, equivalently to a continuous execution.

We apply this technique as a further code processing step, as shown in stage~\step{5} of \figref{fig:pipeline}.
First, we identify the WAR hazards.
For each memory write instruction $I_w$ on a non-volatile memory location with tag $x$, we check if there exists a memory read instruction $I_r$ such that \emph{i)} $I_r$ targets a non-volatile memory location with the same memory tag $x$, and \emph{ii)} $I_r$ may execute before $I_w$, that is, $I_r$ happens before $I_w$ in the control-flow graph.
If  such $I_r$ exists, the pair $(I_w, I_r)$ represents a WAR hazard.

Next, we create the read-only and read-and-write copies by doubling the space that the compiler normally reserves to the data structure $x$ refers to.
As we allocate the two copies in contiguous memory cells, their relative offset is fixed and may be used at compile time to direct the memory operation to either copy.
We then make $I_r$ target the read-only copy, together with every memory read instruction that operate on $x$ and executes before $I_w$.
In contrast, we make $I_w$ target the read-and-write copy of $x$, together with all corresponding memory read instructions that happen after $I_w$.

\fakepar{Compile-time uncertainty}
As memory versioning is applied after program normalization, the compile-time uncertainty in the order of instruction execution or in the span of computation intervals is already resolved at this stage.

Note that when mapping memory read and write operations to non-volatile memory, there are two different versions of a memory location that refer to two different versions of the same data.
We have the same situation when applying memory versioning.
As such, the applicaton of memory versioning may be subjected to the same compile-time uncertainties that we already describe in \secref{sec:uncertainty}, which we can resolve using the same normalization techniques.
However, memory versioning introduces new memory access patterns then we need to account when applying normalization techniques to avoid compile-time uncertainties.
We describe next how we extend the normalization techniques of \secref{sec:uncertainty} to account for such access patterns.

\begin{figure}[t]
    \subfigure[Example of a compile-time uncertainty during the application of memory versioning.]{
        \resizebox{\columnwidth}{!}{\includegraphics{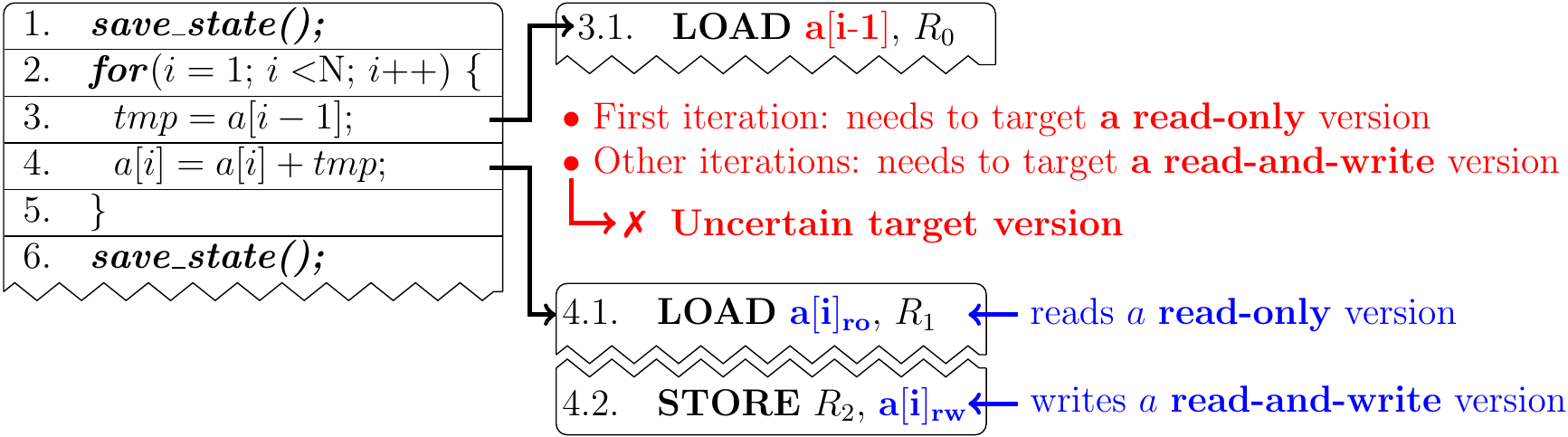}}
        \label{fig:loop-anomaly-uncertainty}
    }
    \subfigure[Normalization of the memory read causing the compile-time uncertainty.]{
        \resizebox{\columnwidth}{!}{\includegraphics{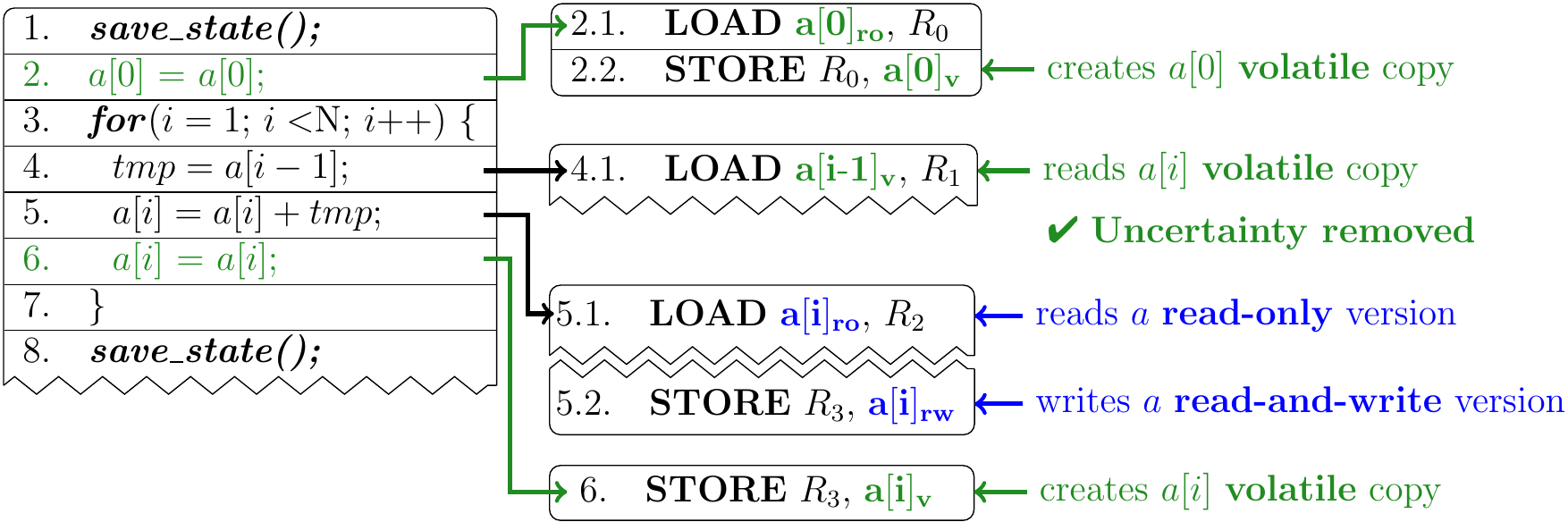}}
        \label{fig:loop-anomaly-uncertainty-fix}
    }
    \caption{Example of a compile-time uncertainty in loops during the application of memory versioning to avoid intermittence anomalies.}
\end{figure}

\subsubsection{Compile-time Uncertainty $\rightarrow$ Loops}
Memory versioning may introduce a new access pattern inside loops that may lead to a compile-time similar to the one described in \secref{sec:uncertainty-loops}.

\fakepar{Example}
\figref{fig:loop-anomaly-uncertainty} exemplifies the situation, where we already applied the mapping of read and write operations, described in \secref{sec:memory-principles}.
Say that we are to apply memory versioning to avoid the WAR hazard of line $4$.
We need to make all the memory read operations before the \code{STORE} of line $4$ target $a$ read-only version.
During the first loop iteration, the \code{LOAD} of line $3$ needs to target $a$ read-only version.
Instead, during all the other loop iterations, the \code{LOAD} of line $3$ needs to target $a$ read-and-write version, as it need to read the data that the \code{STORE} of line $4$ produced during the previous iteration.

This compile-time uncertainty shares the same pattern of the example of \figref{fig:uncertain-last-write}, described in \secref{sec:uncertainty-loops}.
However, differently from such a case, here the uncertainty arises only when memory read and write operations access non-scalar data structures, as the normalization described in \secref{sec:uncertainty-loops} addresses the uncertainty of scalar accesses.

In general, memory versioning introduces a compile-time uncertainty in a loop $L$ whenever 
\begin{mylist}
    \item $L$ contains no state-saving operation,
    \item $L$ controls the execution of a memory read instruction $I_r$ targeting a cell of a non-scalar data structure $x$ that previous loop iterations may write, and
    \item the instructions executed during an iteration of $L$ contain a WAR hazard on some memory cells of $x$
\end{mylist}.

\fakepar{Normalization}
We use a similar concept to the one of \secref{sec:uncertainty-loops} to avoid such compile-time uncertainty.
We rely on dummy-write operations to create volatile copies of the memory cells of the non-scalar data structure $x$ that $I_r$ reads.

Creating a volatile copy of a non-scalar data structure $x$ requires $n$ dummy-writes, where $n$ is the number of memory cells of $x$.
This introduce a significant energy overhead that we reduce by placing dummy-writes where they introduce the lowest possible energy overhead.
In the example of \figref{fig:loop-anomaly-uncertainty}, we place a dummy-write after the \code{STORE} of line $4$, consisting in a \code{STORE} operation targeting $a[i]$ volatile copy.
Such dummy write can avoid to read $a[i]$ from non-volatile memory, as such value is already present in a register.
Hence, it consists only in a \code{STORE} instruction that writes $a[i]$ volatile copy.
This not only minimizes the energy overhead of such dummy-write, but also ensures that we create a volatile copy only for the memory cells that the uncertain memory read $I_r$ targets.
Note that in the example of \figref{fig:loop-anomaly-uncertainty}, $I_r$ is the \code{LOAD} of line $3$.

We now need to place before the loop one dummy-write for each memory cell that $I_r$ targets, but the instructions in the loop body do not write.
In the example of \figref{fig:loop-anomaly-uncertainty}, we insert a single dummy-write before the loop, which creates a copy of the memory cells of $a$ that the \code{STORE} of line $4$ does not write, that is, $a[0]$.

These transformations ensures that, when $I_r$ executes, a volatile copy of the targeted memory cell is available.
Hence, we can now make $I_r$ target volatile memory, resolving the compile-time uncertainty.
\figref{fig:loop-anomaly-uncertainty-fix} shows the results of the normalization applied to the example of \figref{fig:loop-anomaly-uncertainty}.

\begin{figure}[t]
    \subfigure[Example of a compile-time uncertainty during the application of memory versioning.]{
        \resizebox{0.80\columnwidth}{!}{\includegraphics{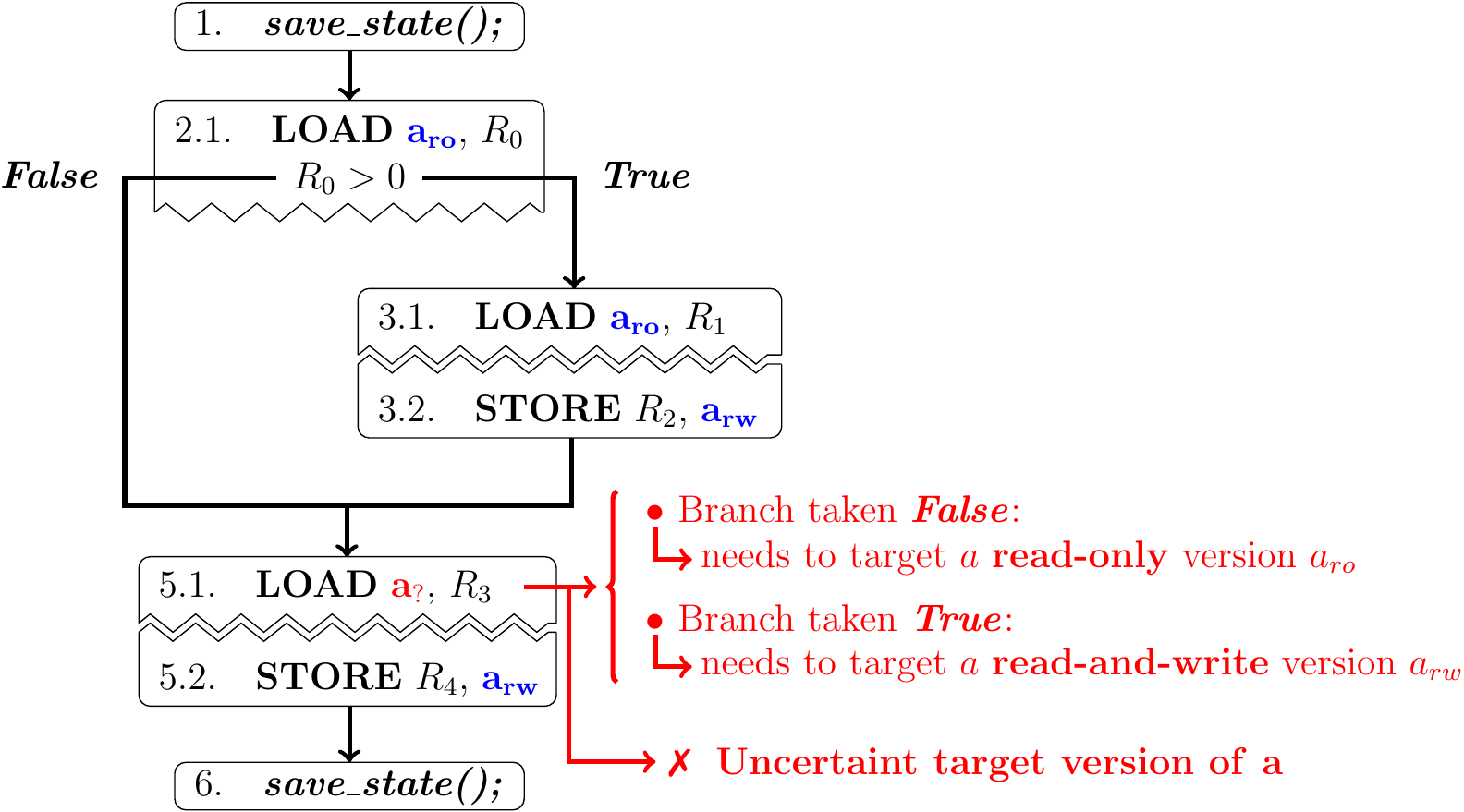}}
        \label{fig:uncertain-versioning-problem}
    }
    \subfigure[Normalization of the memory write causing the compile-time uncertainty.]{
        \resizebox{0.75\columnwidth}{!}{\includegraphics{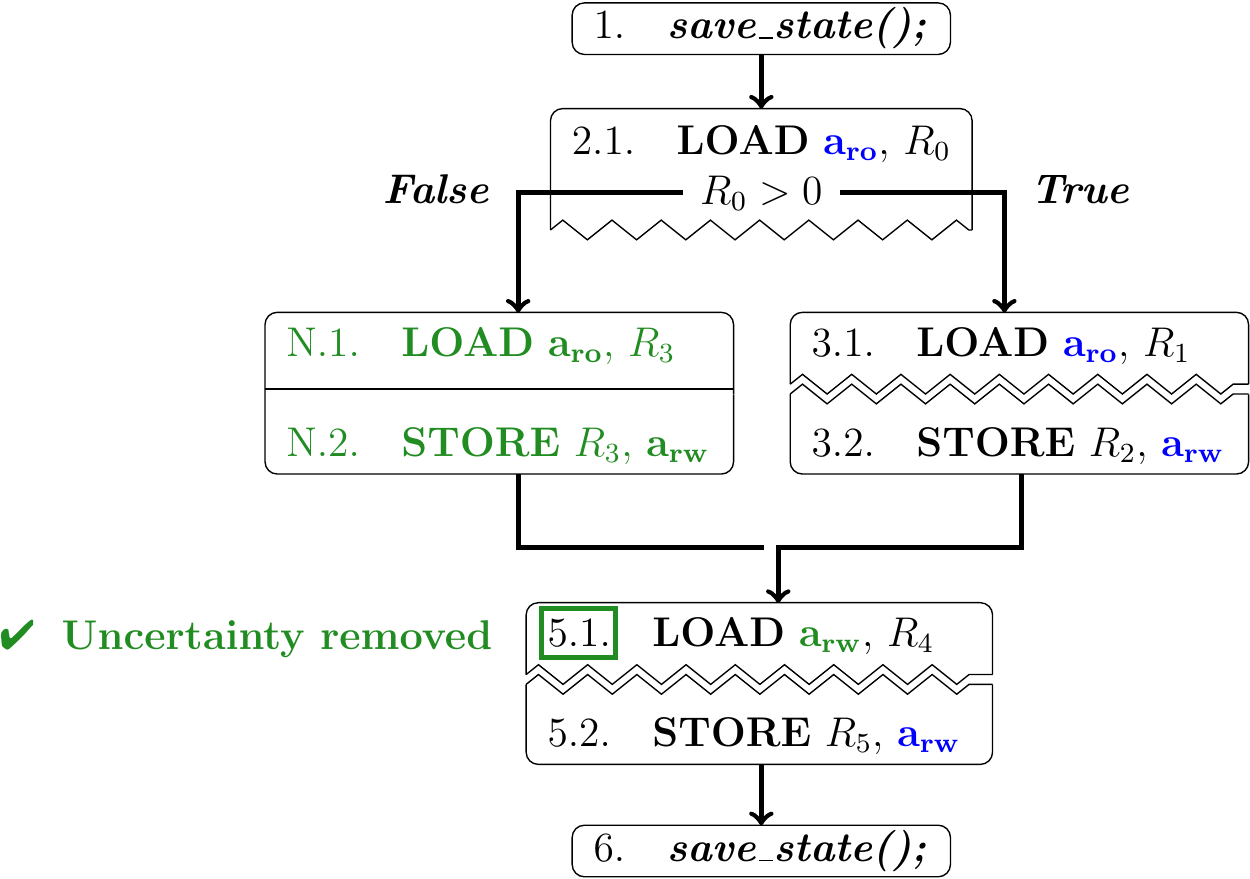}}
        \label{fig:uncertain-versioning-fix}
    }
    \caption{Example of a compile-time uncertainty during the application of memory versioning to avoid intermittence anomalies.}
    \label{fig:uncertain-versioning}
\end{figure}

\subsubsection{Compile-time Uncertainty $\rightarrow$\\Conditionally-executed First Writes}
As we anticipate in \secref{sec:uncertainty-conditionals-map}, we need to address a particular case of compile-time uncertainty that may arise when applying our memory versioning technique to a conditionally-executed memory write operation that targets non-volatile memory and that may execute as first write.

\fakepar{Example}
\figref{fig:uncertain-versioning-problem} exemplifies such situation, which arises when we are to apply memory versioning to the example of \figref{fig:uncertain-conditional}.
Note that the \code{STORE} of line $3$ targets non-volatile memory to address the compile-time uncertainty that was present during the mapping of memory read operations, for which we already provide a detailed description in \secref{sec:uncertainty-conditionals-map}.
When the $if$ statement evaluates to $true$, the \code{LOAD} of line $5$ needs to target $a$ read-only version, as it needs to read the value of $a$ produced during a previous computation interval.
Instead, when the $if$ statement evaluates to $false$, the \code{LOAD} of line $5$ needs to target $a$ read-and-write version, as it needs to read the value of $a$ that the \code{STORE} of line $3$ writes.

This compile-time uncertainty shares the same pattern of the one of the example of \figref{fig:uncertain-conditional}, described in \secref{sec:uncertainty-conditionals-map}.
Moreover, it arises as a consequence of the application of the conservative strategy to deal with the compile-time uncertainty on the \code{LOAD} of line $5$.

In general, such compile-time uncertainty arises whenever in a computation interval 
\begin{mylist}
    \item a memory location $x$ requires two non-volatile versions due to the presence of a WAR hazard,
    \item a memory write operation $I_w$ that targets $x$ may execute as first write, and
    \item a memory read operation $I_r$ can both read the result of $I_w$ or of another operation
\end{mylist}.
Note that $I_r$ is the same memory read instruction involved in the compile-time uncertainty that we describe in \secref{sec:uncertainty-conditionals-map}.

\fakepar{Normalization}
To avoid such uncertainty we need to apply the non-conservative strategy, instead of the conservative one, when dealing with the normalization of $I_r$, as the uncertainty is a consequence of the application of the conservative strategy.
This ensures that $I_r$ always reads the result of $I_w$, removing the compile-time uncertainty on the memory location it needs to target.
Hence, when we map the memory read and write operations in the example of \figref{fig:uncertain-versioning-problem}, instead of making the \code{STORE} of line $3$ and the \code{LOAD} of line $5$ target non-volatile memory, that is, the conservative strategy, we insert a dummy-write targeting $a$ in the $false$ branch of the $if$ statement of line $2$, that is, the non-conservative strategy.
\figref{fig:uncertain-versioning-fix} shows the result.

\subsubsection{Compile-time Uncertainty $\rightarrow$ Partial Updates to Non-scalar Data Structures}
\label{sec:partial-updates}
Partial updates to non-scalar data structures, such as arrays or structs, may leave the read-and-write version of the data structure in an uncertain state, as the non-updated memory cells may contain old or invalid data.
This introduces a compile-time uncertainty on the version of the data structure that computation intervals need to consider as read-only and read-and-write versions.

\begin{figure}[t]
    \resizebox{\columnwidth}{!}{\includegraphics{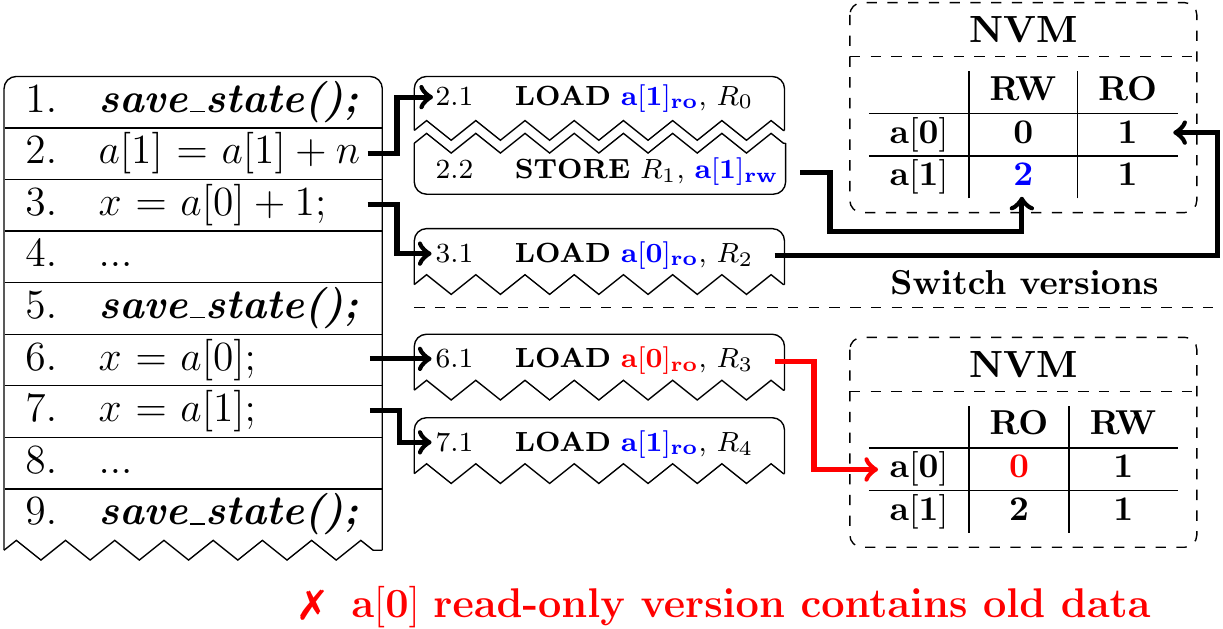}}
    \caption{Example of a compile-time uncertainty due to a partial update of a non-scalar data structure.}
    \label{fig:partial-update}
\end{figure}

\fakepar{Example}
\figref{fig:partial-update} exemplifies this situation.
Say that we already mapped the memory read and write operations, as described in \secref{sec:memory-principles}, and that we are to apply memory versioning.
We need to create two versions of $a$, as there is a WAR hazard between the \code{LOAD} and \code{STORE} instructions of line $2$.
We make the \code{LOAD} of line $2$ target $a$ read-only version and the \code{STORE} of line $2$ target $a$ read-and-write version.
Then, we make the \code{LOAD} of line $3$ target $a$ read-only version, as no prior instruction writes the memory cell of $a$ that it targets.
The instructions in the computation interval of lines $2$-$4$ update only $a[1]$ and the read-and-write version of $a[0]$ contains a previous value of $a[0]$.
Before applying memory versioning to the computation interval of lines $6$-$9$, we need to switch the read-only and read-and-write versions of $a$, so that the \code{LOAD} instructions of lines $6$ and $7$ target the version that contains the most updated value of $a$.
However, being $a[0]$ not modified in the computation interval of lines $2$-$4$, the \code{LOAD} of lines $6$-$7$ need to target two different versions of $a$.
The \code{LOAD} of line $6$ needs to read the data that is now contained in the read-and-write version of $a$, whereas the \code{LOAD} of line $7$ needs to read the data that is now contained in the read-only version of $a$.
Note that this situation happens as, in the computation interval of lines $6$-$8$, $a$ read-only version does not contain the updated value of each cell, as it is spread across $a$ read-only and read-and-write versions.

In general, such compile-time uncertainty happens whenever the instructions of a computation interval $C_1$ write a subset of a non-scalar data structure $x$ and a computation interval $C_2$, which executes after $C_1$, contains
\begin{mylist}
    \item a memory read instruction $I_{r1}$ that reads a memory cell of $x$ that is updated in $C_1$ and
    \item a memory read instruction $I_{r2}$ that reads a memory cell of $x$ that is not updated in $C_1$
\end{mylist}.
Note that if $C_1$ is part of a loop and the instructions it contains end up in writing all the cells of $x$, no instruction of a computation interval $C_2$ outside the loop can read a not-updated memory cell of $x$.
In the example of \figref{fig:partial-update}, $C_1$ ($C_2$) is the computation interval of lines $2$-$4$ ($6$-$8$).
Moreover, the \code{LOAD} of line $7$ ($6$) is $I_{r1}$ ($I_{r2}$).

\begin{figure}[t]
    \subfigure[Copy forward strategy]{
        \resizebox{\columnwidth}{!}{\includegraphics{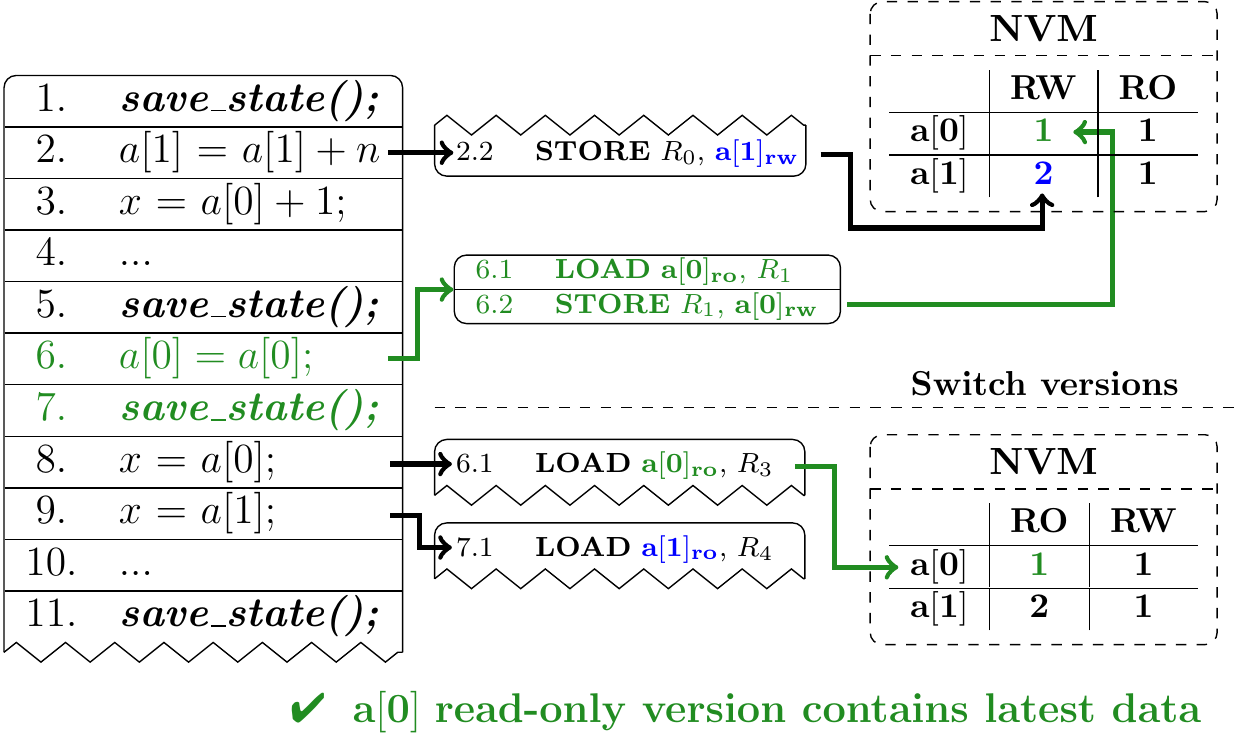}}
        \label{fig:partial-update-fix-copy-forward}
    }
    \subfigure[Copy back strategy]{
        \resizebox{\columnwidth}{!}{\includegraphics{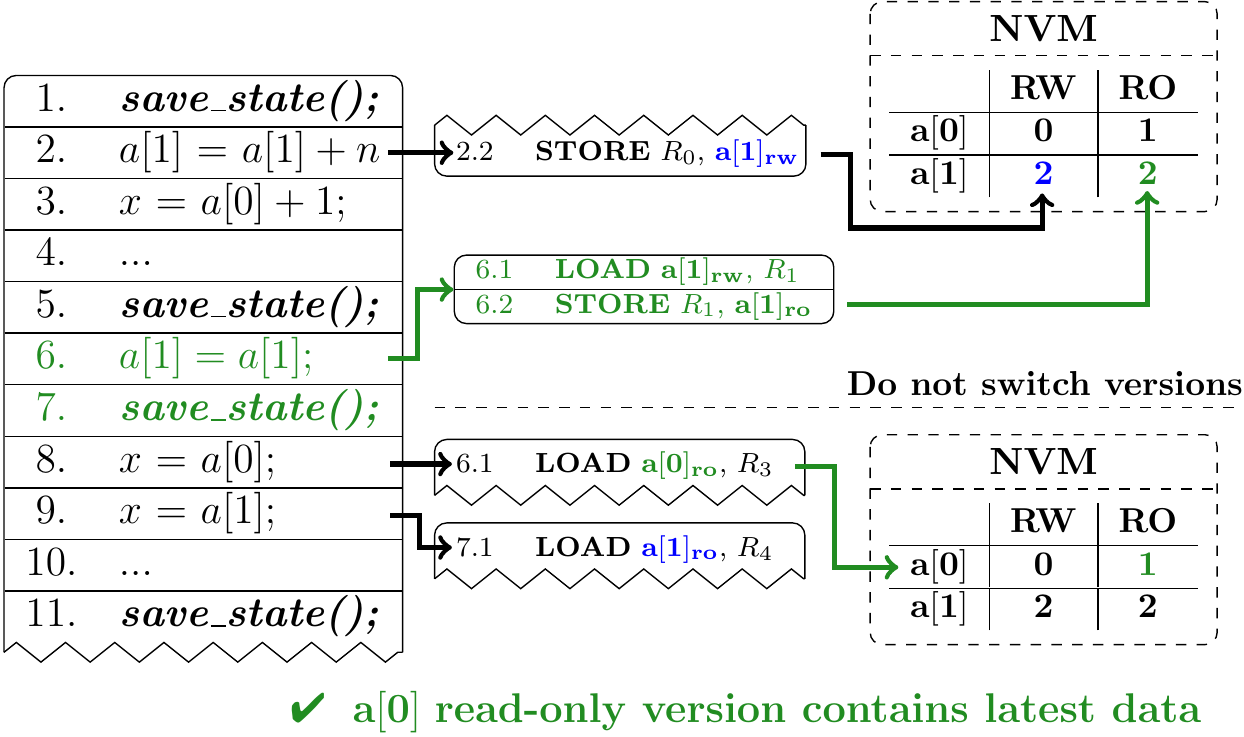}}
        \label{fig:partial-update-fix-copy-back}
    }
    \caption{Normalization of the example of \figref{fig:partial-update} to avoid the compile-time uncertainty due to a partial update of a non-scalar data structure.}
    \label{fig:partial-update-fix}
\end{figure}

\fakepar{Normalization}
To address this issue, we normalize the version of the non-scalar data structure $x$ that a computation interval considers as read-only by copying from the other version the data that it lacks.
We consider two possible strategies: a \textbf{copy forward} and \textit{copy back}.

The \textbf{copy forward} strategy uses a set of dummy-write operations to copy the value of the non-updated memory cells from the read-only version of $x$ to the read-and-write version of $x$.
In the example of \figref{fig:partial-update}, we add a dummy-write after the computation interval of lines $2$-$4$ to copy the value of $a[0]$ from $a$ read-only version to $a$ read-and-write version.
\figref{fig:partial-update-fix-copy-forward} shows the result.

Instead, the \textbf{copy back} strategy uses a set of dummy-write operations to copy the value of the updated memory cells from the read-and-write version of $x$ to the read-only version of $x$.
Note that in such a case, we need not to switch the read-only and read-and-write versions of $x$, as the updated data resides in the read-only version of $x$.
In the example of \figref{fig:partial-update}, we add a dummy-write after the computation interval of lines $2$-$4$ to copy the value of $a[1]$ from $a$ read-and-write version to $a$ read-only version.
Then, the computation interval of lines $2$-$4$ considers as $a$ read-only (read-and-write) version the same read-only (read-and-write) version that the computation interval of lines $6$-$8$ considers.
\figref{fig:partial-update-fix-copy-back} shows the result.

As \figref{fig:partial-update-fix} shows, idependently of the normalization strategy, we place dummy-write operations in a custom computation interval that execute after the computation interval $C_1$ that partially updates the non-scalar data structure.
This ensures that the dummy-write operations required for the normalization
\begin{mylist}
    \item cannot introduce any intermittence anomaly, and
    \item cannot increase the energy consumption of the computation interval where the execuute in such a way that makes the program unable to reach the next state-saving operation~\cite{CleanCut}
\end{mylist}.
Note that the state-saving operation that ends the custom computation interval introduces a low energy and computation overhead, as it needs only to save the program counter.
In fact, the general purpose register used by the dummy-write operations does not contain data read by the program and consequently need not to be preserved.

\fakepar{Strategy selection}
We select the strategy that minimizes the overhead of dummy-write operations, that is, the one that minimizes the number of dummy-write operations.
We identify as $n$ the number of memory cells of a non-scalar data structure $x$ and as $n_{new}$ the number of memory cells of $x$ that a computation interval writes.
When $n_{new} \ge \frac{n}{2}$, we apply the \textit{copy forward} strategy.
Otherwise, we apply the \textit{copy back} strategy.
In the example of \figref{fig:partial-update}, we apply the \textit{copy forward} strategy, as $n$ is $2$ and $n_{new}$ is $1$.

Note that when the updated memory cells are not predictable at compile-time, we instrument the program to track the updated memory cells, using a technique similar to differential checkpoints~\cite{DICE}.
Then, at runtime, we evaluate and execute the strategy that introduces the lowest overhead.

\begin{figure}[b]
  \vspace{-3mm}
  \resizebox{.7\columnwidth}{!}{
    \begin{tabular}{|p{0.19\textwidth}|c|}
        \hline
        \textbf{Dimension} & \textbf{Possible instances} \\
        \hline
        Memory configuration & \baselineV, \baselineNV\\
        \hline
        \shortstack{~ \\ Checkpoint call placement \\ ~ \\ ~}  & \shortstack{\latch, \freturn, \\ \idempotent} \\
        \hline
        Checkpoint execution & \trigger, \execute\\
        \hline
    \end{tabular}}
    \vspace{-2.5mm}
    \caption{Design dimensions for baselines.}
    \label{fig:dimensions}
\end{figure}

\begin{figure}[t]
    \subfigtopskip = -2pt
    \subfigure{
        \centering
        \resizebox{\columnwidth}{!}{\includegraphics{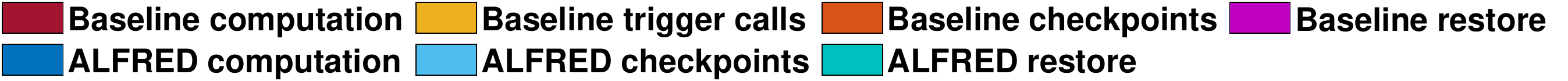}}
    }
    \setcounter{subfigure}{0}
    \subfigure[Energy consumption]{
        \centering
        \resizebox{0.8\columnwidth}{!}{\includegraphics{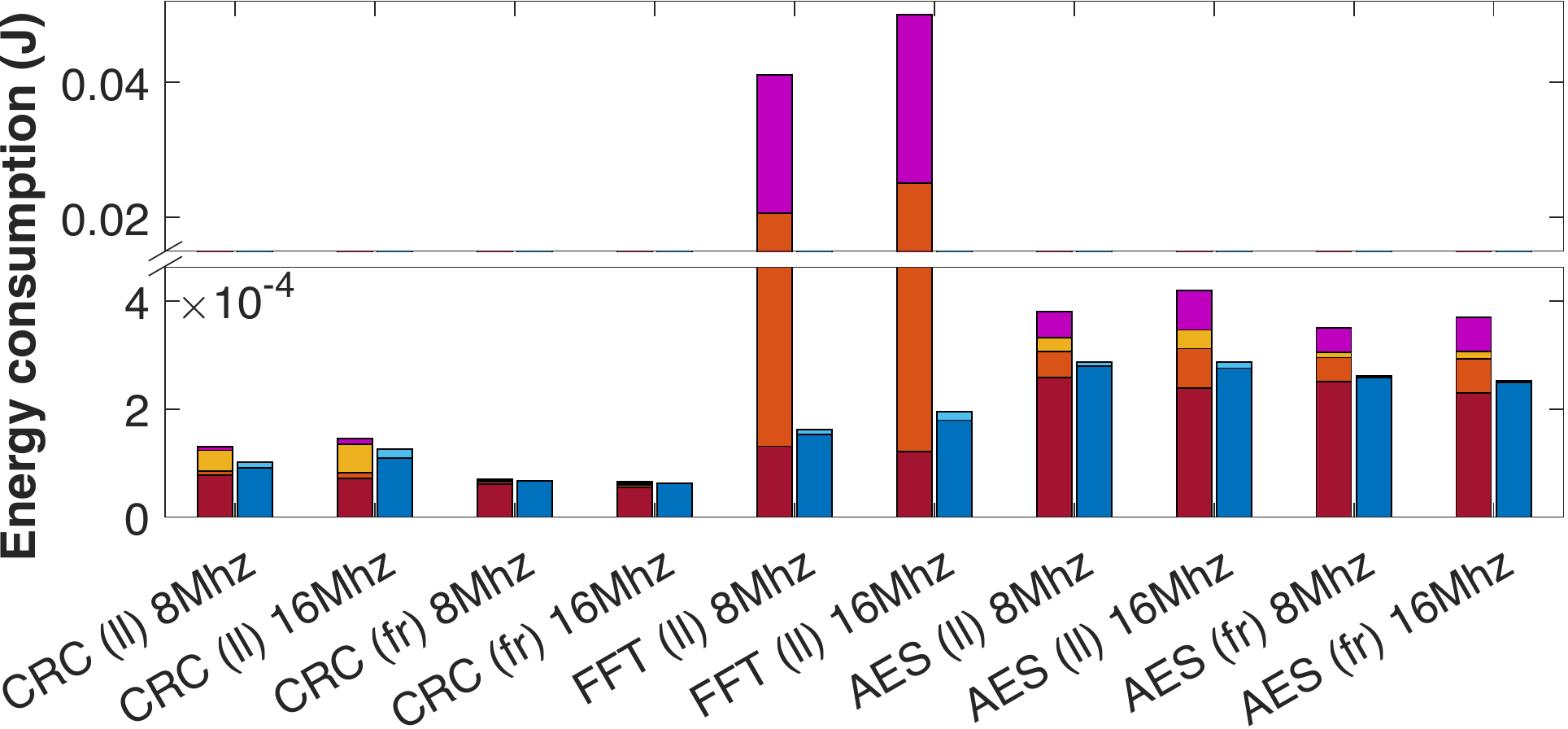}}
        \label{fig:mementos_power_failures_energy}
    }
    \subfigure[Clock cycles]{
        \centering
        \resizebox{0.8\columnwidth}{!}{\includegraphics{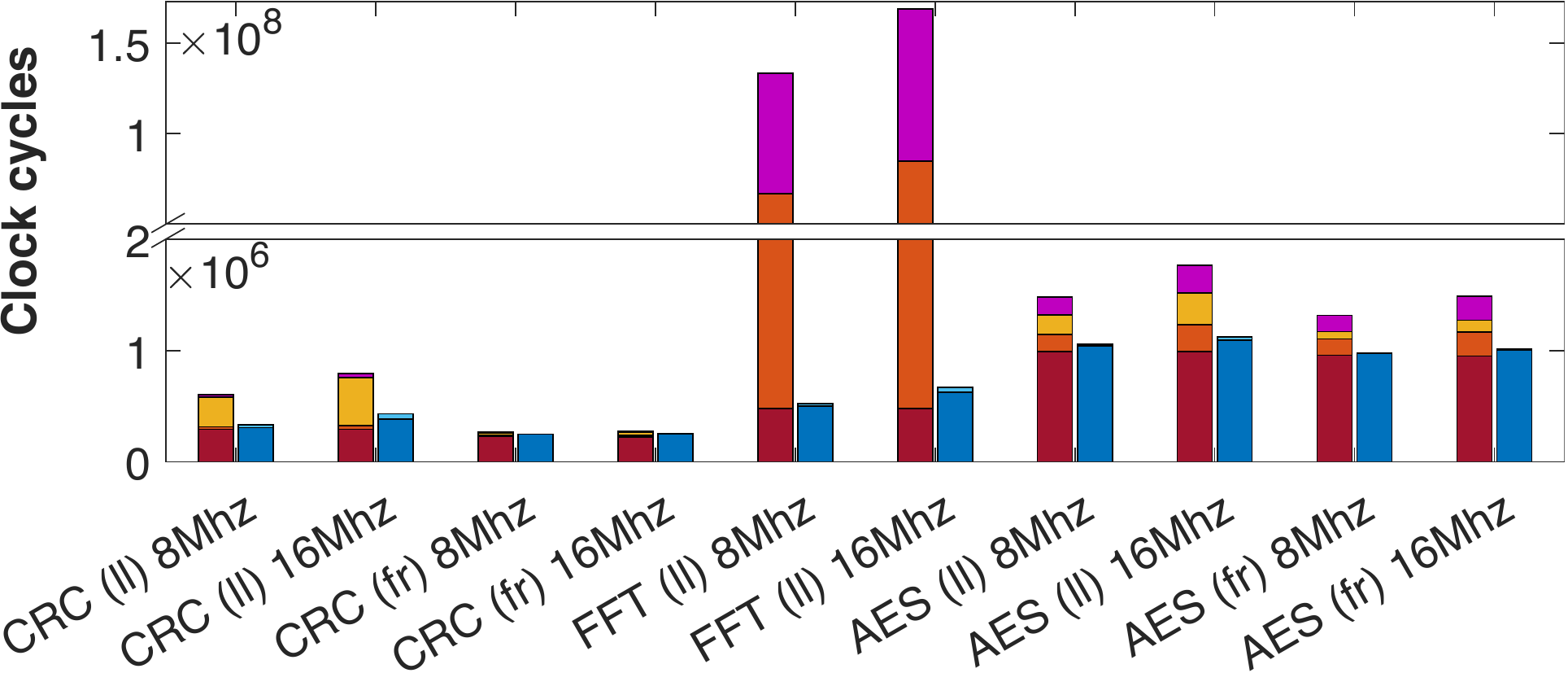}}
        \label{fig:mementos_power_failures_cc}
    }
    \vspace{-3mm}
    \caption{Energy consumption and number of clock cycles comparing \name with a baseline using \baselineV, \trigger, and either \latch or \freturn. \capt{For a baseline, 'll' or 'fr' indicate \latch or \freturn.}}
    \label{fig:mementos_power_failures_energy_cc}
\vspace{-3mm}
  \end{figure}

\section{Evaluation}
\label{sec:eval}
Our evaluation of \name considers multiple dimensions.
We describe next the experimental setup and the corresponding results.

\subsection{Setting}

We opt for system emulation over hardware-based experimentation, as it enables better control on experiment parameters and allows us to carefully reproduce program execution and energy patterns across \name and the baselines we consider.
Because of the highly non-deterministic behavior of energy sources, achieving perfect reproducibility is extremely challenging using real devices~\cite{EKHO}.

\fakepar{Tool and implementation} We use \sceptic~\cite{maioli21ewsn, sceptic-repo}, an open-source extensible emulation tool for intermittent programs. 
\revision{\sceptic emulates the execution of the LLVM Intermediate Representation (IR)~\cite{llvm} of a source code and provides bindings for implementing custom extensions to \emph{i)} apply program transformations and \emph{ii)} map specific performance metrics of the IR to those of machine-specific code, for example, to measure energy consumption.}

\sceptic organizes the LLVM IR into a set of Abstract Syntax Trees (ASTs), one for each function in source code.
Each of these ASTs is generated by augmenting the original LLVM AST with dedicated \sceptic elements, which represent information on the emulated instructions and architectural elements, such as I/O operations and registers.
We implement the pipeline of \figref{fig:pipeline} from stage~\step{3} onwards as a set of further transformations of these ASTs.
\revision{An open-source prototype release of our \sceptic extension implementing \name transformations is available~\cite{alfred-repo}.}

We create a \name module inside \sceptic, which implements part of the pipeline of \figref{fig:pipeline} as a set of transformations and analysis of these ASTs, each one contained in a dedicated submodule.

The \textbf{ASTParser} submodule reorganizes the instructions of ASTs into basic blocks, and groups the basic blocks that represent a loop (conditional statement) onto a \textit{LoopBlock} (\textit{ConditionalBlock}).
Such instruction organization ease the logic complexity of further transformations and analysis passes, as not only conditional statements and loops correspond to dedicated objects, but also the ASTs no longer have backward edges or bifurcations.
Next, the \textbf{ComputationIntervalsNormalizer} submodule applies the normalization techniques of \secref{sec:uncertainty-checkpoints} to all the instances of \textit{LoopBlocks} and \textit{ConditionalBlocks}, removing the compile-time uncertainty in the span of computation intervals.
The \textbf{ComputationIntervalsManager} submodule now extracts the basic blocks contained in the instances of \textit{LoopBlocks} and \textit{ConditionalBlocks} that include a state-saving operation, and then splits each ASTs into computation intervals.
Then, the \textbf{MemoryTagParser} submodule associates to each memory read and write operation the memory tag of the targeted memory location.
Note that the memory tag is identified from the metadata that \sceptic associates to each instruction, which consists in debug symbols.
The resulting ASTs represent stage~\step{3} of the pipeline shown in \figref{fig:pipeline}.

The \textbf{VirtualMemoryTransformation} submodule applies the techniques of \secref{sec:memory-principles}.
First, it normalizes each computation interval, using the techniques of \secref{sec:uncertainty-loops}, \secref{sec:uncertainty-conditionals-map}, and \secref{sec:uncertainty-function-calls} to address the compile-time uncertainty of loops, conditional statements, and function calls, respectively.
Then, it maps memory read and write operations, as described in \secref{sec:implicit-memory-save} and \secref{sec:implicit-memory-restore}.
Finally, it consolidates memory reads operations, as described in \secref{sec:reads}, using the iterative approach of \secref{sec:uncertainty-conditionals-reads} to address compile-time uncertainty.
Note that when normalizing the ASTs, the VirtualMemoryTransformation module accounts also for the normalization extensions of \secref{sec:dealing-anomalies}, necessary for memory versioning.
The resulting ASTs represent stage~\step{4} of the pipeline shown in \figref{fig:pipeline}.

As \sceptic abstracts the emulated MCU memory, we need not reproduce the memory layout and versioning of \secref{sec:impl}, applied during stage~\step{5} of \figref{fig:pipeline}.
However, to differentiate volatile and non-volatile memory accesses, we flag each memory read and write, specifying whether it targets volatile or non-volatile memory.
Finally, the \textbf{VirtualMemoryTransformation} submodule simply joins the computation intervals of each AST, reaching stage~\step{6} of \figref{fig:pipeline}.

\revision{We also implement a machine-specific \sceptic extension to map the execution of the IR to the energy consumption of the MSP430-FR5969~\cite{msp430fr5969}, a low-power MCU that features an internal and directly-addressable FRAM as non-volatile memory.
The MSP430-FR5969 is often employed for intermittent computing~\cite{Mementos, Hibernus, ratchet, alpaca, DINO}.
Our extension takes as configuration parameters the energy consumption per clock cycle of various operating modes of the MSP430-FR5969~\cite{msp430fr5969}, such as regular computation, (non-)vo\-latile memory read/write operations, and peripheral accesses.}

\fakepar{Baselines and benchmarks}
\revision{We compare \name with checkpoint mechanisms that instrument programs automatically~\cite{chinchilla, Mementos, HarvOS, ratchet} by placing calls to a checkpoint library at specific places in the code.}
\revision{We do not consider, instead, checkpoints mechanisms that use interrupts to trigger the execution of checkpoints~\cite{Hibernus, Hibernus++, QuickRecall, jayakumar-hybrid-nvm-mapping, TICS}, including TICS~\cite{TICS} and the work of Jayakumar et. al~\cite{jayakumar-hybrid-nvm-mapping}, as checkpoints do not execute at pre-defined places in the code and thus boundaries of computation intervals cannot be identified.}
\revision{The latter is required for \name to apply the transformations of \secref{sec:memory-principles}.}

\revision{Due to the variety of existing compile-time checkpoint systems, we abstract the key design dimensions in a framework that allows us to instantiate baselines that correspond to existing works, while retaining the ability to explore configurations not strictly corresponding to available systems.}
\revision{\figref{fig:dimensions} summarizes these dimensions.}

On such design dimension is the \emph{memory configuration}.
We consider two possible instances, \baselineV and \baselineNV.
\baselineV allocates the entire main memory onto volatile memory.
To ensure forward progress, each checkpoint must therefore save the content of main memory, register file, and special registers onto non-volatile memory.
This is the case, for example, in Mementos~\cite{Mementos} and HarvOS~\cite{HarvOS}.
Instead, the \baselineNV instance allocates the entire main memory onto non-volatile memory.
Here checkpoints may be limited to saving the content of the register file and program counter onto non-volatile memory, as main memory is already non-volatile.
This is the case of Ratchet~\cite{ratchet}.

\begin{figure}[t]
    \subfigtopskip = -2pt
    \subfigure{
        \hspace{10pt}
    }
    \subfigure{
        \includegraphics[width=0.65\columnwidth]{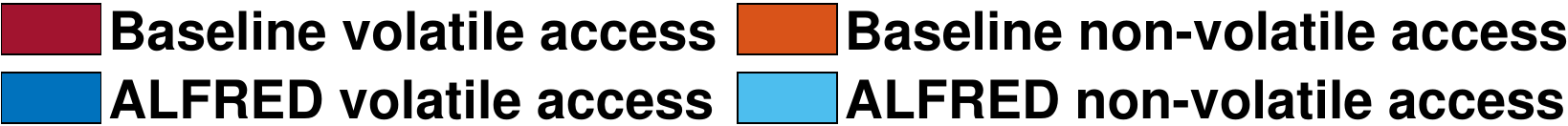}
    }
    \subfigure{
        \includegraphics[width=0.8\columnwidth]{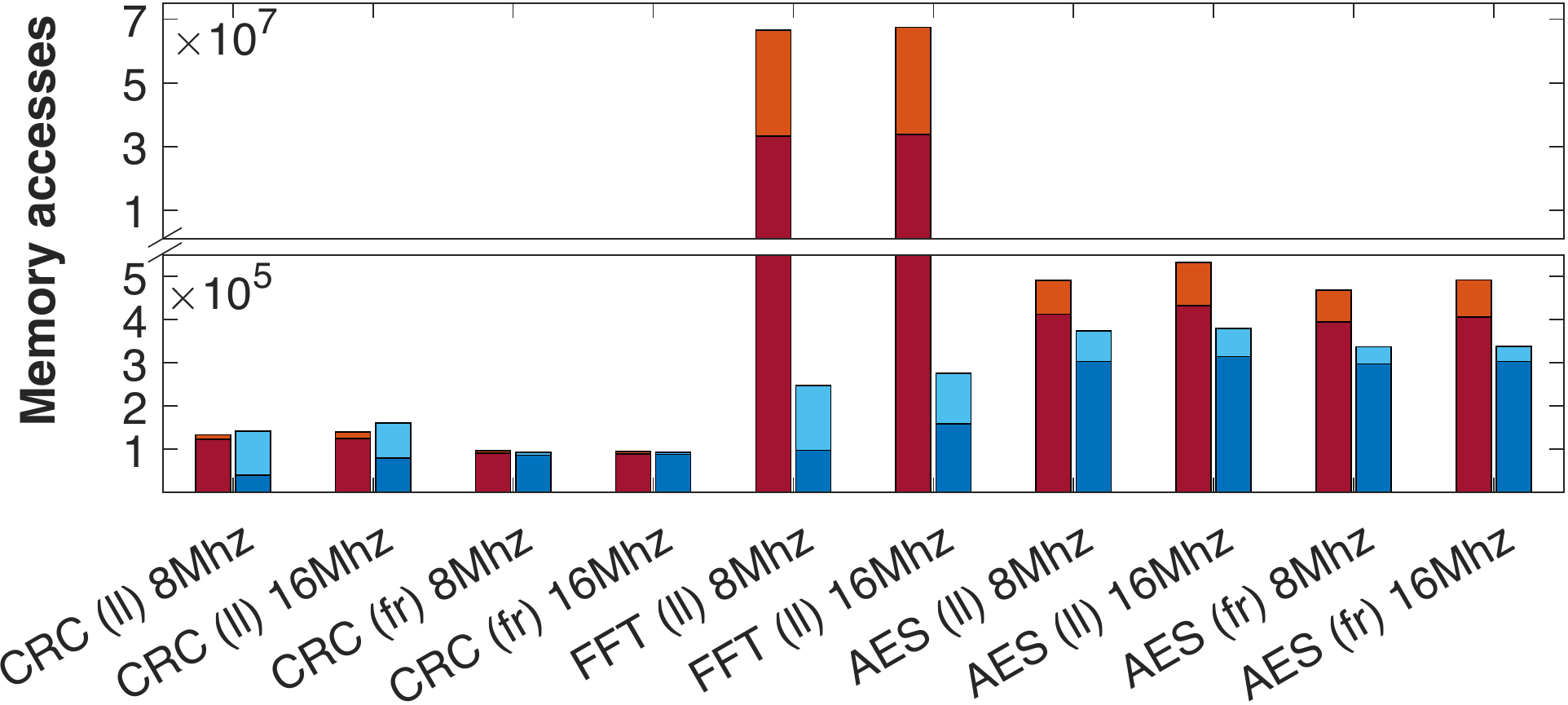}
    }
    \vspace{-3mm}
    \caption{Memory accesses comparing \name with a baseline using \baselineV, \trigger, and either \latch or \freturn.}
    \label{fig:mementos_power_failures_memory}
    \vspace{-3mm}
\end{figure}

\begin{figure}
    \resizebox{\columnwidth}{!}{
        \begin{tabular}{|c|c|c|c|c|}
            \hline
            \textbf{Benchmark} & \makecell{\textbf{Baseline VM} \\ \textbf{(bytes)}} & \makecell{\textbf{Baseline NVM} \\ \textbf{(bytes)}} & \makecell{\textbf{\name VM} \\ \textbf{(bytes)}} & \makecell{\textbf{\name NVM} \\ \textbf{(bytes)}}\\
            \hline
            CRC (ll) 8Mhz & 812 & 1688 & 6 & 850 \\
            \hline
            CRC (ll) 16Mhz & 812 & 1688 & 26 & 850 \\
            \hline
            CRC (fr) 8Mhz & 812 & 1636 & 26 & 810 \\
            \hline
            CRC (fr) 16Mhz & 812 & 1636 & 30 & 810 \\
            \hline
            FFT (ll) 8Mhz & 16708 & 33514 & 64 & 29082 \\
            \hline
            FFT (ll) 16Mhz & 16708 & 33514 & 2188 & 29082 \\
            \hline
            AES (ll) 8Mhz & 1276 & 2614 & 40 & 1334 \\
            \hline
            AES (ll) 16Mhz & 1276 & 2614 & 42 & 1334 \\
            \hline
            AES (fr) 8Mhz & 1276 & 2614 & 58 & 1338 \\
            \hline
            AES (fr) 16Mhz & 1276 & 2614 & 62 & 1338 \\
            \hline
        \end{tabular}
    }
    \vspace{-3mm}
    \caption{Volatile memory (VM) and non-volatile memory (NVM) in \name against a baseline using \baselineV, \trigger, and either \latch or \freturn.}
    \vspace{-3mm}
    \label{fig:memory-usage-volatile}
\end{figure}

A given memory configuration is typically coupled to a dedicated \emph{strategy for placing checkpoint calls} in the code. 
Systems that only use volatile main memory, as in \baselineV, may place checkpoints using the \latch or \freturn placement strategies of Mementos~\cite{Mementos}.
Systems that only use non-volatile main memory, as in \baselineNV, place checkpoints using the strategy of Ratchet~\cite{ratchet}.
This entails identifying idempotent sections of the code and placing checkpoint calls at their boundaries.
We accordingly call this strategy \idempotent.
This ensures that intermittence anomalies are solved by construction, as re-execution of code only occurs across idempotent sections of code.

Once checkpoint calls are placed in the code, the \emph{checkpoint execution policy} dictates the conditions that possibly determine the actual execution of a checkpoint.
Indeed, a checkpoint call may systematically cause a checkpoint to be written on non-volatile memory, or rather probe the current energy levels first, for example, through an ADC query, and postpone the execution of a checkpoint if energy is deemed sufficient to continue without it.
\revision{The former kind of behavior, which we call \execute, is the case of Ratchet~\cite{ratchet}, Chinchilla~\cite{chinchilla}, and TICS~\cite{TICS} when it relies on checkpoints manually placed by developers}; the latter kind of behavior we call \trigger and reflects HarvOS~\cite{HarvOS} and Mementos~\cite{Mementos}.

\begin{figure}[t]
    \subfigtopskip = -2pt
    \subfigure{
        \hspace{-2mm}
    }
    \subfigure{
        \resizebox{.85\columnwidth}{!}{\includegraphics{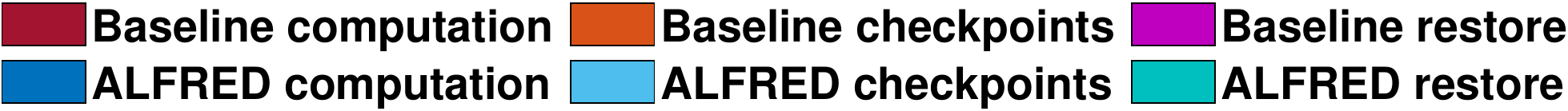}}
    }
    \setcounter{subfigure}{0}
    \subfigure[Energy consumption]{
        \centering
        \hspace{-5pt}
        \resizebox{0.8\columnwidth}{!}{\includegraphics{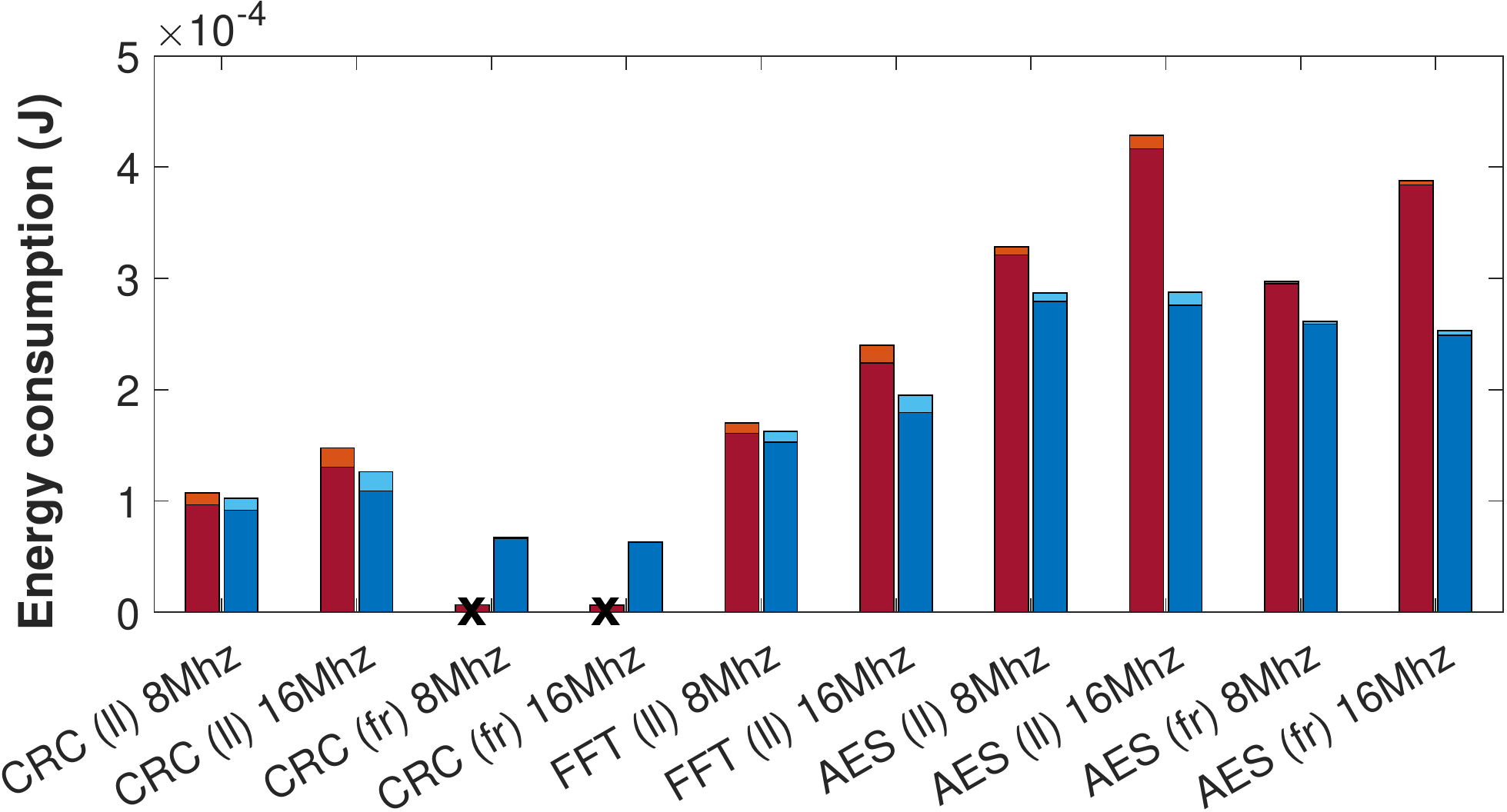}}
        \label{fig:mementos_nv_power_failures_energy}
    }
    \subfigure[Clock cycles]{
        \centering
        \resizebox{0.8\columnwidth}{!}{\includegraphics{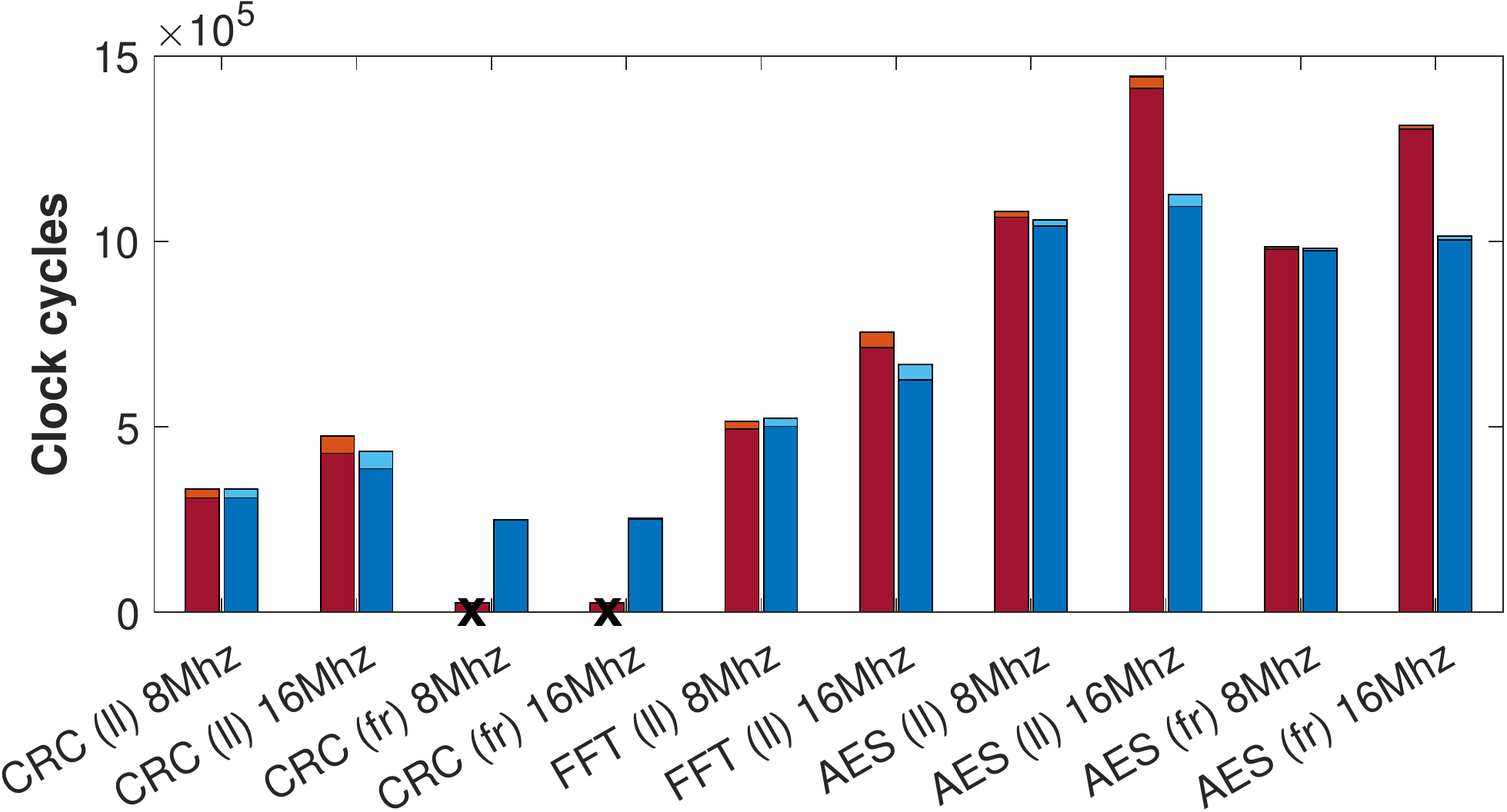}}
        \label{fig:mementos_nv_power_failures_cc}
    }
    \vspace{-3mm}
    \caption{Energy consumption and number of clock cycles comparing \name with a baseline using \baselineNV, \execute, and either \latch or \freturn. }
    \label{fig:mementos_nv_power_failures_energy_cc}
    \vspace{-3mm}
\end{figure}

A combination of memory configuration, strategy for placing checkpoint calls, and checkpoint execution policy represents the single baseline.
Note that not all combinations of these dimensions are necessarily meaningful.
For instance a \baselineNV memory configuration necessarily requires checkpoints to behave in an \execute manner, or the risk of intermittence anomalies would be too high and the overhead to address them correspondingly prohibitive~\cite{brokenTM}.
As \name requires as input a placement of state-saving operations, when comparing with a certain baseline we use the same such placement.
Moreover, being the FRAM performance and energy consumption affected by the MCU operating frequency~\cite{msp430fr5969}, we consider both $8Mhz$ and $16Mhz$ clock configurations.

\revision{Applications deployed onto battery-less devices typically consist in a sense-process-transmit loop~\cite{sensys20deployment, water-deployment-microbial-fuel-cell, soil-termoelectric}.
Checkpoint techniques and memory configurations mainly affect processing, whereas sensing and transmissions impose  the same overhead regardless of the former.}
\revision{For this reason, similar to related literature, we focus on processing functionality and consider a diverse set of benchmarks commonly used in intermittent computing~\cite{ratchet, clank, Hibernus, Hibernus++, Mementos, QuickRecall, maioli21ewsn}:} Cyclic Redundancy Check (CRC) for data integrity, Advanced Encryption Standard (AES) for data encryption, and Fast Fourier Transform (FFT) for signal analysis.
We use Clang version $7.1.0$ to compile their open-source implementations, as available in the MiBench2~\cite{MiBench2} suite, using the default compiler settings.
\revision{The binaries output by the compiler never exceed $30kB$.}

\begin{figure}[t]
    \subfigtopskip = -2pt
    \subfigure{
        \hspace{5pt}
    }
    \subfigure{
        \includegraphics[width=0.8\columnwidth]{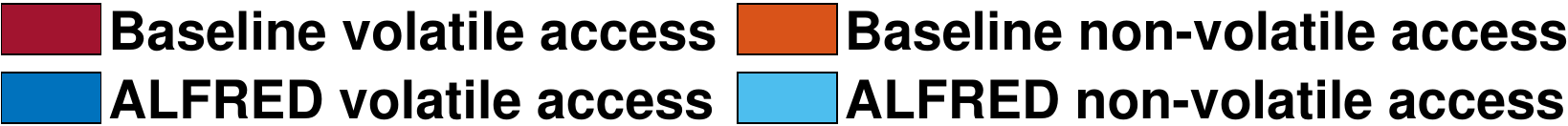}
    }
    \subfigure{
        \includegraphics[width=0.8\columnwidth]{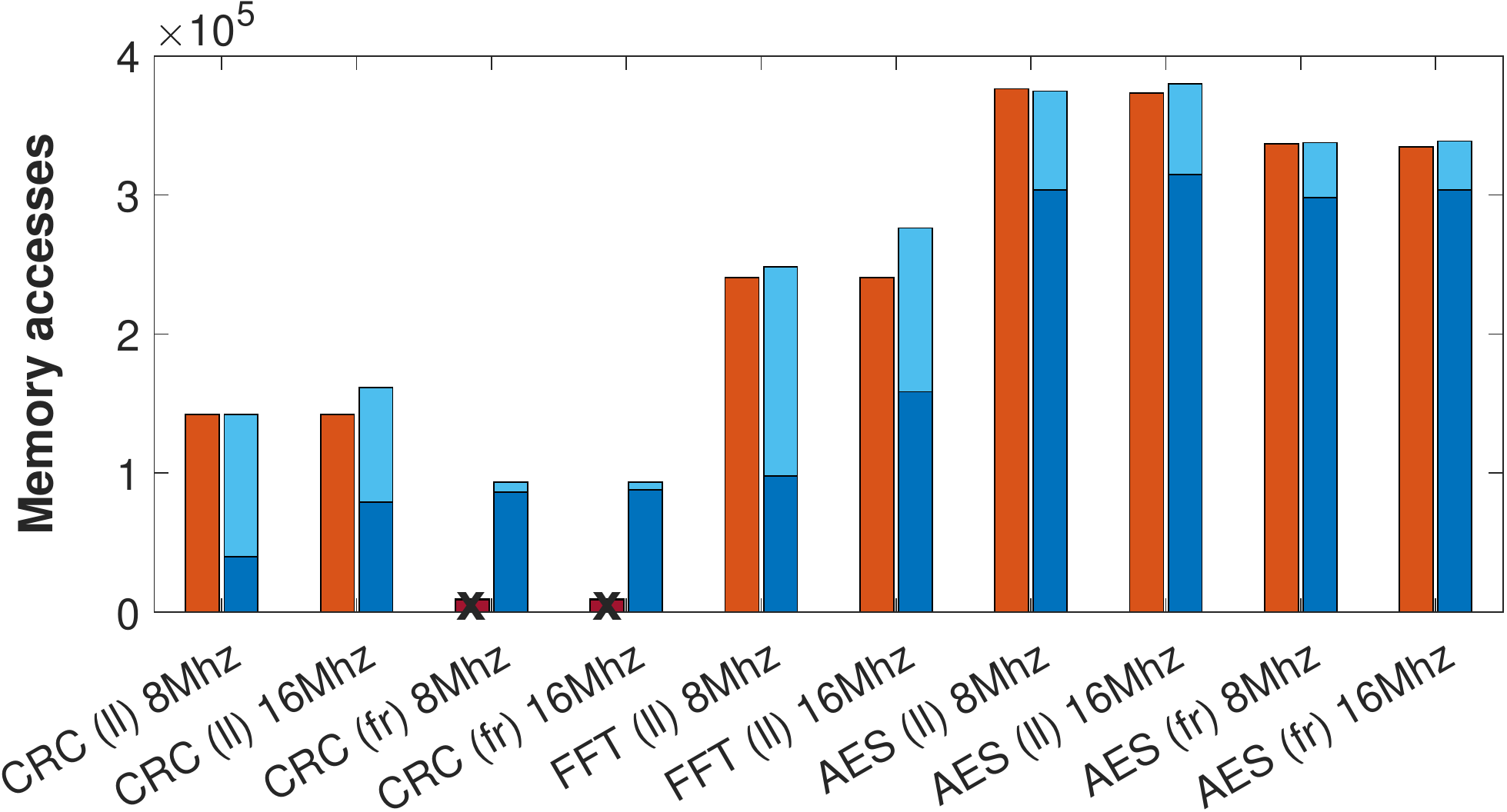}
    }
    \vspace{-3mm}
    \caption{Memory accesses comparing \name with a baseline using \baselineNV, \execute, and either \latch or \freturn. }
    \label{fig:mementos_nv_power_failures_memory}
    \vspace{-3mm}
\end{figure}

\fakepar{Metrics and energy patterns}
We focus on \emph{energy consumption} and \emph{number of clock cycles} necessary to complete a fixed workload.
\revision{Being harvested energy scarce, the former captures how battery-less devices perform when deployed and represents an indication of the perceived end-user performance~\cite{sensys20deployment, water-deployment-microbial-fuel-cell, soil-termoelectric}.}
\revision{The latter allows us to identify how the overhead of \name affects performance, as it mainly consists in the additional instructions required to address the compile-time uncertainties, as described in \secref{sec:uncertainty}.}
Note that the two metrics are not necessarily proportional, because non-volatile memory accesses may require extra clock cycles and consume more energy than accesses to volatile memory~\cite{msp430fr5969}.
\revision{\name may also introduce an overhead in the form of additional memory occupation, as the same data may need space in both volatile and non-volatile memory.
  To measure this, we keep track of the use of \emph{volatile/non-volatile memory spaces} during the execution.}
To gain a deeper insight into the performance trends we also record \emph{volatile/non-volatile memory accesses}, and the execution of \emph{checkpoint and restore operations}. 

\revision{Patterns of ambient energy harvesting may be simulated using IV surfaces~\cite{EKHO, SIREN} or by repetitively simulating power failures after a pre-determined number of executed clock cycles~\cite{ratchet, maioli20enssys}.}
\revision{The former makes simulated power failures happen at arbitrary points in times and provides little control on experiment executions, making it difficult to sweep the parameter space. 
The latter may be tuned according to statistical models, and offers better control on experiment executions by properly tuning model parameters.}
\revision{The behavior of \name is largely independent of the specific number of executed clock cycles between consecutive power failures; we therefore opt for the second option.}

\revision{We model an RF energy source.
To determine the number of executed clock cycles between two power failures, we rely on the existing measurements from ten real RF energy sources used for the evaluation of Mementos~\cite{Mementos}, which features a MCU configuration compatible with our setup.}
\revision{To evaluate multiple scenarios, including the worst and best possible ones, we execute each benchmark considering the minimum, average, and maximum number of executed clock cycles between power failures, modeled after the aforementioned real measurements.}
\revision{We report on the results obtained in the average scenario, as there is no sensible difference among the three scenarios.}
Note that, when using the \trigger strategy, we make sure that the last checkpoint call before a power failure is the one that does save a checkpoint, as this represents the same behavior of real scenarios.

\subsection{Results}

We consider three combinations of the design dimensions of \figref{fig:dimensions}.

\fakepar{Checkpointing from volatile memory} We begin comparing with a baseline configuration using \baselineV, \trigger, and either \latch or \freturn.
This configuration represents Mementos~\cite{Mementos} and solutions inspired by its design~\cite{HarvOS,chinchilla}.

\begin{figure}[t]
    \subfigtopskip = -2pt
    \subfigure{
        \centering
        \resizebox{0.85\columnwidth}{!}{\includegraphics{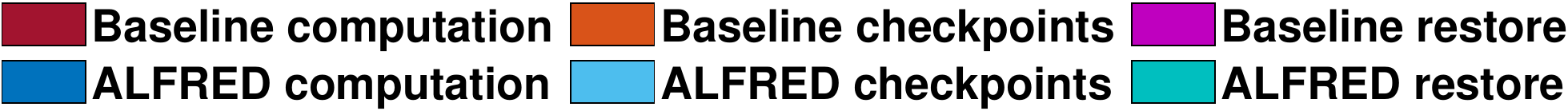}}
    }
    \setcounter{subfigure}{0}
    \subfigure[Energy consumption]{
        \centering
        \hspace{-2.5mm}
        \resizebox{.70\columnwidth}{!}{\includegraphics{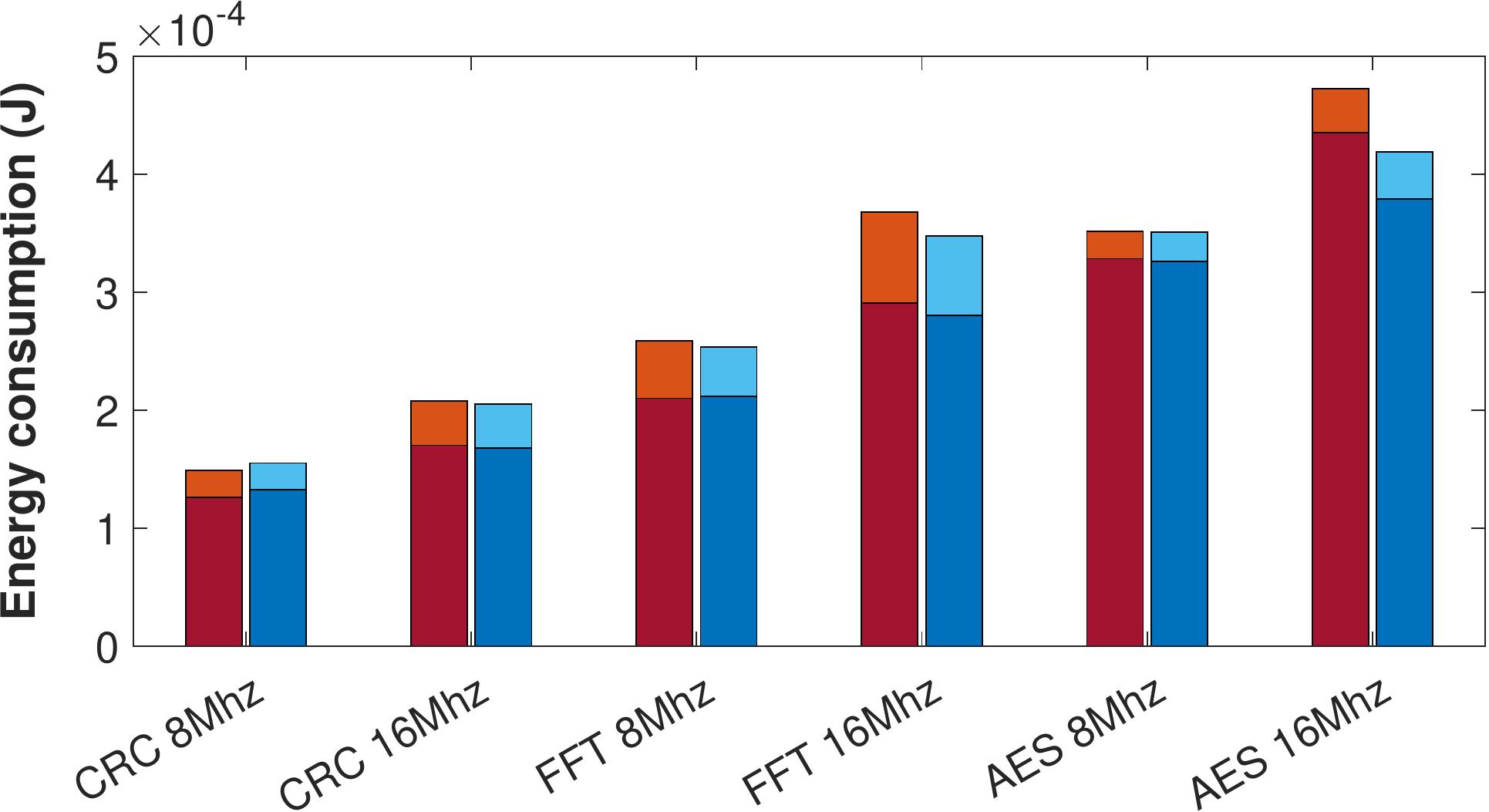}}
        \label{fig:ratchet_power_failures_energy}
    }
    \subfigure[Clock cycles]{
        \centering
        \resizebox{.70\columnwidth}{!}{\includegraphics{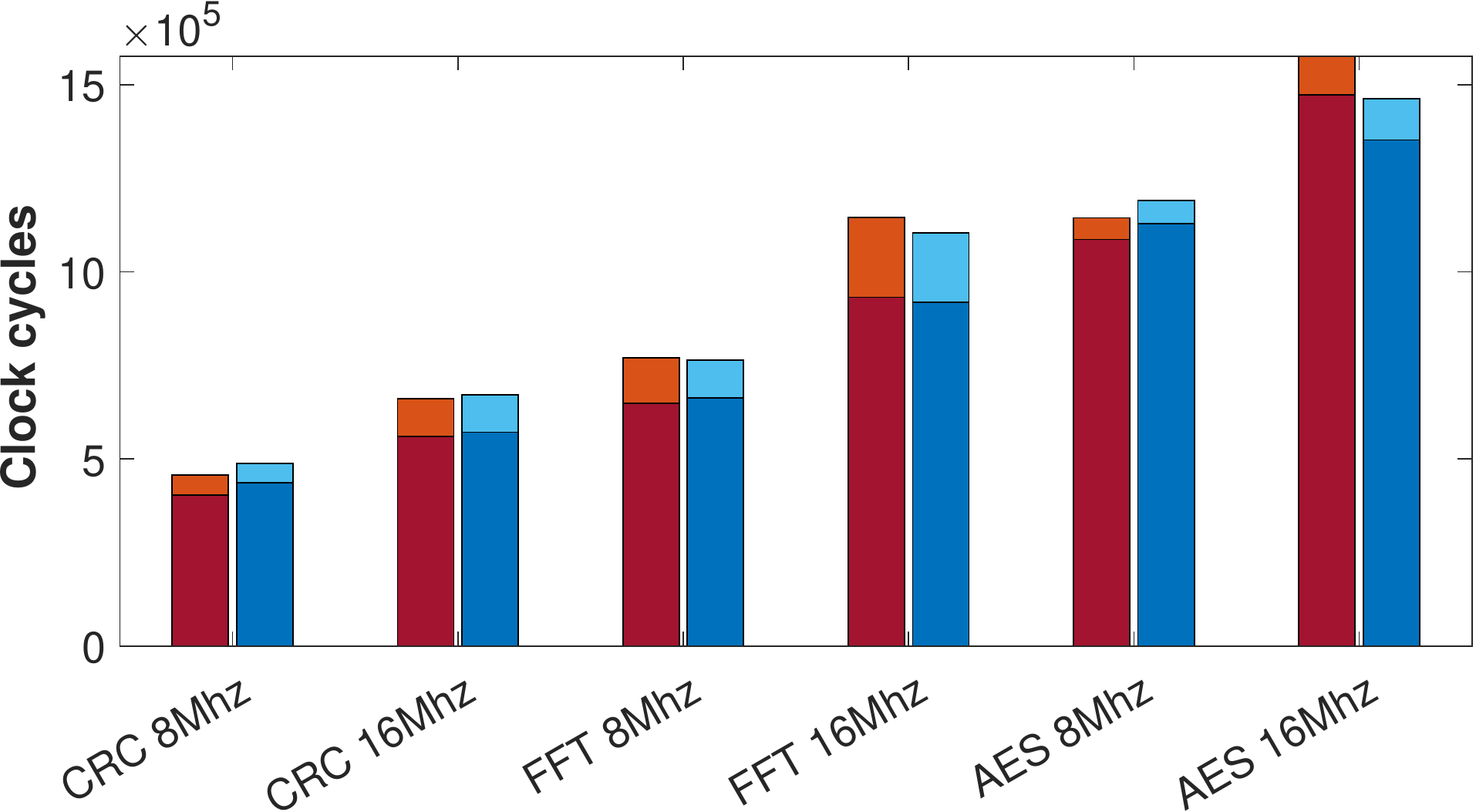}}
        \label{fig:ratchet_power_failures_cc}
    }
    \vspace{-3mm}
    \caption{Energy consumption and number of clock cycles comparing \name with a baseline using \baselineNV, \execute, and \idempotent.}
    \label{fig:ratchet_power_failures_energy_cc}
 \end{figure}

\begin{figure}[tb]
    \subfigtopskip = -2pt
    \subfigure{
        \includegraphics[width=0.75\columnwidth]{evaluation/figures/bar_memory/legend_baseline_nv.pdf}
    }
    \subfigure{
        \includegraphics[width=0.65\columnwidth]{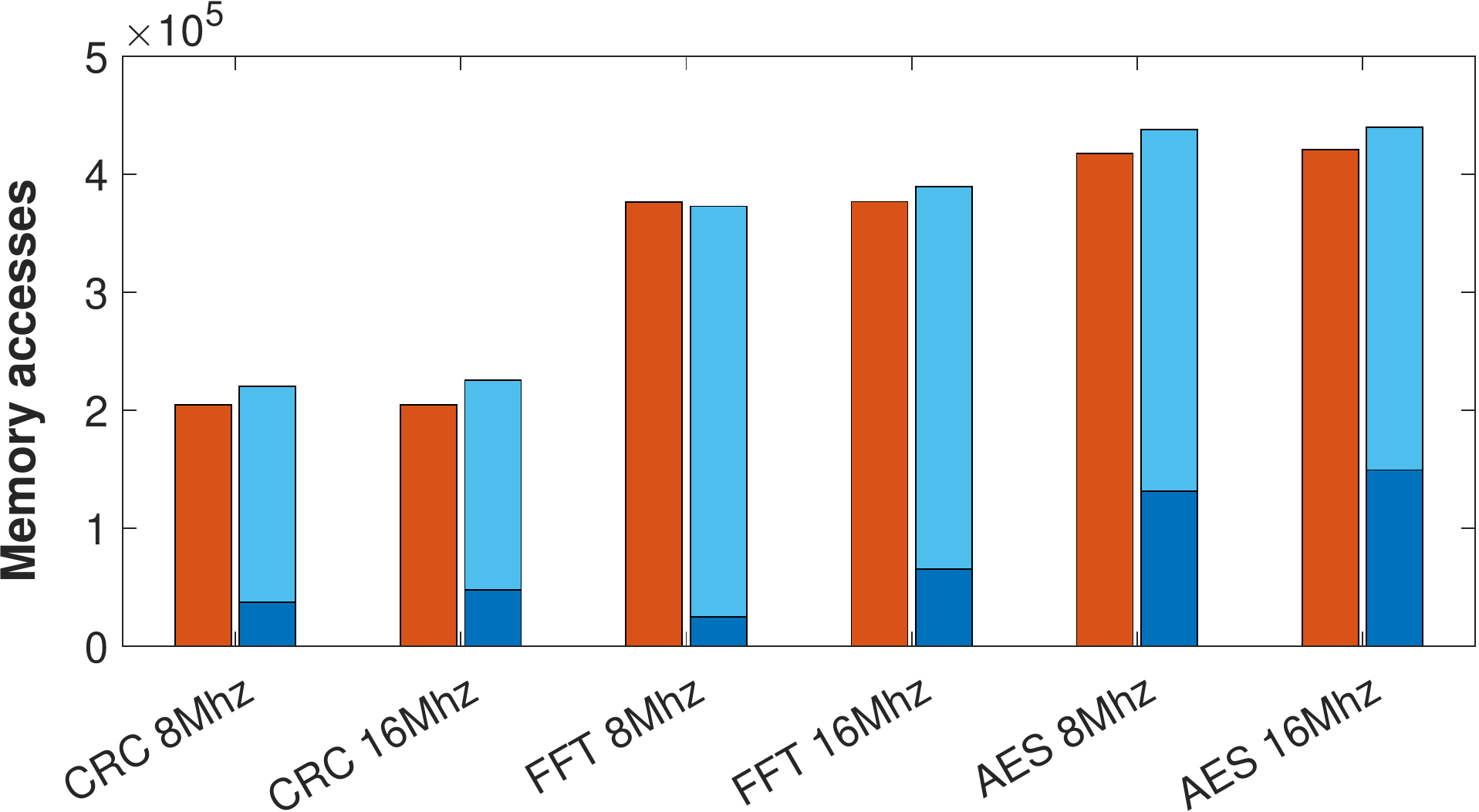}
    }
    \vspace{-3mm}
    \caption{Memory accesses comparing \name with a baseline using \baselineNV, \execute, and \idempotent.}
    \label{fig:ratchet_power_failures_memory}
    \vspace{-3mm}
\end{figure}

\figref{fig:mementos_power_failures_energy_cc} shows the results we obtain.
\figref{fig:mementos_power_failures_energy} shows how, depending on the benchmark, \name provides up to several-fold improvements in energy consumption to complete the fixed workload.
CRC computation is the simplest benchmark and has little state to make persistent.
The improvements are marginal here, especially when using \freturn as the checkpoint placement strategy, which is unsuited to the structure of the code in the first place.
The improvements grow as the complexity of the code increases.
Computing FFTs is the most complex benchmark we consider, and the improvements are largest in this case.
These observations are confirmed by the measurements of clock cycles, shown in \figref{fig:mementos_power_failures_cc}. 

 \figref{fig:mementos_power_failures_memory} provides a finer-grained view on the results in this specific setting.
 The small state in CRC corresponds to the fewest number of memory accesses, especially in volatile memory, as little data is to be made persistent to cross power failures.
 In both AES and FFT, \name greatly reduces the number of memory accesses.
 \revision{Checkpoint operations in these benchmarks must load a significant amount of data from volatile memory and copy it to non-volatile memory for creating the necessary persistent state.}
\revision{These accesses are not necessary in \name, as data is made persistent as soon as it becomes final; therefore, checkpoint operations do not process main memory, but only register file and program counter.}
 As for the nature of memory accesses, \name can promote, on average, $65\%$ of the accesses the baseline executes on non-volatile memory to volatile memory instead, with a minimum of $20\%$ in $CRC$ at $8Mhz$ with a \latch configuration and a maximum of $95\%$ in $CRC$ with a \freturn configuration.
 This is a key factor that grants \name better energy performance.

\begin{figure}
    \resizebox{\columnwidth}{!}{
        \begin{tabular}{|c|c|c|c|c|}
            \hline
            \textbf{Benchmark} & \makecell{\textbf{Baseline VM} \\ \textbf{Size (Bytes)}} & \makecell{\textbf{Baseline NVM} \\ \textbf{Size (Bytes)}} & \makecell{\textbf{\name VM} \\ \textbf{Size (Bytes)}} & \makecell{\textbf{\name NVM} \\ \textbf{Size (Bytes)}}\\
            \hline
            CRC 8Mhz & 0 & 826 & 6 & 854 \\
            \hline
            CRC 16Mhz & 0 & 826 & 16 & 854 \\
            \hline
            FFT 8Mhz & 0 & 16730 & 40 & 29074 \\
            \hline
            FFT 16Mhz & 0 & 16730 & 1116 & 29074 \\
            \hline
            AES 8Mhz & 0 & 1294 & 24 & 1342 \\
            \hline
            AES 16Mhz & 0 & 1294 & 40 & 1342 \\
            \hline
        \end{tabular}
    }
    \vspace{-3mm}
    \caption{Volatile memory (VM) and non-volatile memory (NVM) in \name against a baseline using \baselineNV, \execute, and \idempotent}
        \vspace{-4mm}
    \label{fig:memory-usage-non-volatile}
\end{figure}

\begin{figure*}[t]
    \subfigtopskip = -2pt
    \subfigure{
        \hspace{138pt}
    }
    \subfigure{
        \resizebox{0.32\columnwidth}{!}{\includegraphics{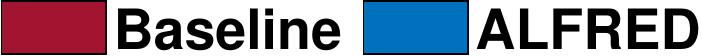}}
    }
    \subfigure{
        \hspace{120pt}
    }
    \setcounter{subfigure}{0}
    \subfigure[Baseline using \baselineV, \trigger, \latch \hspace{9mm} or \freturn]{
        \centering
        \resizebox{0.68\columnwidth}{!}{\includegraphics{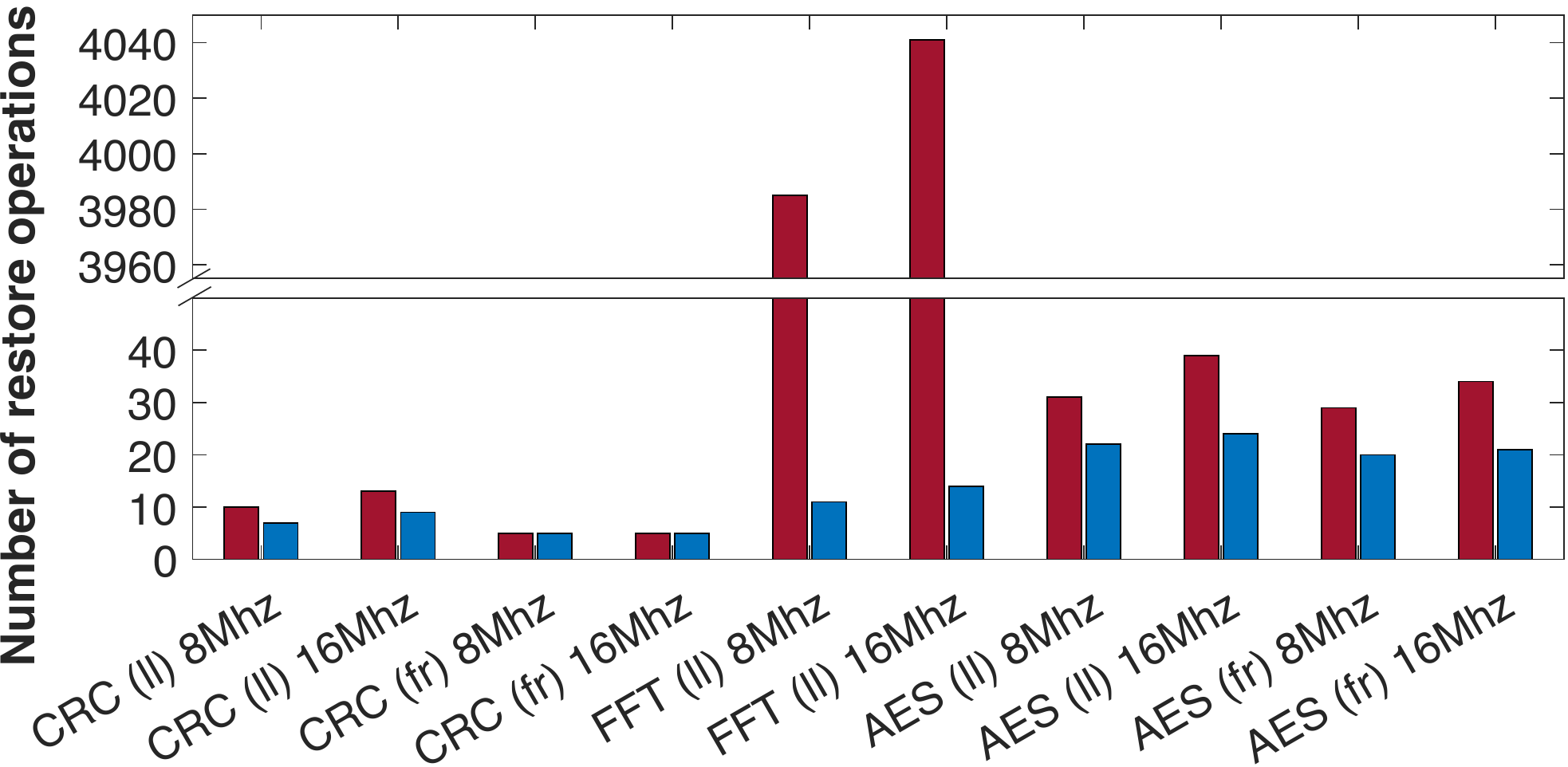}}
        \label{fig:mementos_n_power_failures}
    }
    \subfigure[Baseline using \baselineNV, \execute, \latch \hspace{9mm} or \freturn]{
        \centering
        \resizebox{0.67\columnwidth}{!}{\includegraphics{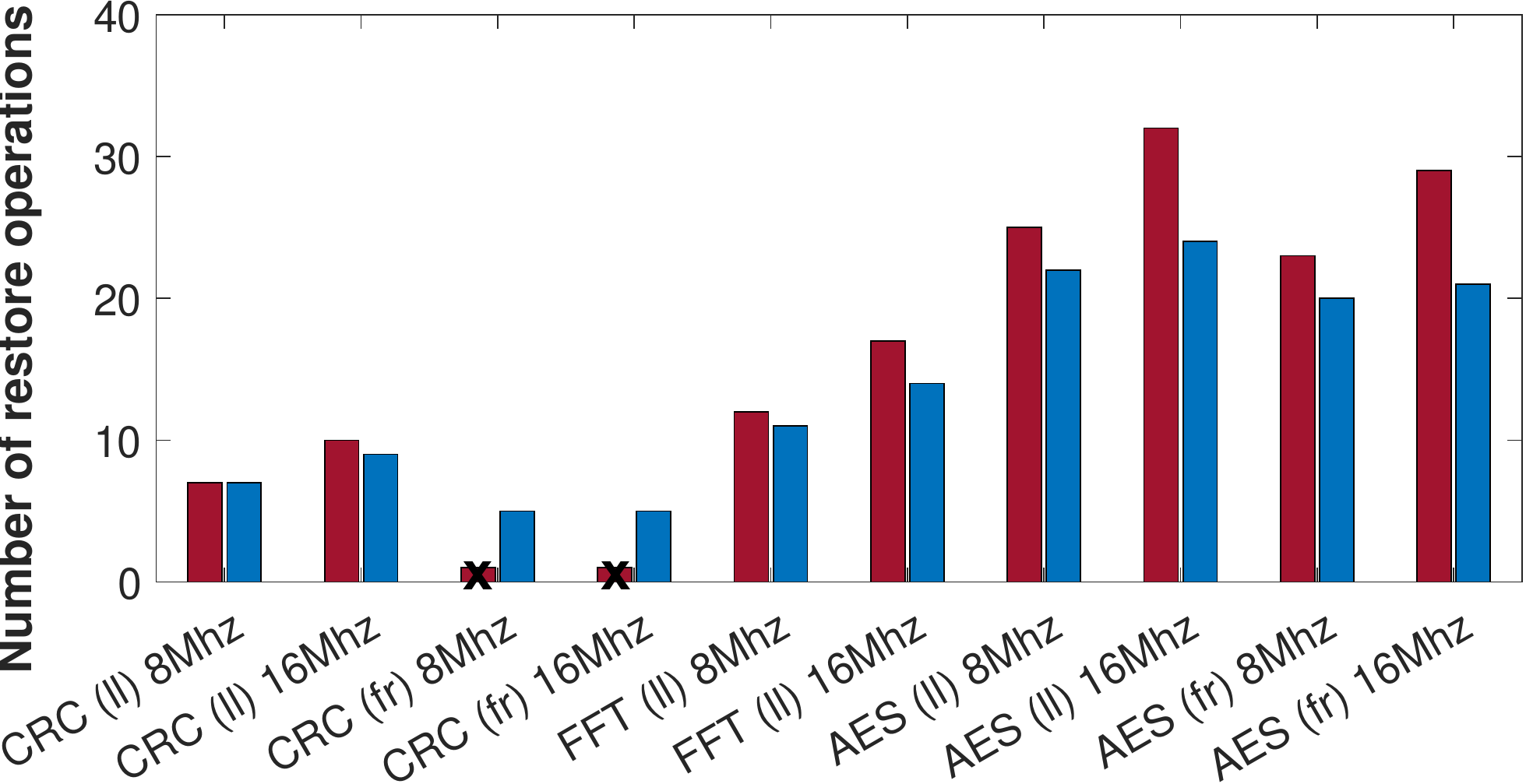}}
        \label{fig:mementos_nv_n_power_failures}
    }
    \subfigure[Baseline using \baselineNV, \execute, \idempotent]{
        \centering
        \resizebox{0.67\columnwidth}{!}{\includegraphics{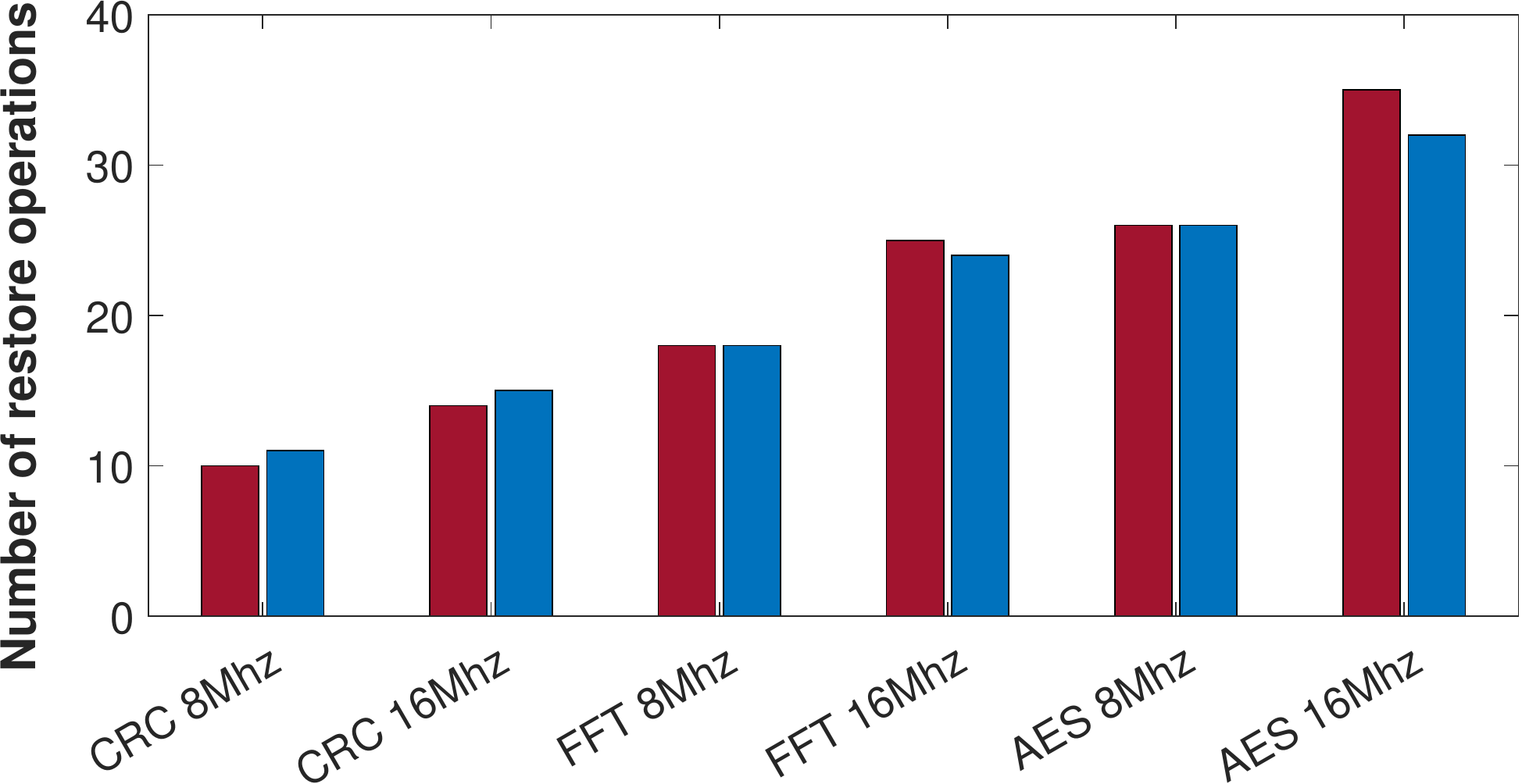}}
        \label{fig:ratchet_n_power_failures}
    }
    \vspace{-4mm}
    \caption{Restore operations to complete the fixed workload in \name compared to the three baselines. }
    \vspace{-3mm}
    \label{fig:n_power_failures}
\end{figure*}

\revision{\figref{fig:memory-usage-volatile} reports on the use of volatile/non-volatile memory.}
\revision{In the baseline, the state to be preserved across power failures includes the entire volatile memory, the register file, and special registers.}
\revision{Requiring to double-buffer the state saved to non-volatile memory, its use in the baseline amounts to more than double the use of volatile memory.}
\revision{In both CRC and AES, \name requires to double buffer less than \emph{4\%} of the program state to avoid intermittence anomalies, resulting in a drastically lower use of non-volatile memory.}
\revision{Interestingly, despite a significant improvement in energy consumption, \name promotes very few memory locations to volatile memory.}
\revision{These correspond to the memory locations that are most frequently accessed, as shown in \figref{fig:mementos_power_failures_memory}.}

\fakepar{Moving to non-volatile memory}
\figref{fig:mementos_nv_power_failures_energy_cc} shows the results we obtain comparing with configuration using \baselineNV, \execute, and either \latch or \freturn.
This combination represents a hybrid solution combining features of several existing systems~\cite{chinchilla, HarvOS, Mementos}.
As \latch and \freturn do not necessarily guarantee that intermittence anomalies cannot occur, we lend our versioning technique, described in \secref{sec:impl}, to the baseline.
The major difference between \name and the baseline, therefore, is in the use of volatile or non-volatile memory.

\figref{fig:mementos_nv_power_failures_energy} shows that the program transformations we devise are effective at improving the energy performance of intermittent programs.
Significant improvements are visible across all benchmarks.
Configurations exist where the baseline cannot complete the workload using the energy patterns we consider, as in the case of the CRC benchmark when using \freturn to place checkpoints.
In contrast, \name reduces energy consumption to an extent that allows the workload to successfully complete.

The corresponding results in the number of executed clock cycles, shown in \figref{fig:mementos_nv_power_failures_cc}, enables a further observation.
When running at $16Mhz$, the baseline shows a significant increase of clock cycles, at least $20\%$ with respect to the same benchmark running at $8Mhz$.
The cause of this increase is in the extra clock cycles required to access the FRAM when the MCU is clocked at $16Mhz$.
In the same scenarios, \name shows a lower increase of clock cycles when comparing the $8Mhz$ and $16Mhz$ configurations, especially in the AES benchmark.
Rather than massively employing non-volatile memory, \name switches to volatile memory whenever possible within a computation interval.
This not only reduces the clock cycles spent waiting for non-volatile memory access, but also enables energy savings in the operations that involve temporary data or intermediate results that do not need persistency.

\figref{fig:mementos_nv_power_failures_memory} confirms this reasoning, showing that \name promotes an average of $65\%$ of the non-volatile memory accesses in the baseline to the more energy-efficient volatile memory.
This functionality grants \name the completion of the CRC benchmark when using the \freturn configuration.
As the baseline directs all memory accesses to non-volatile memory, the resulting energy consumption causes CRC to be stuck in a livelock, as energy is insufficient to reach a checkpoint that would enable forward progress.
This situation is called ``non-termination'' bug~\cite{CleanCut}.
Instead, in the case of CRC, \name promotes more than $95\%$ of the non-volatile memory accesses in the baseline to volatile memory.
This significantly reduces the energy consumption of memory accesses to an extent that allows \name to complete the workload.

\revision{Note that the use non-volatile memory in the baseline is the same as \name, shown in \figref{fig:memory-usage-volatile}, as they employ the same technique to avoid intermittence anomalies.}
\revision{The difference in memory occupation consists in the data that \name allocates onto volatile memory, which ultimately yields lower energy consumption.}

\fakepar{Ruling out intermittence anomalies} We compare the performance of \name with a configuration using \baselineNV, \execute, and \idempotent, as in Ratchet~\cite{ratchet}.
Because of the specific placement of checkpoint calls and the \execute policy, intermittence anomalies cannot occur by construction.
\name and the baseline here only differ in memory management.
  
\figref{fig:ratchet_power_failures_energy_cc} shows the results.
\figref{fig:ratchet_power_failures_energy} illustrates the performance in energy consumption; this time, the improvements of \name are generally less marked than those seen when using \latch or \freturn to place checkpoints.
The results in the number of executed clock cycles are coherent with these trends, as illustrated in \figref{fig:ratchet_power_failures_cc}.
This is because \idempotent tends to create much shorter computation intervals, sometimes solely worth a few instructions; therefore, \name has fewer opportunities to operate on the energy-efficient volatile memory.
\name still improves the energy efficiency overall, especially for the AES benchmark and the configurations running at $16Mhz$.
At this clock frequency, non-volatile memory operations induce higher overhead due to the necessary wait cycles.
Sparing operations on non-volatile memory allows the system not to pay this overhead.

\revision{\figref{fig:ratchet_power_failures_memory} and \figref{fig:memory-usage-non-volatile} provide an assessment on \name's ability to employ volatile memory whenever convenient.}
\name promotes the use of volatile memory from the non-volatile use in the baseline in up to 30\% of the cases.
The impact of this, however, is more limited here because of the shorter computation intervals, as discussed above.
In this plot, it also becomes apparent that sometimes, the total number of memory accesses in \name is higher than in the baseline.
This is a combined effect of the program transformation techniques of \secref{sec:memory-principles} and of the normalization passes in \secref{sec:uncertainty}.
The increase in the total number of memory accesses, however, does not yield a penalty in energy consumption, as a significant fraction of these added accesses operate on volatile memory.

\revision{These results are confirmed in \figref{fig:memory-usage-non-volatile}.}
\revision{Despite being the program partitioned in non-idempotent code sections, our techniques to address compile-time uncertainties introduce intermittence anomalies that require \name to double-buffer a portion of the program state.}
\revision{This situation is particularly evident with FFT.}
\revision{\figref{fig:memory-usage-non-volatile} provides additional evidence of how \name employs volatile memory for frequently-accessed data, which ultimately yields lower energy consumption across all benchmarks executed at $16Mhz$.}

\fakepar{Restore operations}
We complete the discussion by showing in \figref{fig:n_power_failures} the number of restore operations executed in \name compared to those in the three baseline configurations we consider.

The plots demonstrate that the better energy efficiency provided by \name allows the system to restore the state less times.
This trend is especially visible in \figref{fig:mementos_n_power_failures} and \figref{fig:mementos_nv_n_power_failures}.
As a result, \name shifts the available energy budget to useful application processing, leading to workloads that finish sooner compared to the performance offered by the baselines.

\section{Conclusion}
\label{sec:conclusion}

\name is a virtual memory abstraction for intermittent computing that relieves programmers from the burden of explicitly managing application state across memory facilities, and efficiently employ volatile and non-volatile memory to improve energy consumption, while ensuring forward progress across power failures.
The mapping from virtual to volatile or non-volatile memory is decided at compile time through a series of program transformations.
These aim at using volatile memory whenever possible because of the lower energy consumption, resorting to non-volatile memory whenever necessary to ensure forward progress.
In contrast to existing works, the memory mapping is not fixed at variable level, but is adjusted at different places in the code for the same data item, based on read/write patterns and program structure.
To apply the transformations to arbitrary programs, we use code normalization techniques, a dedicated memory layout, and a tightly integrated solution to address possible intermittence anomalies.
Our evaluation indicates that, depending on the workload, \name provides several-fold improvements in energy consumption compared to the multiple baselines we consider, leading to a similar improvement in the number of restore operations required to complete a fixed workload.

\clearpage

\balance
\bibliographystyle{ACM-Reference-Format}
\bibliography{bibliography}

\end{document}